\documentclass[12pt,a4paper]{article}
\usepackage{jheppub}
\usepackage[dvipsnames]{xcolor}
\usepackage{tikz}
\usepackage{amsthm,amsbsy,amsfonts,mathrsfs,enumerate,float,wrapfig,amsmath,amssymb,cleveref,hyperref}
\usepackage[colorinlistoftodos,prependcaption,textsize=scriptsize]{todonotes}
\allowdisplaybreaks

\newcommand{\nn}{\nonumber}

\tikzset{>=stealth}
\usetikzlibrary{decorations.pathreplacing,decorations.markings}
\usetikzlibrary{arrows, arrows.meta}
\title{Weyl symmetry in $(D_4,D_4)$ conformal matter on a circle}
\author{Xing-Yue Wei}
\affiliation{Yau Mathematical Sciences Center, Tsinghua University, Beijing, 100084, China}
\emailAdd{xingyue\_wei@mail.tsinghua.edu.cn}

\abstract{We study Weyl symmetry in quadrivalently glued 5-brane webs of rank $N$ $(D_4,D_4)$ conformal matter theories on a circle. We find that these theories all have affine $E_8$ Weyl symmetry in their brane webs, which indicates that they all have affine $E_8$ global symmetry. When $N\geq 2$, the theory has 64 different sets of affine $E_8$ invariant Coulomb branch parameters. }

\begin{document}
\maketitle

\section{Introduction}\label{sec:intro}
$(p,q)$ 5-brane web \cite{Aharony:1997bh} in type IIB string theory is a powerful tool to construct 5d $\mathcal N=1$ SCFTs and 6d $\mathcal N=(1,0)$ SCFTs compactified on a circle. The 5-brane web describes the low energy effective gauge theories of the SCFTs on the Coulomb branch, and the global symmetry of the SCFTs is broken in the low energy theories, or in other words the global symmetry of the low energy theories is enhanced in the SCFTs in the UV. The typical examples are 5d $SU(2)$ gauge theories with $N_f=0,\cdots,7$ flavors whose UV completions are 5d SCFTs with $E_{N_f+1}$ global symmetry \cite{Seiberg:1996bd}, the global symmetry of the low energy gauge theories is $SO(2N_f)\times U(1)$ where $SO(2N_f)$ is the flavor symmetry and $U(1)$ is the instanton symmetry, and this global symmetry is enhanced to $E_{N_f+1}$ in the UV. Direct check of the enhanced $E_{N_f+1}$ symmetry in the superconformal index was performed in \cite{Kim:2012gu} by localization method. In the $(p,q)$ 5-brane web construction, such global symmetry enhancement can also be observed.
Although the global symmetry is broken in the low energy gauge theories, the Weyl symmetry of the global symmetry group still remains. Thus we can deduce the global symmetry of the SCFTs by figuring out the corresponding Weyl symmetry in the low energy theories. The $(p,q)$ 5-brane web of the low energy gauge theory exactly captures the information of the Weyl symmetry of the global symmetry group of the SCFT. In \cite{Mitev:2014jza}, the enhanced $E_{N_f+1}$ symmetry was observed in the brane web by combination of the $SO(2N_f)$ flavor symmetry and ``fiber-base symmetry", we take 5d $SU(2)+2\mathbf F$ as an example to illustrate how to figure out the $E_3$ Weyl symmetry in the brane web. 

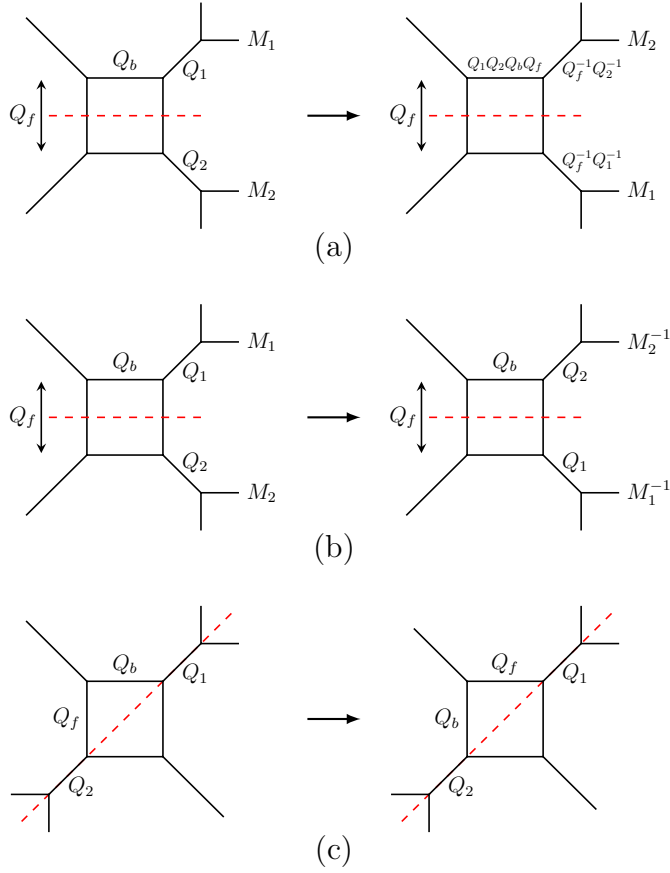
\begin{figure}[htbp]
\centering
\begin{tikzpicture}[scale=0.5]
	\draw[semithick] (-1,1)--(1,1)
	(1,1)--(1,-1)
	(1,-1)--(-1,-1)
	(-1,-1)--(-1,1)
	(1,1)--(2,2)
	(2,2)--(3,2)
	(2,2)--(2,3)
	(1,-1)--(2,-2)
	(2,-2)--(3,-2)
	(2,-2)--(2,-3)
	(-1,1)--(-2.6,2.6)
	(-1,-1)--(-2.6,-2.6);
	\node[scale=0.7] at (1.85,1.2) {$Q_1$};
	\node[scale=0.7] at (1.85,-1.2) {$Q_2$};
	\node[scale=0.7] at (0,1.45) {$Q_b$};
	\node[scale=0.7] at (-2.68,0) {$Q_f$};
	\node[scale=0.7] at (3,2) [right]{$M_1$};
	\node[scale=0.7] at (3,-2) [right]{$M_2$};
	\draw[red,dashed,semithick]
	(-2,0)--(2,0);
	
	\draw[<->,semithick]
	(-2.2,0.95)--(-2.2,-0.95)
	;
	
	\draw [>=latex,->,thick] (4.8,0)--(6.2,0);
	\node at (5.5,-3.5) {(a)};
	
	\draw[semithick] (-1+10,1)--(1+10,1)
	(1+10,1)--(1+10,-1)
	(1+10,-1)--(-1+10,-1)
	(-1+10,-1)--(-1+10,1)
	(1+10,1)--(2+10,2)
	(2+10,2)--(3+10,2)
	(2+10,2)--(2+10,3)
	(1+10,-1)--(2+10,-2)
	(2+10,-2)--(3+10,-2)
	(2+10,-2)--(2+10,-3)
	(-1+10,1)--(-2.6+10,2.6)
	(-1+10,-1)--(-2.6+10,-2.6);
	\node[scale=0.55] at (2.3+10,1.2) {$Q_f^{-1}Q_2^{-1}$};
	\node[scale=0.55] at (2.3+10,-1.2) {$Q_f^{-1}Q_1^{-1}$};
	\node[scale=0.5] at (0+10,1.35) {$Q_1Q_2Q_bQ_f$};
	\node[scale=0.7] at (-2.68+10,0) {$Q_f$};
	\node[scale=0.7] at (3+10,2) [right]{$M_2$};
	\node[scale=0.7] at (3+10,-2) [right]{$M_1$};
	\draw[red,dashed,semithick]
	(-2+10,0)--(2+10,0);
	\draw[<->,semithick]
	(-2.2+10,0.95)--(-2.2+10,-0.95)
	;
	
	\begin{scope}[shift={(0,-8)}]
	\draw[semithick] (-1,1)--(1,1)
	(1,1)--(1,-1)
	(1,-1)--(-1,-1)
	(-1,-1)--(-1,1)
	(1,1)--(2,2)
	(2,2)--(3,2)
	(2,2)--(2,3)
	(1,-1)--(2,-2)
	(2,-2)--(3,-2)
	(2,-2)--(2,-3)
	(-1,1)--(-2.6,2.6)
	(-1,-1)--(-2.6,-2.6);
	\node[scale=0.7] at (1.85,1.2) {$Q_1$};
	\node[scale=0.7] at (1.85,-1.2) {$Q_2$};
	\node[scale=0.7] at (0,1.45) {$Q_b$};
	\node[scale=0.7] at (-2.68,0) {$Q_f$};
	\node[scale=0.7] at (3,2) [right]{$M_1$};
	\node[scale=0.7] at (3,-2) [right]{$M_2$};
	
	\draw [>=latex,->,thick] (4.8,0)--(6.2,0);
	\node at (5.5,-3.5) {(b)};
	\draw[red,dashed,semithick]
	(-2,0)--(2,0);
	\draw[<->,semithick]
	(-2.2,0.95)--(-2.2,-0.95)
	;
	
	\draw[semithick] (-1+10,1)--(1+10,1)
	(1+10,1)--(1+10,-1)
	(1+10,-1)--(-1+10,-1)
	(-1+10,-1)--(-1+10,1)
	(1+10,1)--(2+10,2)
	(2+10,2)--(3+10,2)
	(2+10,2)--(2+10,3)
	(1+10,-1)--(2+10,-2)
	(2+10,-2)--(3+10,-2)
	(2+10,-2)--(2+10,-3)
	(-1+10,1)--(-2.6+10,2.6)
	(-1+10,-1)--(-2.6+10,-2.6);
	\node[scale=0.7] at (1.85+10,1.2) {$Q_2$};
	\node[scale=0.7] at (1.85+10,-1.2) {$Q_1$};
	\node[scale=0.7] at (0+10,1.45) {$Q_b$};
	\node[scale=0.7] at (-2.68+10,0) {$Q_f$};
	\node[scale=0.7] at (3+10,2) [right]{$M_2^{-1}$};
	\node[scale=0.7] at (3+10,-2) [right]{$M_1^{-1}$};
	\draw[red,dashed,semithick]
	(-2+10,0)--(2+10,0);
	\draw[<->,semithick]
	(-2.2+10,0.95)--(-2.2+10,-0.95)
	;
	\end{scope}
	
	\begin{scope}[shift={(0,-16)}]
	\draw[dashed,red,semithick]
	(2.8,2.8)--(-2.8,-2.8);
	\draw[semithick] (-1,1)--(1,1)
	(1,1)--(1,-1)
	(1,-1)--(-1,-1)
	(-1,-1)--(-1,1)
	(1,1)--(2,2)
	(2,2)--(3,2)
	(2,2)--(2,3)
	(1,-1)--(2.6,-2.6)
	(-1,1)--(-2.6,2.6)
	(-1,-1)--(-2,-2)
	(-2,-2)--(-2,-3)
	(-2,-2)--(-3,-2);
	\node[scale=0.7] at (1.85,1.2) {$Q_1$};
	\node[scale=0.7] at (-1.15,-1.8) {$Q_2$};
	\node[scale=0.7] at (0,1.45) {$Q_b$};
	\node[scale=0.7] at (-1.53,0) {$Q_f$};

	\draw [>=latex,->,thick] (4.8,0)--(6.2,0);
	\node at (5.5,-3.5) {(c)};
	
	\draw[dashed,red,semithick]
	(2.8+10,2.8)--(-2.8+10,-2.8);
	
	\draw[semithick] (-1+10,1)--(1+10,1)
	(1+10,1)--(1+10,-1)
	(1+10,-1)--(-1+10,-1)
	(-1+10,-1)--(-1+10,1)
	(1+10,1)--(2+10,2)
	(2+10,2)--(3+10,2)
	(2+10,2)--(2+10,3)
	(1+10,-1)--(2.4+10,-2.4)
	(-2+10,-2)--(-3+10,-2)
	(-2+10,-2)--(-2+10,-3)
	(-1+10,1)--(-2.4+10,2.4)
	(-1+10,-1)--(-2+10,-2);
	\node[scale=0.7] at (1.85+10,1.2) {$Q_1$};
	\node[scale=0.7] at (-1.15+10,-1.8) {$Q_2$};
	\node[scale=0.7] at (0+10,1.45) {$Q_f$};
	\node[scale=0.7] at (-1.53+10,0) {$Q_b$};
	
	\end{scope}
\end{tikzpicture}
\caption{Weyl reflections in 5-brane web of 5d $SU(2)+2\mathbf F$. }
\label{fig:SU2with2F}
\end{figure}

The 5-brane web of 5d $SU(2)+2\mathbf F$ is depicted in figure \ref{fig:SU2with2F}. In figure \ref{fig:SU2with2F}(a), the transformation on the brane web from left to right corresponds to a flop transition that exchanges\footnote{Move the flavor brane labelled by the mass parameter $M_1$ down and the flavor brane labelled by the mass parameter $M_2$ up. } the two flavor branes, the K\"ahler parameters are transformed in the following way:
\begin{equation}
	\mathbf{R}_1=\big\{Q_1\to Q_f^{-1}Q_2^{-1},\ Q_2\to Q_f^{-1}Q_1^{-1},\ Q_b\to Q_1Q_2Q_bQ_f\big\}. 
\end{equation}
Figure \ref{fig:SU2with2F}(b) corresponds to the global manipulation that reflect the brane web by 180 degree along the red dashed line which is the center of Coulomb branch, the corresponding transformation\footnote{
Using the physical parameters $M_1=Q_1\sqrt{Q_f},M_2=\frac{1}{Q_2\sqrt{Q_f}},u=\sqrt{\frac{Q_1 Q_2 Q_b^2}{Q_f}},A=\sqrt{Q_f}$ where $M_1,M_2$ are the mass parameters, $u$ is the instanton factor and $A$ is the Coulomb branch parameter, 
the transformations in figure \ref{fig:SU2with2F}(a) and \ref{fig:SU2with2F}(b) can be expressed as $\mathbf{R}_1=\{M_1\to M_2,M_2\to M_1\}$ and $\mathbf{R}_2=\{M_1\to M_2^{-1},M_2\to M_1^{-1}\}$ respectively, which together form the complete Weyl symmetry of the $SO(4)$ flavor symmetry. } on the K\"ahler parameters is
\begin{equation}
	\mathbf{R}_2=\big\{Q_1\to Q_2,\ Q_2\to Q_1\big\}. 
\end{equation}
Note that the two transformations on the brane webs in figure \ref{fig:SU2with2F}(a) and \ref{fig:SU2with2F}(b) both keep the shape of the brane web invariant, which leads to the K\"ahler parameter transformations. 

Figure \ref{fig:SU2with2F}(c) is another brane web for 5d $SU(2)+2\mathbf F$ obtained by moving one of the flavor branes to the left via Hanany-Witten transition, the transformation on the brane web is a global manipulation that reflects the brane web by 180 degree along the diagonal red dashed line. This manipulation also keeps the shape of the brane web invariant and it can be seen as a combination of a 90 degree rotation and a 180 degree reflection along the horizontal line. The corresponding transformation on the K\"ahler parameters is
\begin{equation}
	\mathbf{R}_3=\big\{Q_b\to Q_f,\ Q_f\to Q_b\big\}. 
\end{equation}

In order to see the manifest $E_3$ symmetry, we define the following parameters:
\begin{equation}
	P_1\equiv(Q_1Q_2Q_f)^{2/3},\ P_2\equiv\left(\frac{Q_b}{Q_f}\right)^{2/3},\ P_3\equiv\frac{Q_2}{Q_1},\ A\equiv\sqrt{Q_f}, 
\end{equation}
in which $P_1,P_2,P_3$ are the fugacities for the enhanced global symmetry, $A$ is the Coulomb branch parameter. 

The transformations $\mathbf{R}_1,\mathbf{R}_2,\mathbf{R}_3$ represented by the new parameters are the following:
\begin{align}
	&\mathbf{R}_1=\big\{P_1\to P_1^{-1},\ P_2\to P_1P_2\big\},\\
	&\mathbf{R}_2=\big\{P_3\to P_3^{-1}\big\},\\
	&\mathbf{R}_3=\big\{P_1\to P_1P_2,\ P_2\to P_2^{-1}\ \ ; \ \ A\to P_2^{3/4}A\big\},\label{eq:R3inE3}
\end{align}
where the semicolon in equation \eqref{eq:R3inE3} is used to distinguish between transformations on the global symmetry parameters and transformations on the local symmetry parameters. 
\begin{figure}[htbp]
\centering
	\begin{tikzpicture}
	\draw 
	(6,0) circle (0.2)
	(7,0) circle (0.2)
	(5,1) circle (0.2)
	
	(6.2,0) to (6.8,0);

	\node at (6,-0.5) {1};
	\node at (7,-0.5) {2};
	\node at (5.5,1) {3};
\end{tikzpicture}
\caption{Dynkin diagram of $E_3=SU(2)\times SU(3)$. }
\label{fig:E3-DK}
\end{figure}
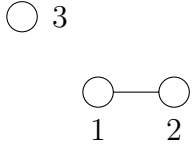

Figure \ref{fig:E3-DK} is the Dynkin diagram of $E_3=SU(2)\times SU(3)$, the $E_3$ Weyl reflections in $\alpha$-basis exactly correspond to the $P_i$ parameter transformations in $\mathbf{R}_1,\mathbf{R}_2,\mathbf{R}_3$, and $P_i$ is the fugacity corresponding to the $i$-th simple root in the Dynkin diagram of $E_3$. 

With the complete $E_3$ Weyl reflections\footnote{Precisely speaking, the $E_3$ Weyl reflections only involve the transformations on $P_i$, however $\mathbf{R}_3$ also involves the transformation on the Coulomb branch parameter $A$. But we still call $\mathbf{R}_3$ as a Weyl reflection. } $\mathbf{R}_1,\mathbf{R}_2,\mathbf{R}_3$, we can further compute the so-called invariant Coulomb branch parameter \cite{Mitev:2014jza}, which is a modification of the usual Coulomb branch parameter such that it is invariant under all the Weyl reflections. In the current example, the usual Coulomb branch parameter $A$ will change under the Weyl reflection $\mathbf{R}_3$ in equation \eqref{eq:R3inE3}. 
We use the following ansatz for the invariant Coulomb branch parameter $\tilde{A}$:
\begin{equation}
	\tilde{A}=P_1^{\alpha_1}P_2^{\alpha_2}P_3^{\alpha_3}A.
\end{equation}
By requiring $\tilde{A}$ to be invariant under $\mathbf{R}_1,\mathbf{R}_2,\mathbf{R}_3$, we can determine $\alpha_1,\alpha_2,\alpha_3$, and we obtain
\begin{equation}
	\tilde{A}=P_1^{1/4}P_2^{1/2}A. 
\end{equation}
In terms of $P_1,P_2,P_3,\tilde{A}$, the $E_3$ Weyl reflections are
\begin{align}
	&\mathbf{R}_1=\big\{P_1\to P_1'=P_1^{-1},\ \ P_2\to P_2'=P_1P_2,\ \ P_3\to P_3'=P_3\ \ ;\ \ \tilde{A}\to\tilde{A}\big\},\label{eq:R1new}\\
	&\mathbf{R}_2=\big\{P_1\to P_1'=P_1,\ \ P_2\to P_2'=P_2,\ \ P_3\to P_3'=P_3^{-1}\ \ ;\ \ \tilde{A}\to\tilde{A}\big\},\label{eq:R2new}\\
	&\mathbf{R}_3=\big\{P_1\to P_1'=P_1P_2,\ \ P_2\to P_2'=P_2^{-1},\ \ P_3\to P_3'=P_3\ \ ; \ \ \tilde{A}\to \tilde{A}\big\},\label{eq:R3new}
\end{align}
where we have explicitly written all the parameters $P_i, \tilde{A}$ and the transformed parameters $P_i'$ for illustration purpose in the following paragraphs.

The BPS partition functions of 5d $\mathcal N=1$ gauge theories with 5-brane web construction can be computed by topological vertex \cite{Aganagic:2003db,Iqbal:2007ii}, and the partition function has the property that it is invariant up to analytic continuation or extra factor under flop transitions in the 5-brane web \cite{Iqbal:2004ne,Konishi:2006ev,Taki:2008hb}. 
Flop transitions are local manipulations on the brane web that involve only part of the brane web, Hanany-Witten transition is another local manipulation on the brane web which also keep the partition function invariant up to analytic continuation or extra factor. 
Reflection along vertical or horizontal direction on the brane web is a global manipulation which also does not change the partition function. A 90 degree rotation on the brane web is another global manipulation which maps the theory to its S-dual \cite{Aharony:1997bh}, it is also the fiber-base duality in M-theory setup \cite{Katz:1997eq}, and the partition function is invariant under the rotation. Thus the combination of a 90 degree rotation and a 180 degree reflection along vertical or horizontal direction which leads to a diagonal reflection is also a global manipulation on the brane web which keeps the partition function invariant. 

So for the 5d $SU(2)+2\mathbf F$ theory the three manipulations on the brane web in figure \ref{fig:SU2with2F} all leave the partition function invariant, which can be represented by the following formula:
\begin{equation}
	Z_{\text{before}}=Z_{\text{after}}. 
\end{equation}
 On the other hand, these three manipulations also do not change the shape of the brane web, the shape invariance under the diagonal reflection in figure \ref{fig:SU2with2F}(c) is called fiber-base symmetry in \cite{Mitev:2014jza}, thus we have
 \begin{align}
 	Z_{\text{before}}&=Z(P_1,P_2,P_3,\tilde{A}),\\
 	Z_{\text{after}}&=Z(P_1',P_2',P_3',\tilde{A}),
 \end{align} 
 where the functional form of $Z$ on the right hand side of the above two equations are the same. 
 
 Thus the partition function $Z(P_1,P_2,P_3,\tilde{A})$ is invariant under the $E_3$ Weyl reflections in equations \eqref{eq:R1new}-\eqref{eq:R3new}. If we expand $Z$ with respect to $\tilde{A}$, the coefficients will be combinations of characters of $E_3$ in some representations with the fugacities being $P_1,P_2,P_3$, so the partition function will have manifest $E_3$ symmetry. 

In the previous example, identifying the fiber-base symmetry is an important step toward uncovering the $E_3$ symmetry in the brane web. However 5-brane webs contain much more symmetries which are sometimes ignored, such symmetries are also important toward the uncovering of the full symmetry of the theory. All these symmetries in the brane web which keep the shape of the web invariant correspond to some Weyl symmetry.
In figure \ref{fig:SU2with2F}(a), the flop transition that exchanges the two flavor branes corresponds to the Weyl symmetry of exchanging $M_1$ with $M_2$. More generally, the exchanging of any parallel branes in a 5-brane web that keeps the shape of the web invariant also corresponds to some Weyl symmetry, this property is used in \cite{Wei:2022hjx} to figure out the $D_4,D_5,E_6$ invariant Coulomb branch parameters of the 6d $\mathcal N=(1,0)$ $D_4,D_5,E_6$-type little string theories. 

As the Weyl symmetry is a very general property of a 5-brane web, it is interesting to study it in more examples. In this paper, we study the Weyl symmetry in 6d $(D_4,D_4)$ conformal matter theory with general rank $N$ compactified on a circle. The rank $N$ 6d $(D_4,D_4)$ conformal matter theory is realized by $N$ M5-branes probing a $D_4$-type singularity, which has $D_4\times D_4$ flavor symmetry, it is dual to 5d affine $D_4$ quiver gauge theory after compactified on a circle \cite{DelZotto:2014hpa} and can be realized in 5-brane web by the method called quadrivalent gluing \cite{Hayashi:2021pcj,Hayashi:2017jze}. In section \ref{sec:rank1}, we study the rank 1 case in which the theory is also known as E-string theory \cite{Ganor:1996mu,Seiberg:1996vs} on a circle which is known to have affine $E_8$ global symmetry. We figure out the complete affine $E_8$ Weyl symmetry in the quadrivalently glued brane webs and also the affine $E_8$ invariant Coulomb branch parameter. In section \ref{sec:rank2}, we study the rank 2 case, we find that the theory still have affine $E_8$ Weyl symmetry, but due to the increase of the number of Coulomb branch parameters there are 64 different ways to form the affine $E_8$ symmetry, and correspondingly there are 64 different sets of affine $E_8$ invariant Coulomb branch parameters. In section \ref{sec:rankN}, we further study the rank $N\geq 3$ case, we find that the theory also has affine $E_8$ symmetry, and there are also 64 different ways to form the affine $E_8$ symmetry, we also obtain 64 different sets of affine $E_8$ invariant Coulomb branch parameters.

\section{Weyl symmetry in rank 1 $(D_4,D_4)$ conformal matter on a circle}\label{sec:rank1}
The rank 1 $(D_4,D_4)$ conformal matter on a circle is also known as E-string theory on a circle, the global symmetry of the theory is affine $E_8$. Before we study the Weyl symmetry in the corresponding brane webs, we first list the Weyl reflections of affine $E_8$.

\begin{figure}[htbp]
\centering
	\begin{tikzpicture}
	\draw 
	(0,0) circle (0.2)
	(1,0) circle (0.2)
	(2,0) circle (0.2)
	(3,0) circle (0.2)
	(4,0) circle (0.2)
	(5,0) circle (0.2)
	(6,0) circle (0.2)
	(7,0) circle (0.2)
	(5,1) circle (0.2)
	(0.2,0) to (0.8,0)
	(1.2,0) to (1.8,0)
	(2.2,0) to (2.8,0)
	(3.2,0) to (3.8,0)
	(4.2,0) to (4.8,0)
	(5.2,0) to (5.8,0)
	(6.2,0) to (6.8,0)
	(5,0.2) to (5,0.8);
	\node at (0,-0.5) {0};
	\node at (1,-0.5) {1};
	\node at (2,-0.5) {2};
	\node at (3,-0.5) {3};
	\node at (4,-0.5) {4};
	\node at (5,-0.5) {5};
	\node at (6,-0.5) {6};
	\node at (7,-0.5) {7};
	\node at (5.5,1) {8};
\end{tikzpicture}
\caption{Dynkin diagram of affine $E_8$. }
\label{fig:affineE8-DK}
\end{figure}
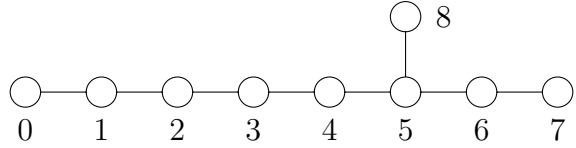

Figure \ref{fig:affineE8-DK} is the Dynkin diagram of affine $E_8$, we define parameters $P_i$ as the fugacities corresponding to the affine $E_8$ simple roots in $\alpha$-basis, then the Weyl reflections corresponding to the nine simple roots in terms of $P_i$ are the following:
\begin{align}	
	&\mathbf{W}_0=\big{\{}P_0\rightarrow P_0^{-1},\ P_1\rightarrow P_1P_0\big{\}},\nn\\
	&\mathbf{W}_1=\big{\{}P_1\rightarrow P_1^{-1},\ P_0\rightarrow P_0P_1,\ P_2\rightarrow P_2P_1\big{\}},\nn\\
	&\mathbf{W}_2=\big{\{}P_2\rightarrow P_2^{-1},\ P_1\rightarrow P_1P_2,\ P_3\rightarrow P_3P_2\big{\}},\nn\\
	&\mathbf{W}_3=\big{\{}P_3\rightarrow P_3^{-1},\ P_2\rightarrow P_2P_3,\ P_4\rightarrow P_4P_3\big{\}},\nn\\
	&\mathbf{W}_4=\big{\{}P_4\rightarrow P_4^{-1},\ P_3\rightarrow P_3P_4,\ P_5\rightarrow P_5P_4\big{\}},\nn\\
	&\mathbf{W}_5=\big{\{}P_5\rightarrow P_5^{-1},\ P_4\rightarrow P_4P_5,\ P_6\rightarrow P_6P_5,\ P_8\rightarrow P_8P_5\big{\}},\nn\\
	&\mathbf{W}_6=\big{\{}P_6\rightarrow P_6^{-1},\ P_5\rightarrow P_5 P_6,\ P_7\rightarrow P_7P_6\big{\}},\nn\\
	&\mathbf{W}_7=\big{\{}P_7\rightarrow P_7^{-1},\ P_6\rightarrow P_6P_7\big{\}},\nn\\
	&\mathbf{W}_8=\big{\{}P_8\rightarrow P_8^{-1},\ P_5\rightarrow P_5P_8\big{\}}.
	\label{eq:affineE8WQ}
\end{align}

The affine $E_8$ has the embedding $\widehat{E}_8\supset E_8$, and it is known that the Weyl reflections of $E_8$ in orthonormal basis have a very simple form, so we can utilize the orthonormal basis of $E_8$ to transform equation \eqref{eq:affineE8WQ} into a simple form. We use the following parameterization:
\begin{align}
	&P_0=\frac{y_8}{y_1}q,\quad P_1=\frac{y_1}{y_2},\quad P_2=\frac{y_2}{y_3},\quad P_3=\frac{y_3}{y_4},\quad P_4=\frac{y_4}{y_5},\quad P_5=\frac{y_5}{y_6},\nn\\
	&P_6=y_6 y_7,\quad P_7=\frac{1}{\sqrt{y_1 y_2 y_3 y_4 y_5 y_6 y_7 y_8}},\quad P_8=\frac{y_6}{y_7}.
\end{align}
In the above equations, the second one to the ninth one are in the orthonormal basis of $E_8$, the first one can be derived from the other eight ones by
\begin{equation}
	q=P_0P_1^2P_2^3P_3^4P_4^5P_5^6P_6^4P_7^2P_8^3, 
\end{equation}
where $q$ is the period due to the affine Lie algebra.

In terms of $y_i$ and $q$, the affine $E_8$ Weyl reflections are in the following simple form:
\begin{align}
	&\mathbf{W}_0=\big{\{}y_1\rightarrow y_8\,q,\ y_8\rightarrow y_1\,q^{-1}\big{\}},\nn\\
	&\mathbf{W}_1=\big{\{}y_1\rightarrow y_2,\ y_2\rightarrow y_1\big{\}},\nn\\
	&\mathbf{W}_2=\big{\{}y_2\rightarrow y_3,\ y_3\rightarrow y_2\big{\}},\nn\\
	&\mathbf{W}_3=\big{\{}y_3\rightarrow y_4,\ y_4\rightarrow y_3\big{\}},\nn\\
	&\mathbf{W}_4=\big{\{}y_4\rightarrow y_5,\ y_5\rightarrow y_4\big{\}},\nn\\
	&\mathbf{W}_5=\big{\{}y_5\rightarrow y_6,\ y_6\rightarrow y_5\big{\}},\nn\\
	&\mathbf{W}_6=\big{\{}y_6\rightarrow y_7^{-1},\ y_7\rightarrow y_6^{-1}\big{\}},\nn\\
	&\mathbf{W}_7=\big{\{}y_i\rightarrow \frac{y_i}{(y_1y_2y_3y_4y_5y_6y_7y_8)^{1/4}}\ \forall i\in \{1,\cdots,8\}\big{\}},\nn\\
	&\mathbf{W}_8=\big{\{}y_6\rightarrow y_7,\ y_7\rightarrow y_6\big{\}}.
	\label{eq:affineE8Wy}
\end{align}

\begin{figure}[htbp]
	\centering
	\begin{tikzpicture}[scale=0.8]
	\draw [semithick]

	(-0.7,1.1)--(0.7,1.1)

	;
	\draw [ semithick]

	(-0.7,3.9)--(0.7,3.9)

	;
	\draw [semithick,cyan]
	(-1.4,1.8)--(-2.1-0.2,1.8)
	(1.4,1.8)--(2.1+0.2,1.8)
	;
	\draw [semithick,red]
	(-1.4,3.2)--(-2.1-0.2,3.2)
	(1.4,3.2)--(2.1+0.2,3.2)
	;
	\draw[semithick,BurntOrange]
	(0.7,0.1)--(0.7,1.1)
	(0.7,1.1)--(1.4,1.8)
	(1.4,1.8)--(1.4,3.2)
	(0.7,3.9)--(1.4,3.2)
	(0.7,4.9)--(0.7,3.9)
	;
	\draw[semithick,LimeGreen]
	(-0.7,0.1)--(-0.7,1.1)
	(-0.7,1.1)--(-1.4,1.8)
	(-0.7,4.9)--(-0.7,3.9)
	(-0.7,3.9)--(-1.4,3.2)
	(-1.4,3.2)--(-1.4,1.8)
	;

	\draw [ semithick]
	(-4,5.8)--(-3.2,5.8)

	(-4.7,4.2)--(-3.2,4.2)
	(-5.6,5.1)--(-5.6,6.1)
	(-5.6,5.1)--(-6.4,4.3)
	
	;
	\draw [semithick,red]
	(-4.7,5.1)--(-5.6,5.1)
	;
	\draw[semithick,LimeGreen]
	(-4,6.8)--(-4,5.8)
	(-4,5.8)--(-4.7,5.1)
	(-4.7,5.1)--(-4.7,4.2)
	(-4.7,4.2)--(-5.5,3.4)
	;

	\draw [dashed,lightgray,semithick]
	(-5.6,5.1)--(-5.5,5.2)
	(-5,5.7)--(-4,6.7)
	(-4,6.7)--(-4.5,6.7)
	(-5.2,6.7)--(-6.2,6.7)
	(-5.6,5.1)--(-6.5,5.1);
	
	\draw [dashed,lightgray,semithick]
	(5.6,5.1)--(5.5,5.2)
	(5,5.7)--(4,6.7)
	(4,6.7)--(4.5,6.7)
	(5.2,6.7)--(6.2,6.7)
	(5.6,5.1)--(6.5,5.1);
	
	\draw [dashed,lightgray,semithick]
	(-5.6,5-5.1)--(-5.5,5-5.2)
	(-5,5-5.7)--(-4,5-6.7)
	(-4,5-6.7)--(-4.5,5-6.7)
	(-5.2,5-6.7)--(-6.2,5-6.7)
	(-5.6,5-5.1)--(-6.5,5-5.1);
	
	\draw [dashed,lightgray,semithick]
	(5.6,5-5.1)--(5.5,5-5.2)
	(5,5-5.7)--(4,5-6.7)
	(4,5-6.7)--(4.5,5-6.7)
	(5.2,5-6.7)--(6.2,5-6.7)
	(5.6,5-5.1)--(6.5,5-5.1);
	
	\draw [lightgray,semithick]
	(-2.8,2.5)--(2.8,2.5)
	(2.8,3.9)--(2.5,3.9);

	\draw [ semithick]
	
	(-4,5-5.8)--(-3.2,5-5.8)

	(-4.7,5-4.2)--(-3.2,5-4.2)
	(-5.6,5-5.1)--(-5.6,5-6.1)
	(-5.6,5-5.1)--(-6.4,5-4.3)
	
	;
	\draw [semithick,cyan]
	(-4.7,5-5.1)--(-5.6,5-5.1)
	;
	\draw[semithick,LimeGreen]
	(-4,5-6.8)--(-4,5-5.8)
	(-4,5-5.8)--(-4.7,5-5.1)
	(-4.7,5-5.1)--(-4.7,5-4.2)
	(-4.7,5-4.2)--(-5.5,5-3.4)
	;
	
	\draw [ semithick]
	
	(4,5.8)--(3.2,5.8)

	(4.7,4.2)--(3.2,4.2)
	(5.6,5.1)--(5.6,6.1)
	(5.6,5.1)--(6.4,4.3)
	
	;
	\draw [semithick,red]
	(4.7,5.1)--(5.6,5.1)
	;
	\draw[semithick,BurntOrange]
	(4,6.8)--(4,5.8)
	(4,5.8)--(4.7,5.1)
	(4.7,5.1)--(4.7,4.2)
	(4.7,4.2)--(5.5,3.4)
	;
	
	\draw [ semithick]
	
	(4,5-5.8)--(3.2,5-5.8)

	(4.7,5-4.2)--(3.2,5-4.2)
	(5.6,5-5.1)--(5.6,5-6.1)
	(5.6,5-5.1)--(6.4,5-4.3)
	
	;
	\draw [semithick,cyan]
	(4.7,5-5.1)--(5.6,5-5.1)
	;
	\draw[semithick,BurntOrange]
	(4,5-6.8)--(4,5-5.8)
	(4,5-5.8)--(4.7,5-5.1)
	(4.7,5-5.1)--(4.7,5-4.2)
	(4.7,5-4.2)--(5.5,5-3.4)
	;
	
	\node at (6.2,6.7) [right,gray] {$M_1$};
	\node at (6.5,5.1) [right,gray] {$M_2$};
	\node at (6.2,5-6.7) [right,gray] {$M_4$};
	\node at (6.5,5-5.1) [right,gray] {$M_3$};
	\node at (-6.2,6.7) [left,gray] {$M_5$};
	\node at (-6.5,5.1) [left,gray] {$M_6$};
	\node at (-6.2,5-6.7) [left,gray] {$M_8$};
	\node at (-6.5,5-5.1) [left,gray] {$M_7$};
	\node at (0,2.15) [below] {$Q_b$};
	\node at (0.45,2.8) [right] {$Q_f$};
	\node at (1.3,3.73) {$Y_1$};
	\node at (-1.3,3.73) {$Y_3$};
	\node at (1.3,1.27) {$Z_2$};
	\node at (-1.3,1.27) {$Z_4$};
	\node at (0,0) {$Q_D$};
	\node at (0,5) {$Q_U$};
	\node at (-5.15,5.4) {$X_3$};
	\node at (-4.4,4.65) {$Z_3$};
	\node at (-4.05,5.25) {$Y_3$};
	\node at (-5.15,5-5.4) {$X_4$};
	\node at (-4.4,5-4.65) {$Y_4$};
	\node at (-4.05,5-5.25) {$Z_4$};
	\node at (5.15,5-5.4) {$X_2$};
	\node at (4.4,5-4.65) {$Y_2$};
	\node at (4.05,5-5.25) {$Z_2$};
	\node at (5.15,5.4) {$X_1$};
	\node at (4.4,4.65) {$Z_1$};
	\node at (4.05,5.25) {$Y_1$};
	\node at (-4.8,6.7) {$Q_5$};
	\node at (-5.9,3.6) {$Q_6$};
	\node at (-3.05,5) {$Q_f$};
	\node at (4.8,6.7) {$Q_1$};
	\node at (5.9,3.6) {$Q_2$};
	\node at (3.05,5) {$Q_f$};
	\node at (-4.8,5-6.7) {$Q_8$};
	\node at (-5.9,5-3.6) {$Q_7$};
	\node at (-3.05,0) {$Q_f$};
	\node at (4.8,5-6.7) {$Q_4$};
	\node at (5.9,5-3.6) {$Q_3$};
	\node at (3.05,0) {$Q_f$};
	\node at (2.95,3.2) [gray] {$A$};
	\draw [<->,semithick] (-1.35,2.1)--(1.35,2.1);
	\draw [<->,semithick] (0.5,1.15)--(0.5,3.85);
	\draw [<->,semithick] (-0.65,0.3)--(0.65,0.3);
	\draw [<->,semithick] (-0.65,5-0.3)--(0.65,5-0.3);
	\draw [<->,semithick] (-5.55,6.4)--(-4.05,6.4);
	\draw [<->,semithick] (-6.45-0.3,4.2-0.3)--(-4.75-0.3,4.2-0.3);
	\draw [<->,semithick] (-3.4,4.25)--(-3.4,5.75);
	
	\draw [<->,semithick] (-5.55,5-6.4)--(-4.05,5-6.4);
	\draw [<->,semithick] (-6.45-0.3,5-4.2+0.3)--(-4.75-0.3,5-4.2+0.3);
	\draw [<->,semithick] (-3.4,5-4.25)--(-3.4,5-5.75);
	
	\draw [<->,semithick] (5.55,6.4)--(4.05,6.4);
	\draw [<->,semithick] (6.45+0.3,4.2-0.3)--(4.75+0.3,4.2-0.3);
	\draw [<->,semithick] (3.4,4.25)--(3.4,5.75);
	
	\draw [<->,semithick] (5.55,5-6.4)--(4.05,5-6.4);
	\draw [<->,semithick] (6.45+0.3,5-4.2+0.3)--(4.75+0.3,5-4.2+0.3);
	\draw [<->,semithick] (3.4,5-4.25)--(3.4,5-5.75);
	\draw [<->,lightgray,semithick] (2.65,3.85)--(2.65,2.55);
	\end{tikzpicture}
\caption{Rank 1 $(D_4,D_4)$ conformal matter on a circle in terms of brane webs with quadrivalent gluing. }
\label{fig:D4D4Quad}
\end{figure}

Figure \ref{fig:D4D4Quad} is the quadrivalently glued 5-brane web realization of rank 1 $(D_4,D_4)$ conformal matter theory on a circle \cite{Hayashi:2021pcj} which corresponds to the following affine $D_4$ quiver:
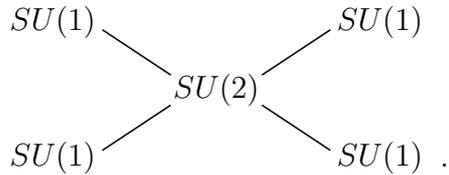
\begin{align*}
	\begin{tikzpicture}
	\node at (0,0) {$SU(2)$};
	\node at (2.15,0.9){$SU(1)$};
	\node at (2.15,-0.9){$SU(1)$};
	\node at (-2.15,0.9){$SU(1)$};
	\node at (-2.15,-0.9){$SU(1)$};
	\node at (3,-1){.};
	\draw[semithick]
	(0.6,0.2)--(1.5,0.8)
	(0.6,-0.2)--(1.5,-0.8)
	(-0.6,0.2)--(-1.5,0.8)
	(-0.6,-0.2)--(-1.5,-0.8)
	;
	\end{tikzpicture}
\end{align*}

\begin{figure}
	\centering
	\begin{tikzpicture}
		\draw [semithick]
	(-0.7,-0.5)--(-0.7,1.1)
	(0.7,-0.5)--(0.7,1.1)
	(-0.7,1.1)--(0.7,1.1)
	(-0.7,1.1)--(-1.4,1.8)
	(0.7,1.1)--(1.4,1.8)
	(1.4,1.8)--(1.4,3.2)
	(2.1,1.8)--(2.1,-0.5)
	(-2.1,1.8)--(-2.1,-0.5)
	(2.1,1.8)--(2.8,2.5)
	(2.8,2.5)--(2.1,3.2)
	(-2.1,1.8)--(-2.8,2.5)
	(-2.8,2.5)--(-2.1,3.2)
	(-2.1,3.2)--(-2.1,5.5)
	(2.1,3.2)--(2.1,5.5)
	;
	\draw [semithick]
	(-0.7,5.5)--(-0.7,3.9)
	(0.7,5.5)--(0.7,3.9)
	(-0.7,3.9)--(0.7,3.9)
	(-0.7,3.9)--(-1.4,3.2)
	(0.7,3.9)--(1.4,3.2)
	(-1.4,3.2)--(-1.4,1.8)
	;
	\node at (0,2.15) [below] {$Q_b$};
	\node at (0.45,2.8) [right] {$Q_f$};
	\node at (-2.8,6) [above] {ON$^-$};
	\node at (2.8,6) [above] {ON$^-$};
	\draw [<->,semithick] (-1.35,2.1)--(1.35,2.1);
	\draw [<->,semithick] (0.5,1.15)--(0.5,3.85);
	\draw [dashed, semithick]
	(-2.8,-1)--(-2.8,6)
	(2.8,-1)--(2.8,6);
	\draw [semithick,red]
	(-1.4,3.2)--(-2.1,3.2)
	(1.4,3.2)--(2.1,3.2)
	;
	\draw [semithick,cyan]
	(-1.4,1.8)--(-2.1,1.8)
	(1.4,1.8)--(2.1,1.8)
	;
	\node at (1.32,3.73) {$Y_1$};
	\node at (-1.35,3.73) {$Y_3$};
	\node at (1.33,1.27) {$Z_2$};
	\node at (-1.3,1.27) {$Z_4$};
	\node at (0,0) {$Q_D$};
	\node at (0,5) {$Q_U$};
	\node at (1.84,2.88) {$X_1$};
	\node at (1.84,2.07) {$X_2$};
	\node at (-1.8,2.88) {$X_3$};
	\node at (-1.8,2.07) {$X_4$};
	\node at (-1.35,5) {$Q_5$};
	\node at (1.45,5) {$Q_1$};
	\node at (1.45,0) {$Q_4$};
	\node at (-1.35,0) {$Q_8$};
	\node[scale=0.8] at (2.48,5.33) {$Q_3$};
	\node[scale=0.8] at (-2.48,5.33) {$Q_7$};
	\node[scale=0.8] at (-2.48,-4.12+3.8) {$Q_6$};
	\node[scale=0.8] at (2.48,-4.12+3.8) {$Q_2$};
	\draw [<->,semithick] (-0.65,0.3)--(0.65,0.3);
	\draw [<->,semithick] (-0.65,5-0.3)--(0.65,5-0.3);
	\draw [<->,semithick] (-2.05,0.3)--(-0.75,0.3);
	\draw [<->,semithick] (2.05,0.3)--(0.75,0.3);
	\draw [<->,semithick] (-2.05,5-0.3)--(-0.75,5-0.3);
	\draw [<->,semithick] (2.05,5-0.3)--(0.75,5-0.3);
	\draw [<->,semithick] (-2.75,-1.2)--(2.75,-1.2);
	\node at (0,6.28-7.17) {$q^{1/2}$};
	
	\draw [->,semithick] (2.5,6-0.3)--(0.75,6-0.3);
	\draw [->,semithick] (2.5,5.85-0.3)--(2.15,5.85-0.3);
	\draw [semithick] (2.5,5.85-0.3) arc(-90:90:0.3cm and 0.075cm);
	\draw [->,semithick] (-2.5,6-0.3)--(-0.75,6-0.3);
	\draw [->,semithick] (-2.5,5.85-0.3)--(-2.15,5.85-0.3);
	\draw [semithick] (-2.5,5.85-0.3) arc(270:90:0.3cm and 0.075cm);

	\draw [->,semithick] (2.5,-0.7)--(0.75,-0.7);
	\draw [->,semithick] (2.5,-0.55)--(2.15,-0.55);
	\draw [semithick] (2.5,-0.7) arc(-90:90:0.3cm and 0.075cm);
	\draw [->,semithick] (-2.5,-0.7)--(-0.75,-0.7);
	\draw [->,semithick] (-2.5,-0.55)--(-2.15,-0.55);
	\draw [semithick] (-2.5,-0.7) arc(270:90:0.3cm and 0.075cm);
	\end{tikzpicture}
	\caption{Rank 1 $(D_4,D_4)$ conformal matter on a circle in terms of brane web with ON-planes. }
	\label{fig:R1D4D4ONcorres}
\end{figure}
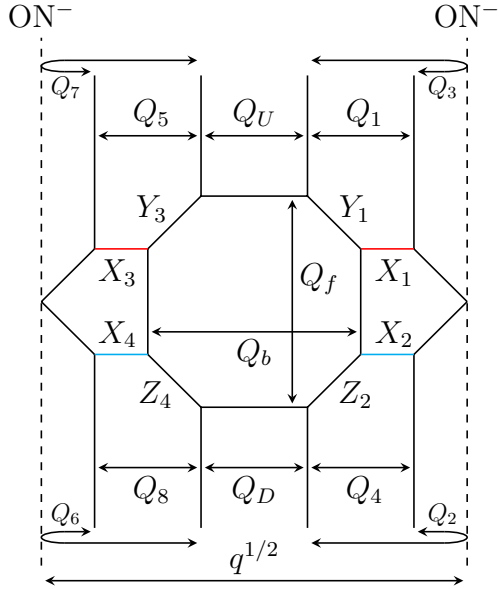
The figure contains five subdiagrams, the middle subdiagram corresponds to $SU(2)$ node and the four subdiagrams in the corners correspond to $SU(1)$ nodes. These five subdiagrams are quadrivalently glued together, and there are overlaps between them, for example, the red brane in the upper right $SU(1)$ subdiagram is the same red brane on the right side of the middle $SU(2)$ subdiagram, and roughly speaking the three yellow strips of NS-charged branes are the same brane.

We choose $Q_1,\cdots,Q_8,Q_b,Q_f$ as the basic K\"ahler parameters, then we can deduce that
\begin{align}
	&X_1=\sqrt{\frac{Q_1 Q_2}{Q_f}},\quad Y_1=\sqrt{\frac{Q_1 Q_f}{Q_2}},\quad Z_1=\sqrt{\frac{Q_f Q_2}{Q_1}},\quad X_2=\sqrt{\frac{Q_3 Q_4}{Q_f}},\nn\\ &Y_2=\sqrt{\frac{Q_3 Q_f}{Q_4}},\quad Z_2=\sqrt{\frac{Q_f Q_4}{Q_3}},\quad X_3=\sqrt{\frac{Q_5 Q_6}{Q_f}},\quad Y_3=\sqrt{\frac{Q_5 Q_f}{Q_6}},\nn\\
	&Z_3=\sqrt{\frac{Q_f Q_6}{Q_5}},\quad X_4=\sqrt{\frac{Q_7 Q_8}{Q_f}},\quad Y_4=\sqrt{\frac{Q_7 Q_f}{Q_8}},\quad Z_4=\sqrt{\frac{Q_f Q_8}{Q_7}},\nn\\
	&Q_U=\frac{Q_b}{Q_f}\sqrt{\frac{Q_2 Q_6}{Q_1 Q_5}},\quad Q_D=\frac{Q_b}{Q_f}\sqrt{\frac{Q_3 Q_7}{Q_4 Q_8}}. 
\end{align}

The rank 1 theory on a circle can also be viewed as 5d $SU(2)+8\mathbf{F}$, where each of the $SU(1)$ nodes in figure \ref{fig:D4D4Quad} contributes two flavors and we have labelled the corresponding mass parameters $M_1,\cdots,M_8$ as well as the Coulomb branch parameter $A\equiv\sqrt{Q_f}$ in the figure. The mass parameters $M_1,\cdots,M_8$ have the following relations to the K\"ahler parameters: 
\begin{align}
	&M_1=\sqrt{Q_1Q_2},\quad M_2=\sqrt{\frac{Q_2}{Q_1}},\quad M_3=\sqrt{\frac{Q_4}{Q_3}},\quad M_4=\sqrt{\frac{1}{Q_3 Q_4}},\nn\\
	&M_5=\sqrt{Q_5Q_6},\quad M_6=\sqrt{\frac{Q_6}{Q_5}},\quad M_7=\sqrt{\frac{Q_8}{Q_7}},\quad M_8=\sqrt{\frac{1}{Q_7 Q_8}}. 
	\label{eq:MtoQ}
\end{align}
The theory has a period $q$ due to the affine $D_4$ quiver structure:
\begin{align}	q&=(\sqrt{Q_UQ_D})^2\sqrt{Q_1Q_2}\sqrt{Q_3Q_4}\sqrt{Q_5Q_6}\sqrt{Q_7Q_8}\nn\\
	&=Q_2Q_3Q_6Q_7\frac{Q_b^2}{Q_f^2}. \label{eq:periodofR1}
\end{align}

From the expression of $q$ in terms of the K\"ahler parameters in equation \eqref{eq:periodofR1}, we find that it is more convenient to define
\begin{equation}
	Q_0\equiv \frac{Q_b}{Q_f}
\end{equation}
and use it to replace $Q_b$ as the proper parameter that is responsible for the global symmetry. Then we have
\begin{equation}
	q=Q_2 Q_3 Q_6 Q_7 Q_0^2. 
	\label{eq:period_inR1}
\end{equation}

The rank 1 theory on a circle can also be realized by 5-brane web with two ON-planes as depicted in figure \ref{fig:R1D4D4ONcorres} in which we have labelled all the parameters that are in one-to-one correspondence with the quadrivalent gluing in figure \ref{fig:D4D4Quad}. This two diagrams not only describe the same theory but also have the same practical way of doing the computation of partition functions by topological vertex \cite{Kim:2022dbr,Wei:2022hjx}. However the quadrivalent gluing has the advantage of uncovering hidden Weyl symmetries that are not easy to see in the usual brane web, so we mainly use quadrivalent gluing to study the Weyl symmetry in $(D_4,D_4)$ conformal matter on a circle.

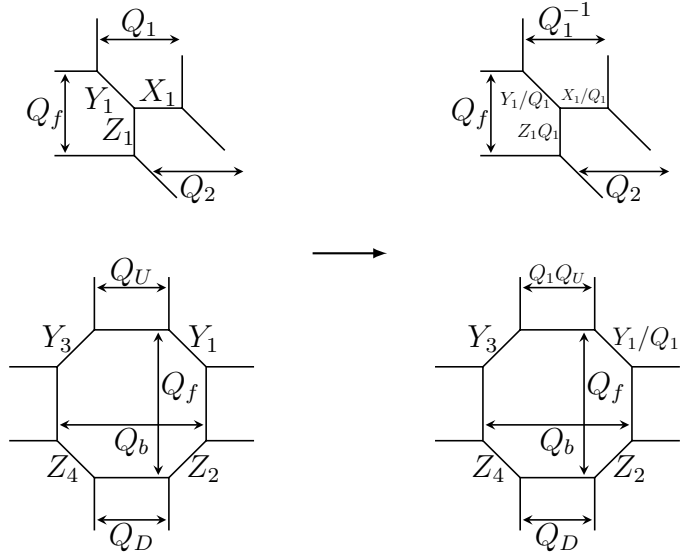
\begin{figure}[htbp]
	\centering
	\begin{tikzpicture}[scale=0.7]
	\draw [semithick]
	(-0.7,0.1)--(-0.7,1.1)
	(0.7,0.1)--(0.7,1.1)
	(-0.7,1.1)--(0.7,1.1)
	(-0.7,1.1)--(-1.4,1.8)
	(0.7,1.1)--(1.4,1.8)
	(-1.4,1.8)--(-2.1-0.2,1.8)
	(1.4,1.8)--(2.1+0.2,1.8)
	(1.4,1.8)--(1.4,3.2);
	\draw [semithick]
	(-0.7,4.9)--(-0.7,3.9)
	(0.7,4.9)--(0.7,3.9)
	(-0.7,3.9)--(0.7,3.9)
	(-0.7,3.9)--(-1.4,3.2)
	(0.7,3.9)--(1.4,3.2)
	(-1.4,3.2)--(-2.1-0.2,3.2)
	(-1.4,3.2)--(-1.4,1.8)
	(1.4,3.2)--(2.1+0.2,3.2);
	\node at (0,2.25) [below] {$Q_b$};
	\node at (0.34,2.8) [right] {$Q_f$};
	\node at (1.35,3.73) {$Y_1$};
	\node at (-1.4,3.73) {$Y_3$};
	\node at (1.35,1.27) {$Z_2$};
	\node at (-1.3,1.27) {$Z_4$};
	\node at (0,-0.03) {$Q_D$};
	\node at (0,5) {$Q_U$};
	\draw [<->,semithick] (-1.35,2.1)--(1.35,2.1);
	\draw [<->,semithick] (0.5,1.15)--(0.5,3.85);
	\draw [<->,semithick] (-0.65,0.3)--(0.65,0.3);
	\draw [<->,semithick] (-0.65,5-0.3)--(0.65,5-0.3);

	\draw [semithick]
	(4-4.65,6.8+3)--(4-4.65,5.8+3)
	(4-4.65,5.8+3)--(3.2-4.65,5.8+3)
	(4-4.65,5.8+3)--(4.7-4.65,5.1+3)
	(4.7-4.65,5.1+3)--(4.7-4.65,4.2+3)
	(4.7-4.65,4.2+3)--(3.2-4.65,4.2+3)
	(4.7-4.65,5.1+3)--(5.6-4.65,5.1+3)
	(5.6-4.65,5.1+3)--(5.6-4.65,6.1+3)
	(5.6-4.65,5.1+3)--(6.4-4.65,4.3+3)
	(4.7-4.65,4.2+3)--(5.5-4.65,3.4+3);
	\node at (5.15-4.65,5.4+3) {$X_1$};
	\node at (4.4-4.65,4.65+3) {$Z_1$};
	\node at (4.05-4.65,5.25+3) {$Y_1$};
	\node at (4.8-4.65,6.7+3) {$Q_1$};
	\node at (5.9-4.65,3.57+3) {$Q_2$};
	\node at (3.05-4.67,5+3) {$Q_f$};
	\draw [<->,semithick] (5.55-4.65,6.4+3)--(4.05-4.65,6.4+3);
	\draw [<->,semithick] (6.45+0.3-4.65,4.2-0.3+3)--(4.75+0.3-4.65,4.2-0.3+3);
	\draw [<->,semithick] (3.4-4.65,4.25+3)--(3.4-4.65,5.75+3);


	\draw [semithick]
	(4-4.65+8,6.8+3)--(4-4.65+8,5.8+3)
	(4-4.65+8,5.8+3)--(3.2-4.65+8,5.8+3)
	(4-4.65+8,5.8+3)--(4.7-4.65+8,5.1+3)
	(4.7-4.65+8,5.1+3)--(4.7-4.65+8,4.2+3)
	(4.7-4.65+8,4.2+3)--(3.2-4.65+8,4.2+3)
	(4.7-4.65+8,5.1+3)--(5.6-4.65+8,5.1+3)
	(5.6-4.65+8,5.1+3)--(5.6-4.65+8,6.1+3)
	(5.6-4.65+8,5.1+3)--(6.4-4.65+8,4.3+3)
	(4.7-4.65+8,4.2+3)--(5.5-4.65+8,3.4+3);
	\node[scale=0.5] at (5.15-4.65+8,5.4+2.92) {$X_1/Q_1$};
	\node[scale=0.6] at (4.3-4.65+8,4.65+3) {$Z_1Q_1$};
	\node[scale=0.6] at (4.05-4.65+8,5.25+3) {$Y_1/Q_1$};
	\node at (4.8-4.65+8,6.7+3.07) {$Q_1^{-1}$};
	\node at (5.9-4.65+8,3.57+3) {$Q_2$};
	\node at (3.05-4.68+8,5+3) {$Q_f$};
	\draw [<->,semithick] (5.55-4.65+8,6.4+3)--(4.05-4.65+8,6.4+3);
	\draw [<->,semithick] (6.45+0.3-4.65+8,4.2-0.3+3)--(4.75+0.3-4.65+8,4.2-0.3+3);
	\draw [<->,semithick] (3.4-4.65+8,4.25+3)--(3.4-4.65+8,5.75+3);

	\draw [semithick]
	(-0.7+8,0.1)--(-0.7+8,1.1)
	(0.7+8,0.1)--(0.7+8,1.1)
	(-0.7+8,1.1)--(0.7+8,1.1)
	(-0.7+8,1.1)--(-1.4+8,1.8)
	(0.7+8,1.1)--(1.4+8,1.8)
	(-1.4+8,1.8)--(-2.1-0.2+8,1.8)
	(1.4+8,1.8)--(2.1+0.2+8,1.8)
	(1.4+8,1.8)--(1.4+8,3.2);
	\draw [semithick]
	(-0.7+8,4.9)--(-0.7+8,3.9)
	(0.7+8,4.9)--(0.7+8,3.9)
	(-0.7+8,3.9)--(0.7+8,3.9)
	(-0.7+8,3.9)--(-1.4+8,3.2)
	(0.7+8,3.9)--(1.4+8,3.2)
	(-1.4+8,3.2)--(-2.1-0.2+8,3.2)
	(-1.4+8,3.2)--(-1.4+8,1.8)
	(1.4+8,3.2)--(2.1+0.2+8,3.2);
	\node at (0+8,2.25) [below] {$Q_b$};
	\node at (0.34+8,2.8) [right] {$Q_f$};
	\node[scale=0.8] at (1.67+8,3.73) {$Y_1/Q_1$};
	\node at (-1.4+8,3.73) {$Y_3$};
	\node at (1.35+8,1.27) {$Z_2$};
	\node at (-1.3+8,1.27) {$Z_4$};
	\node at (0+8,-0.03) {$Q_D$};
	\node[scale=0.7] at (0+8,4.95) {$Q_1Q_U$};
	\draw [<->,semithick] (-1.35+8,2.1)--(1.35+8,2.1);
	\draw [<->,semithick] (0.5+8,1.15)--(0.5+8,3.85);
	\draw [<->,semithick] (-0.65+8,0.3)--(0.65+8,0.3);
	\draw [<->,semithick] (-0.65+8,5-0.3)--(0.65+8,5-0.3);

	\draw [>=latex,->,thick] (3.4,5.35)--(4.8,5.35);

	\end{tikzpicture}
\caption{Flop transition related to exchanging the parallel branes that sandwich $Q_1$. }
\label{fig:Q1flop}
\end{figure}

\begin{figure}[htbp]
	\centering
	\begin{tikzpicture}[scale=0.7]
	\draw [semithick]
	(-0.7,0.1)--(-0.7,1.1)
	(0.7,0.1)--(0.7,1.1)
	(-0.7,1.1)--(0.7,1.1)
	(-0.7,1.1)--(-1.4,1.8)
	(0.7,1.1)--(1.4,1.8)
	(-1.4,1.8)--(-2.1-0.2,1.8)
	(1.4,1.8)--(2.1+0.2,1.8)
	(1.4,1.8)--(1.4,3.2);
	\draw [semithick]
	(-0.7,4.9)--(-0.7,3.9)
	(0.7,4.9)--(0.7,3.9)
	(-0.7,3.9)--(0.7,3.9)
	(-0.7,3.9)--(-1.4,3.2)
	(0.7,3.9)--(1.4,3.2)
	(-1.4,3.2)--(-2.1-0.2,3.2)
	(-1.4,3.2)--(-1.4,1.8)
	(1.4,3.2)--(2.1+0.2,3.2);
	\node at (0,2.25) [below] {$Q_b$};
	\node at (0.34,2.8) [right] {$Q_f$};
	\node at (1.35,3.73) {$Y_1$};
	\node at (-1.4,3.73) {$Y_3$};
	\node at (1.37,1.27) {$Z_2$};
	\node at (-1.3,1.27) {$Z_4$};
	\node at (0,-0.03) {$Q_D$};
	\node at (0,5) {$Q_U$};
	\draw [<->,semithick] (-1.35,2.1)--(1.35,2.1);
	\draw [<->,semithick] (0.5,1.15)--(0.5,3.85);
	\draw [<->,semithick] (-0.65,0.3)--(0.65,0.3);
	\draw [<->,semithick] (-0.65,5-0.3)--(0.65,5-0.3);

	\draw [semithick]
	(4-4.65,6.8+3)--(4-4.65,5.8+3)
	(4-4.65,5.8+3)--(3.2-4.65,5.8+3)
	(4-4.65,5.8+3)--(4.7-4.65,5.1+3)
	(4.7-4.65,5.1+3)--(4.7-4.65,4.2+3)
	(4.7-4.65,4.2+3)--(3.2-4.65,4.2+3)
	(4.7-4.65,5.1+3)--(5.6-4.65,5.1+3)
	(5.6-4.65,5.1+3)--(5.6-4.65,6.1+3)
	(5.6-4.65,5.1+3)--(6.4-4.65,4.3+3)
	(4.7-4.65,4.2+3)--(5.5-4.65,3.4+3);
	\node at (5.15-4.65,5.4+3) {$X_1$};
	\node at (4.4-4.65,4.65+3) {$Z_1$};
	\node at (4.05-4.65,5.25+3) {$Y_1$};
	\node at (4.8-4.65,6.7+3) {$Q_1$};
	\node at (5.9-4.65,3.56+3) {$Q_2$};
	\node at (3.05-4.7,5+3) {$Q_f$};
	\draw [<->,semithick] (5.55-4.65,6.4+3)--(4.05-4.65,6.4+3);
	\draw [<->,semithick] (6.45+0.3-4.65,4.2-0.3+3)--(4.75+0.3-4.65,4.2-0.3+3);
	\draw [<->,semithick] (3.4-4.65,4.25+3)--(3.4-4.65,5.75+3);


	\draw [semithick]
	(4-4.65+8,6.8+3)--(4-4.65+8,5.8+3)
	(4-4.65+8,5.8+3)--(3.2-4.65+8,5.8+3)
	(4-4.65+8,5.8+3)--(4.7-4.65+8,5.1+3)
	(4.7-4.65+8,5.1+3)--(4.7-4.65+8,4.2+3)
	(4.7-4.65+8,4.2+3)--(3.2-4.65+8,4.2+3)
	(4.7-4.65+8,5.1+3)--(5.6-4.65+8,5.1+3)
	(5.6-4.65+8,5.1+3)--(5.6-4.65+8,6.1+3)
	(5.6-4.65+8,5.1+3)--(6.4-4.65+8,4.3+3)
	(4.7-4.65+8,4.2+3)--(5.5-4.65+8,3.4+3);
	\node[scale=0.5] at (5.15-4.69+8,5.4+2.92) {$X_1/Q_2$};
	\node[scale=0.6] at (4.3-4.75+8,4.65+3) {$Z_1/Q_2$};
	\node[scale=0.6] at (4.05-4.68+8,5.25+3) {$Q_2Y_1$};
	\node at (4.8-4.65+8,6.7+3) {$Q_1$};
	\node at (5.9-4.65+8+0.2,3.6+3-0.06) {$Q_2^{-1}$};
	\node at (3.05-4.7+8,5+3) {$Q_f$};
	\draw [<->,semithick] (5.55-4.65+8,6.4+3)--(4.05-4.65+8,6.4+3);
	\draw [<->,semithick] (6.45+0.3-4.65+8,4.2-0.3+3)--(4.75+0.3-4.65+8,4.2-0.3+3);
	\draw [<->,semithick] (3.4-4.65+8,4.25+3)--(3.4-4.65+8,5.75+3);

	\draw [semithick]
	(-0.7+8,0.1)--(-0.7+8,1.1)
	(0.7+8,0.1)--(0.7+8,1.1)
	(-0.7+8,1.1)--(0.7+8,1.1)
	(-0.7+8,1.1)--(-1.4+8,1.8)
	(0.7+8,1.1)--(1.4+8,1.8)
	(-1.4+8,1.8)--(-2.1-0.2+8,1.8)
	(1.4+8,1.8)--(2.1+0.2+8,1.8)
	(1.4+8,1.8)--(1.4+8,3.2);
	\draw [semithick]
	(-0.7+8,4.9)--(-0.7+8,3.9)
	(0.7+8,4.9)--(0.7+8,3.9)
	(-0.7+8,3.9)--(0.7+8,3.9)
	(-0.7+8,3.9)--(-1.4+8,3.2)
	(0.7+8,3.9)--(1.4+8,3.2)
	(-1.4+8,3.2)--(-2.1-0.2+8,3.2)
	(-1.4+8,3.2)--(-1.4+8,1.8)
	(1.4+8,3.2)--(2.1+0.2+8,3.2);
	\node[scale=0.8] at (0+8-0.1,2.2) [below] {$Q_2Q_b$};
	\node at (0.34+8,2.8) [right] {$Q_f$};
	\node[scale=0.8] at (1.55+8,3.73) {$Q_2Y_1$};
	\node at (-1.4+8,3.73) {$Y_3$};
	\node at (1.37+8,1.27) {$Z_2$};
	\node at (-1.3+8,1.27) {$Z_4$};
	\node[scale=0.8] at (0+8.05,0.03) {$Q_2Q_D$};
	\node at (0+8,5) {$Q_U$};
	\draw [<->,semithick] (-1.35+8,2.1)--(1.35+8,2.1);
	\draw [<->,semithick] (0.5+8,1.15)--(0.5+8,3.85);
	\draw [<->,semithick] (-0.65+8,0.3)--(0.65+8,0.3);
	\draw [<->,semithick] (-0.65+8,5-0.3)--(0.65+8,5-0.3);

	\draw [>=latex,->,thick] (3.4,5.35)--(4.8,5.35);

	\end{tikzpicture}
\caption{Flop transition related to exchanging the parallel branes that sandwich $Q_2$. }
\label{fig:Q2flop}
\end{figure}

\begin{figure}[htbp]
	\centering
	\begin{tikzpicture}[scale=0.8]
	\draw [semithick]
	(-0.7,0.1)--(-0.7,1.1)
	(0.7,0.1)--(0.7,1.1)
	(-0.7,1.1)--(0.7,1.1)
	(-0.7,1.1)--(-1.4,1.8)
	(0.7,1.1)--(1.4,1.8)
	(-1.4,1.8)--(-2.1-0.2,1.8)
	(1.4,1.8)--(2.1+0.2,1.8)
	(1.4,1.8)--(1.4,3.2);
	\draw [semithick]
	(-0.7,4.9)--(-0.7,3.9)
	(0.7,4.9)--(0.7,3.9)
	(-0.7,3.9)--(0.7,3.9)
	(-0.7,3.9)--(-1.4,3.2)
	(0.7,3.9)--(1.4,3.2)
	(-1.4,3.2)--(-2.1-0.2,3.2)
	(-1.4,3.2)--(-1.4,1.8)
	(1.4,3.2)--(2.1+0.2,3.2);
	\draw [semithick]
	(-4,6.8)--(-4,5.8)
	(-4,5.8)--(-3.2,5.8)
	(-4,5.8)--(-4.7,5.1)
	(-4.7,5.1)--(-4.7,4.2)
	(-4.7,4.2)--(-3.2,4.2)
	(-4.7,5.1)--(-5.6,5.1)
	(-5.6,5.1)--(-5.6,6.1)
	(-5.6,5.1)--(-6.4,4.3)
	(-4.7,4.2)--(-5.5,3.4);

	\draw [semithick]
	(-4,5-6.8)--(-4,5-5.8)
	(-4,5-5.8)--(-3.2,5-5.8)
	(-4,5-5.8)--(-4.7,5-5.1)
	(-4.7,5-5.1)--(-4.7,5-4.2)
	(-4.7,5-4.2)--(-3.2,5-4.2)
	(-4.7,5-5.1)--(-5.6,5-5.1)
	(-5.6,5-5.1)--(-5.6,5-6.1)
	(-5.6,5-5.1)--(-6.4,5-4.3)
	(-4.7,5-4.2)--(-5.5,5-3.4);
	\draw [semithick]
	(4,6.8)--(4,5.8)
	(4,5.8)--(3.2,5.8)
	(4,5.8)--(4.7,5.1)
	(4.7,5.1)--(4.7,4.2)
	(4.7,4.2)--(3.2,4.2)
	(4.7,5.1)--(5.6,5.1)
	(5.6,5.1)--(5.6,6.1)
	(5.6,5.1)--(6.4,4.3)
	(4.7,4.2)--(5.5,3.4);
	\draw [semithick]
	(4,5-6.8)--(4,5-5.8)
	(4,5-5.8)--(3.2,5-5.8)
	(4,5-5.8)--(4.7,5-5.1)
	(4.7,5-5.1)--(4.7,5-4.2)
	(4.7,5-4.2)--(3.2,5-4.2)
	(4.7,5-5.1)--(5.6,5-5.1)
	(5.6,5-5.1)--(5.6,5-6.1)
	(5.6,5-5.1)--(6.4,5-4.3)
	(4.7,5-4.2)--(5.5,5-3.4);

	\node at (0,2.15) [below] {$Q_b$};
	\node[scale=0.6] at (0.43,2.8) [right] {$Q_UQ_f$};
	\node[scale=0.8] at (1.3+0.2,3.73) {$Q_UY_1$};
	\node[scale=0.8] at (-1.3-0.25,3.73) {$Q_UY_3$};
	\node at (1.3,1.27) {$Z_2$};
	\node at (-1.3,1.27) {$Z_4$};
	\node at (0,0) {$Q_D$};
	\node at (0,5.05) {$Q_U^{-1}$};
	\node at (-5.15,5.4) {$X_3$};
	\node at (-4.4,4.65) {$Z_3$};
	\node[scale=0.6] at (-4.05,5.25) {$Q_UY_3$};
	\node at (-5.15,5-5.4) {$X_4$};
	\node[scale=0.8] at (-4.4+0.25,5-4.65) {$Q_UY_4$};
	\node at (-4.05,5-5.25) {$Z_4$};
	\node at (5.15,5-5.4) {$X_2$};
	\node[scale=0.8] at (4.4-0.25,5-4.65) {$Q_UY_2$};
	\node at (4.05,5-5.25) {$Z_2$};
	\node at (5.15,5.4) {$X_1$};
	\node at (4.4,4.65) {$Z_1$};
	\node[scale=0.6] at (4.05,5.25) {$Q_UY_1$};
	\node[scale=0.8] at (-4.8,6.7) {$Q_UQ_5$};
	\node at (-5.9,3.6) {$Q_6$};
	\node at (-3.05+0.4,5) {$Q_UQ_f$};
	\node[scale=0.8] at (4.8,6.7) {$Q_UQ_1$};
	\node at (5.9,3.6) {$Q_2$};
	\node at (2.7,5) {$Q_UQ_f$};
	\node at (-4.8,5-6.7) {$Q_8$};
	\node[scale=0.8] at (-5.9-0.1,5-3.6) {$Q_UQ_7$};
	\node at (-3.05+0.4,0) {$Q_UQ_f$};
	\node at (4.8,5-6.7) {$Q_4$};
	\node[scale=0.8] at (5.9+0.15,5-3.6) {$Q_UQ_3$};
	\node at (3.05-0.35,0) {$Q_UQ_f$};
	
	\draw [<->,semithick] (-1.35,2.1)--(1.35,2.1);
	\draw [<->,semithick] (0.5,1.15)--(0.5,3.85);
	\draw [<->,semithick] (-0.65,0.3)--(0.65,0.3);
	\draw [<->,semithick] (-0.65,5-0.3)--(0.65,5-0.3);
	\draw [<->,semithick] (-5.55,6.4)--(-4.05,6.4);
	\draw [<->,semithick] (-6.45-0.3,4.2-0.3)--(-4.75-0.3,4.2-0.3);
	\draw [<->,semithick] (-3.4,4.25)--(-3.4,5.75);
	
	\draw [<->,semithick] (-5.55,5-6.4)--(-4.05,5-6.4);
	\draw [<->,semithick] (-6.45-0.3,5-4.2+0.3)--(-4.75-0.3,5-4.2+0.3);
	\draw [<->,semithick] (-3.4,5-4.25)--(-3.4,5-5.75);
	
	\draw [<->,semithick] (5.55,6.4)--(4.05,6.4);
	\draw [<->,semithick] (6.45+0.3,4.2-0.3)--(4.75+0.3,4.2-0.3);
	\draw [<->,semithick] (3.4,4.25)--(3.4,5.75);
	
	\draw [<->,semithick] (5.55,5-6.4)--(4.05,5-6.4);
	\draw [<->,semithick] (6.45+0.3,5-4.2+0.3)--(4.75+0.3,5-4.2+0.3);
	\draw [<->,semithick] (3.4,5-4.25)--(3.4,5-5.75);
	
	\end{tikzpicture}
\caption{Flop transitions related to exchanging the parallel branes that sandwich $Q_U$. }
\label{fig:QAflop}
\end{figure}

\subsection{Flop transitions in the rank 1 quadrivalent gluing web}
As mentioned in the last part of the introduction, the exchanging of any parallel branes which keeps the shape of the brane web invariant corresponds to some Weyl symmetry. 
In the brane web of figure \ref{fig:D4D4Quad}, there are 10 exchangings which transform $Q_1,\cdots,Q_8,Q_U,Q_D$ into their inverses. 
Exchanging the two parallel branes that sandwich $Q_1$ also transforms $Q_U$ into $Q_1Q_U$, but exchanging the two parallel branes that sandwich $Q_2$ does not affect $Q_U$. The corresponding flop transitions are indicated in figure \ref{fig:Q1flop} and figure \ref{fig:Q2flop}. Exchanging the two parallel branes that sandwich $Q_U$ transforms $Q_f$ into $Q_UQ_f$ which also causes $Q_1,Q_3,Q_5,Q_7$ to change, the flop transition of the brane web is indicated in figure \ref{fig:QAflop}. The flop transitions of $Q_3,\cdots,Q_8,Q_D$ can also similarly be derived. 

In summary, the 10 flop transitions in terms of basic K\"ahler parameters are the following:
\begin{align}
	&\mathbf{V}_1=\big{\{}Q_1\rightarrow Q_1^{-1}\big{\}},\nn\\
	&\mathbf{V}_2=\big{\{}Q_2\rightarrow Q_2^{-1},\ Q_0\rightarrow Q_2Q_0\big{\}},\nn\\
	&\mathbf{V}_3=\big{\{}Q_3\rightarrow Q_3^{-1},\ Q_0\rightarrow Q_3Q_0\big{\}},\nn\\
	&\mathbf{V}_4=\big{\{}Q_4\rightarrow Q_4^{-1}\big{\}},\nn\\
	&\mathbf{V}_5=\big{\{}Q_5\rightarrow Q_5^{-1}\big{\}},\nn\\
	&\mathbf{V}_6=\big{\{}Q_6\rightarrow Q_6^{-1},\ Q_0\rightarrow Q_6Q_0\big{\}},\nn\\
	&\mathbf{V}_7=\big{\{}Q_7\rightarrow Q_7^{-1},\ Q_0\rightarrow Q_7Q_0\big{\}},\nn\\
	&\mathbf{V}_8=\big{\{}Q_8\rightarrow Q_8^{-1}\big{\}},\nn\\
	&\mathbf{V}_9=\big{\{}\resizebox{0.65\hsize}{!}{$Q_1\rightarrow \sqrt{\frac{Q_1Q_2Q_6}{Q_5}}Q_0,\ Q_3\rightarrow \sqrt{\frac{Q_2Q_6}{Q_1Q_5}}Q_0 Q_3,\ Q_5\rightarrow \sqrt{\frac{Q_2Q_5Q_6}{Q_1}}Q_0,$}\nn\\
	&\qquad\quad\, \resizebox{0.6\hsize}{!}{$Q_7\rightarrow \sqrt{\frac{Q_2Q_6}{Q_1Q_5}}Q_0 Q_7,\ Q_0\rightarrow\sqrt{\frac{Q_1Q_5}{Q_2Q_6}}\ \ ; \ \ Q_f\rightarrow \sqrt{\frac{Q_2Q_6}{Q_1Q_5}}Q_0 Q_f$}\big{\}},\nn\\
	&\mathbf{V}_{10}=\big{\{}\resizebox{0.65\hsize}{!}{$Q_2\rightarrow \sqrt{\frac{Q_3Q_7}{Q_4Q_8}}Q_0 Q_2,\ Q_4\rightarrow \sqrt{\frac{Q_3Q_4Q_7}{Q_8}}Q_0,\ Q_6\rightarrow \sqrt{\frac{Q_3Q_7}{Q_4Q_8}}Q_0 Q_6,$}\nn\\
	&\qquad\quad\ \, \resizebox{0.6\hsize}{!}{$Q_8\rightarrow \sqrt{\frac{Q_3Q_7Q_8}{Q_4}}Q_0,\ Q_0\rightarrow\sqrt{\frac{Q_4Q_8}{Q_3Q_7}}\ \ ; \ \ Q_f\rightarrow \sqrt{\frac{Q_3Q_7}{Q_4Q_8}}Q_0 Q_f$}\big{\}}. 
	\label{eq:D4D4quadflopsQ}
\end{align}
By equation \eqref{eq:MtoQ}, the 10 flop transitions in terms of physical parameters are the following:
\begin{align}
	&\mathbf{V}_1=\big{\{}M_1\rightarrow M_2,\ M_2\rightarrow M_1\big{\}},\nn\\
	&\mathbf{V}_2=\big{\{}M_1\rightarrow M_2^{-1},\ M_2\rightarrow M_1^{-1}\big{\}},\nn\\
	&\mathbf{V}_3=\big{\{}M_3\rightarrow M_4^{-1},\ M_4\rightarrow M_3^{-1}\big{\}},\nn\\
	&\mathbf{V}_4=\big{\{}M_3\rightarrow M_4,\ M_4\rightarrow M_3\big{\}},\nn\\
	&\mathbf{V}_5=\big{\{}M_5\rightarrow M_6,\ M_6\rightarrow M_5\big{\}},\nn\\
	&\mathbf{V}_6=\big{\{}M_5\rightarrow M_6^{-1},\ M_6\rightarrow M_5^{-1}\big{\}},\nn\\
	&\mathbf{V}_7=\big{\{}M_7\rightarrow M_8^{-1},\ M_8\rightarrow M_7^{-1}\big{\}},\nn\\
	&\mathbf{V}_8=\big{\{}M_7\rightarrow M_8,\ M_8\rightarrow M_7\big{\}},\nn\\
	&\mathbf{V}_9=\big{\{}\resizebox{0.85\hsize}{!}{$M_1\rightarrow\left(\frac{M_1^3M_2M_3M_4M_6M_7M_8q}{M_5}\right)^{\frac14},\ M_2\rightarrow\left(\frac{M_1M_2^3M_5}{M_3M_4M_6M_7M_8q}\right)^{\frac14},\ M_3\rightarrow\left(\frac{M_1M_3^3M_5}{M_2M_4M_6M_7M_8q}\right)^{\frac14},$}\nn\\
	&\qquad\quad\,\resizebox{0.85\hsize}{!}{$M_4\rightarrow\left(\frac{M_1M_4^3M_5}{M_2M_3M_6M_7M_8q}\right)^{\frac14},\ M_5\rightarrow\left(\frac{M_2M_3M_4M_5^3M_6M_7M_8q}{M_1}\right)^{\frac14},\ M_6\rightarrow\left(\frac{M_1M_5M_6^3}{M_2M_3M_4M_7M_8q}\right)^{\frac14},$}\nn\\
	&\qquad\quad\,\resizebox{0.84\hsize}{!}{$M_7\rightarrow\left(\frac{M_1M_5M_7^3}{M_2M_3M_4M_6M_8q}\right)^{\frac14},\ M_8\rightarrow\left(\frac{M_1M_5M_8^3}{M_2M_3M_4M_6M_7q}\right)^{\frac14}\ \ ; \ \ A\rightarrow A\left(\frac{M_2M_3M_4M_6M_7M_8q}{M_1M_5}\right)^{\frac14}$}\big{\}},\nn\\
	&\mathbf{V}_{10}=\big{\{}\resizebox{0.76\hsize}{!}{$M_1\rightarrow\left(\frac{M_1^3M_4M_8q}{M_2M_3M_5M_6M_7}\right)^{\frac14},\ M_2\rightarrow\left(\frac{M_2^3M_4M_8q}{M_1M_3M_5M_6M_7}\right)^{\frac14},\ M_3\rightarrow\left(\frac{M_3^3M_4M_8q}{M_1M_2M_5M_6M_7}\right)^{\frac14},$}\nn\\
	&\qquad\quad\ \,\resizebox{0.82\hsize}{!}{$M_4\rightarrow\left(\frac{M_1M_2M_3M_4^3M_5M_6M_7}{M_8q}\right)^{\frac14},\ M_5\rightarrow\left(\frac{M_4M_5^3M_8q}{M_1M_2M_3M_6M_7}\right)^{\frac14},\ M_6\rightarrow\left(\frac{M_4M_6^3M_8q}{M_1M_2M_3M_5M_7}\right)^{\frac14},$}\nn\\
	&\qquad\quad\ \,\resizebox{0.84\hsize}{!}{$M_7\rightarrow\left(\frac{M_4M_7^3M_8q}{M_1M_2M_3M_5M_6}\right)^{\frac14},\ M_8\rightarrow\left(\frac{M_1M_2M_3M_5M_6M_7M_8^3}{M_4q}\right)^{\frac14}\ \ ; \ \ A\rightarrow A\left(\frac{M_4M_8q}{M_1M_2M_3M_5M_6M_7}\right)^{\frac14}$}\big{\}}.
	\label{eq:D4D4quadflopsM}
\end{align}
The above flop transitions all leave the period $q$ invariant. Aside from these 10 flop transitions, there are also flop transitions due to permutations between the four identical $SU(1)$ subdiagrams:
\begin{align}
	\mathbf{V}_{\text{i}}&=\big{\{}Q_1\to Q_3,\ Q_3\to Q_1,\ Q_2\to Q_4,\ Q_4\to Q_2\big{\}},\nn\\
	\mathbf{V}_{\text{ii}}&=\big{\{}Q_3\to Q_5,\ Q_5\to Q_3,\ Q_4\to Q_6,\ Q_6\to Q_4\big{\}},\nn\\
	\mathbf{V}_{\text{iii}}&=\big{\{}Q_5\to Q_7,\ Q_7\to Q_5,\ Q_6\to Q_8,\ Q_8\to Q_6\big{\}},\nn\\
	\mathbf{V}_{\text{iv}}&=\big{\{}Q_1\to Q_7,\ Q_7\to Q_1,\ Q_2\to Q_8,\ Q_8\to Q_2\big{\}},\nn\\
	\mathbf{V}_{\text{v}}&=\big{\{}Q_1\to Q_5,\ Q_5\to Q_1,\ Q_2\to Q_6,\ Q_6\to Q_2\big{\}},\nn\\
	\mathbf{V}_{\text{vi}}&=\big{\{}Q_3\to Q_7,\ Q_7\to Q_3,\ Q_4\to Q_8,\ Q_8\to Q_4\big{\}}. 
\end{align}

In terms of mass parameters, they are
\begin{align}
	\mathbf{V}_{\text{i}}&=\big{\{}M_1\to M_4^{-1},\ M_4\to M_1^{-1},\ M_2\to M_3,\ M_3\to M_2\big{\}},\nn\\
	\mathbf{V}_{\text{ii}}&=\big{\{}M_4\to M_5^{-1},\ M_5\to M_4^{-1},\ M_3\to M_6,\ M_6\to M_3\big{\}},\nn\\
	\mathbf{V}_{\text{iii}}&=\big{\{}M_5\to M_8^{-1},\ M_8\to M_5^{-1},\ M_6\to M_7,\ M_7\to M_6\big{\}},\nn\\
	\mathbf{V}_{\text{iv}}&=\big{\{}M_1\to M_8^{-1},\ M_8\to M_1^{-1},\ M_2\to M_7,\ M_7\to M_2\big{\}},\nn\\
	\mathbf{V}_{\text{v}}&=\big{\{}M_1\to M_5,\ M_5\to M_1,\ M_2\to M_6,\ M_6\to M_2\big{\}},\nn\\
	\mathbf{V}_{\text{vi}}&=\big{\{}M_3\to M_7,\ M_7\to M_3,\ M_4\to M_8,\ M_8\to M_4\big{\}}.
\end{align}

\subsection{Hidden flop transitions for the $SO(16)$ symmetry}
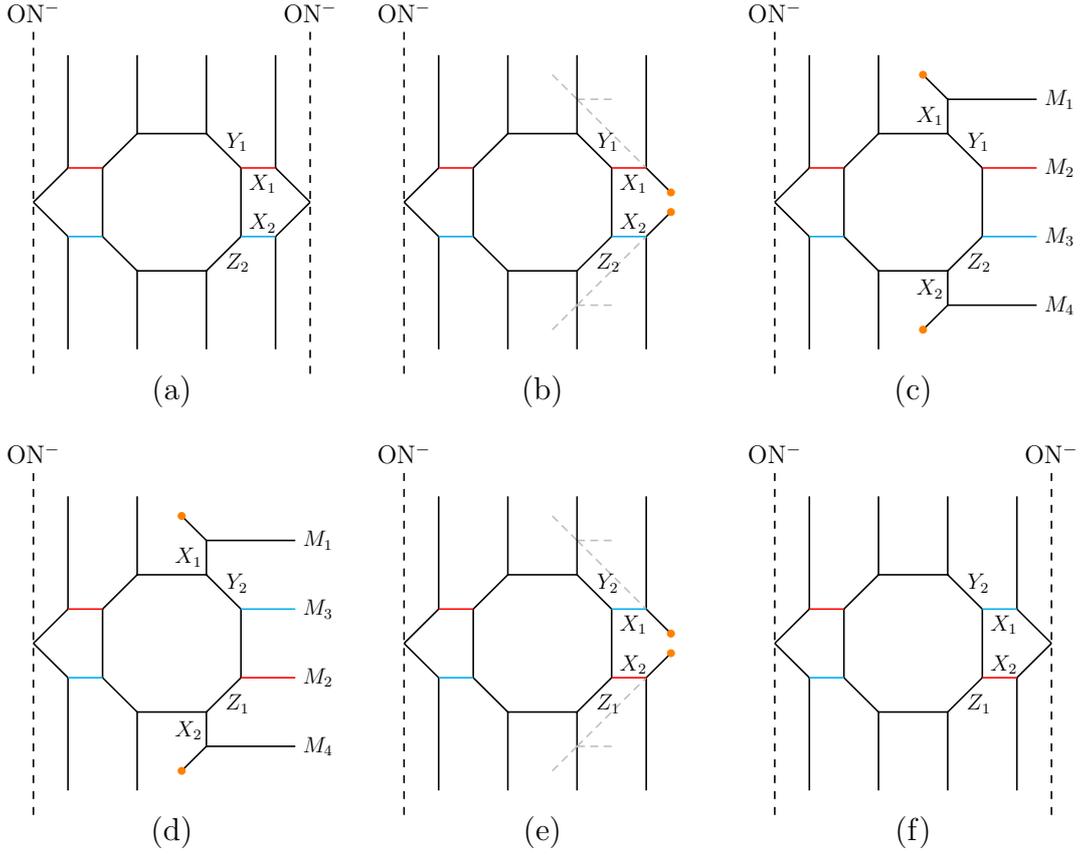
\begin{figure}[htbp]
	\centering
	\begin{tikzpicture}[scale=0.65]
	\draw [semithick]
	(-0.7,-0.5)--(-0.7,1.1)
	(0.7,-0.5)--(0.7,1.1)
	(-0.7,1.1)--(0.7,1.1)
	(-0.7,1.1)--(-1.4,1.8)
	(0.7,1.1)--(1.4,1.8)
	(1.4,1.8)--(1.4,3.2)
	(2.1,1.8)--(2.1,-0.5)
	(-2.1,1.8)--(-2.1,-0.5)
	(2.1,1.8)--(2.8,2.5)
	(2.8,2.5)--(2.1,3.2)
	(-2.1,1.8)--(-2.8,2.5)
	(-2.8,2.5)--(-2.1,3.2)
	(-2.1,3.2)--(-2.1,5.5)
	(2.1,3.2)--(2.1,5.5)
	;
	\draw [semithick]
	(-0.7,5.5)--(-0.7,3.9)
	(0.7,5.5)--(0.7,3.9)
	(-0.7,3.9)--(0.7,3.9)
	(-0.7,3.9)--(-1.4,3.2)
	(0.7,3.9)--(1.4,3.2)
	(-1.4,3.2)--(-1.4,1.8)
	;
	\node[scale=0.8] at (-2.8,6) [above] {ON$^-$};
	\node[scale=0.8] at (2.8,6) [above] {ON$^-$};
	\draw [dashed, semithick]
	(-2.8,-1)--(-2.8,6)
	(2.8,-1)--(2.8,6);
	\draw [semithick,red]
	(-1.4,3.2)--(-2.1,3.2)
	(1.4,3.2)--(2.1,3.2)
	;
	\draw [semithick,cyan]
	(-1.4,1.8)--(-2.1,1.8)
	(1.4,1.8)--(2.1,1.8)
	;
	\node[scale=0.7] at (1.32,3.73) {$Y_1$};
	\node[scale=0.7] at (1.33,1.27) {$Z_2$};
	\node[scale=0.7] at (1.84,2.88) {$X_1$};
	\node[scale=0.7] at (1.84,2.07) {$X_2$};
	\node at (0,-1.35) {(a)};
	
	\begin{scope}[shift={(7.5,0)}]
	\draw [semithick]
	(-0.7,-0.5)--(-0.7,1.1)
	(0.7,-0.5)--(0.7,1.1)
	(-0.7,1.1)--(0.7,1.1)
	(-0.7,1.1)--(-1.4,1.8)
	(0.7,1.1)--(1.4,1.8)
	(1.4,1.8)--(1.4,3.2)
	(2.1,1.8)--(2.1,-0.5)
	(-2.1,1.8)--(-2.1,-0.5)
	(2.1,1.8)--(2.6,2.3)
	(2.6,2.7)--(2.1,3.2)
	(-2.1,1.8)--(-2.8,2.5)
	(-2.8,2.5)--(-2.1,3.2)
	(-2.1,3.2)--(-2.1,5.5)
	(2.1,3.2)--(2.1,5.5)
	;
	\filldraw[orange] (2.6,2.3) circle (2pt);
	\filldraw[orange] (2.6,2.7) circle (2pt);
	\draw [densely dashed,lightgray,semithick]
	(0.7,0.4)--(2.1,1.8)
	(0.7,4.6)--(2.1,3.2)
	;
	\draw [densely dashed,lightgray,semithick]
	(0.2,0.2-0.3)--(0.7,0.7-0.3)	
	(1.4,0.7-0.3)--(0.7,0.7-0.3)
	(0.2,5.1)--(0.7,4.6)
	(1.4,4.6)--(0.7,4.6)
	;
	\draw [semithick]
	(-0.7,5.5)--(-0.7,3.9)
	(0.7,5.5)--(0.7,3.9)
	(-0.7,3.9)--(0.7,3.9)
	(-0.7,3.9)--(-1.4,3.2)
	(0.7,3.9)--(1.4,3.2)
	(-1.4,3.2)--(-1.4,1.8)
	;
	\node[scale=0.8] at (-2.8,6) [above] {ON$^-$};
	\draw [dashed, semithick]
	(-2.8,-1)--(-2.8,6)
	;
	\draw [semithick,red]
	(-1.4,3.2)--(-2.1,3.2)
	(1.4,3.2)--(2.1,3.2)
	;
	\draw [semithick,cyan]
	(-1.4,1.8)--(-2.1,1.8)
	(1.4,1.8)--(2.1,1.8)
	;
	\node[scale=0.7] at (1.84,2.88) {$X_1$};
	\node[scale=0.7] at (1.84,2.07) {$X_2$};
	\node[scale=0.7] at (1.32,3.73) {$Y_1$};
	\node[scale=0.7] at (1.33,1.27) {$Z_2$};
	\node at (0,-1.35) {(b)};
	\end{scope}
	
	\begin{scope}[shift={(15,0)}]
	\draw [semithick]
	(-0.7,-0.5)--(-0.7,1.1)
	(0.7,0.4)--(0.7,1.1)
	(-0.7,1.1)--(0.7,1.1)
	(-0.7,1.1)--(-1.4,1.8)
	(0.7,1.1)--(1.4,1.8)

	(1.4,1.8)--(1.4,3.2)
	(-2.1,1.8)--(-2.1,-0.5)
	(-2.1,1.8)--(-2.8,2.5)
	(-2.8,2.5)--(-2.1,3.2)
	(-2.1,3.2)--(-2.1,5.5)
	;
	\draw [semithick]
	(0.2,0.2-0.3)--(0.7,0.7-0.3)	
	(2.5,0.7-0.3)--(0.7,0.7-0.3)
	(0.2,5.1)--(0.7,4.6)
	(2.5,4.6)--(0.7,4.6)
	;
	\draw [semithick]
	(-0.7,5.5)--(-0.7,3.9)
	(0.7,4.6)--(0.7,3.9)
	(-0.7,3.9)--(0.7,3.9)
	(-0.7,3.9)--(-1.4,3.2)
	(0.7,3.9)--(1.4,3.2)
	
	(-1.4,3.2)--(-1.4,1.8)
	
	;
	\node[scale=0.8] at (-2.8,6) [above] {ON$^-$};
	\draw [dashed, semithick]
	(-2.8,-1)--(-2.8,6)
	;
	\draw [semithick,red]
	(-1.4,3.2)--(-2.1,3.2)
	(1.4,3.2)--(2.5,3.2)
	;
	\draw [semithick,cyan]
	(-1.4,1.8)--(-2.1,1.8)
	(1.4,1.8)--(2.5,1.8)
	;
	\filldraw[orange] (0.2,0.2-0.3) circle (2pt);
	\filldraw[orange] (0.2,5.1) circle (2pt);
	\node[scale=0.7] at (0.34,4.26) {$X_1$};
	\node[scale=0.7] at (0.34,0.73) {$X_2$};
	\node[scale=0.7] at (1.32,3.73) {$Y_1$};
	\node[scale=0.7] at (1.33,1.27) {$Z_2$};
	\node[scale=0.7] at (2.5,4.6) [right] {$M_{1}$};
	\node[scale=0.7] at (2.5,3.2) [right] {$M_{2}$};
	\node[scale=0.7] at (2.5,1.8) [right] {$M_{3}$};
	\node[scale=0.7] at (2.5,0.7-0.3) [right] {$M_{4}$};
	\node at (0,-1.35) {(c)};
	\end{scope}
	
	\begin{scope}[shift={(0,-9)}]
	\draw [semithick]
	(-0.7,-0.5)--(-0.7,1.1)
	(0.7,0.4)--(0.7,1.1)
	(-0.7,1.1)--(0.7,1.1)
	(-0.7,1.1)--(-1.4,1.8)
	(0.7,1.1)--(1.4,1.8)

	(1.4,1.8)--(1.4,3.2)
	(-2.1,1.8)--(-2.1,-0.5)
	(-2.1,1.8)--(-2.8,2.5)
	(-2.8,2.5)--(-2.1,3.2)
	(-2.1,3.2)--(-2.1,5.5)
	;
	\draw [semithick]
	(0.2,0.2-0.3)--(0.7,0.7-0.3)	
	(2.5,0.7-0.3)--(0.7,0.7-0.3)
	(0.2,5.1)--(0.7,4.6)
	(2.5,4.6)--(0.7,4.6)
	;
	\draw [semithick]
	(-0.7,5.5)--(-0.7,3.9)
	(0.7,4.6)--(0.7,3.9)
	(-0.7,3.9)--(0.7,3.9)
	(-0.7,3.9)--(-1.4,3.2)
	(0.7,3.9)--(1.4,3.2)
	
	(-1.4,3.2)--(-1.4,1.8)
	
	;
	\node[scale=0.8] at (-2.8,6) [above] {ON$^-$};
	\draw [dashed, semithick]
	(-2.8,-1)--(-2.8,6)
	;
	\draw [semithick,red]
	(-1.4,3.2)--(-2.1,3.2)
	(1.4,1.8)--(2.5,1.8)
	;
	\draw [semithick,cyan]
	(-1.4,1.8)--(-2.1,1.8)
	(1.4,3.2)--(2.5,3.2)
	;
	\filldraw[orange] (0.2,0.2-0.3) circle (2pt);
	\filldraw[orange] (0.2,5.1) circle (2pt);
	\node[scale=0.7] at (0.34,4.26) {$X_1$};
	\node[scale=0.7] at (0.34,0.73) {$X_2$};
	\node[scale=0.7] at (1.32,3.73) {$Y_2$};
	\node[scale=0.7] at (1.33,1.27) {$Z_1$};
	\node[scale=0.7] at (2.5,4.6) [right] {$M_{1}$};
	\node[scale=0.7] at (2.5,3.2) [right] {$M_{3}$};
	\node[scale=0.7] at (2.5,1.8) [right] {$M_{2}$};
	\node[scale=0.7] at (2.5,0.7-0.3) [right] {$M_{4}$};
	\node at (0,-1.35) {(d)};
	\end{scope}
	
	\begin{scope}[shift={(7.5,-9)}]
	\draw [semithick]
	(-0.7,-0.5)--(-0.7,1.1)
	(0.7,-0.5)--(0.7,1.1)
	(-0.7,1.1)--(0.7,1.1)
	(-0.7,1.1)--(-1.4,1.8)
	(0.7,1.1)--(1.4,1.8)
	(1.4,1.8)--(1.4,3.2)
	(2.1,1.8)--(2.1,-0.5)
	(-2.1,1.8)--(-2.1,-0.5)
	(2.1,1.8)--(2.6,2.3)
	(2.6,2.7)--(2.1,3.2)
	(-2.1,1.8)--(-2.8,2.5)
	(-2.8,2.5)--(-2.1,3.2)
	(-2.1,3.2)--(-2.1,5.5)
	(2.1,3.2)--(2.1,5.5)
	;
	\filldraw[orange] (2.6,2.3) circle (2pt);
	\filldraw[orange] (2.6,2.7) circle (2pt);
	\draw [densely dashed,lightgray,semithick]
	(0.7,0.4)--(2.1,1.8)
	(0.7,4.6)--(2.1,3.2)
	;
	\draw [densely dashed,lightgray,semithick]
	(0.2,0.2-0.3)--(0.7,0.7-0.3)	
	(1.4,0.7-0.3)--(0.7,0.7-0.3)
	(0.2,5.1)--(0.7,4.6)
	(1.4,4.6)--(0.7,4.6)
	;
	\draw [semithick]
	(-0.7,5.5)--(-0.7,3.9)
	(0.7,5.5)--(0.7,3.9)
	(-0.7,3.9)--(0.7,3.9)
	(-0.7,3.9)--(-1.4,3.2)
	(0.7,3.9)--(1.4,3.2)
	(-1.4,3.2)--(-1.4,1.8)
	;
	\node[scale=0.8] at (-2.8,6) [above] {ON$^-$};
	\draw [dashed, semithick]
	(-2.8,-1)--(-2.8,6)
	;
	\draw [semithick,red]
	(-1.4,3.2)--(-2.1,3.2)
	(1.4,1.8)--(2.1,1.8)
	;
	\draw [semithick,cyan]
	(-1.4,1.8)--(-2.1,1.8)
	(1.4,3.2)--(2.1,3.2)
	;
	\node[scale=0.7] at (1.84,2.88) {$X_1$};
	\node[scale=0.7] at (1.84,2.07) {$X_2$};
	\node[scale=0.7] at (1.32,3.73) {$Y_2$};
	\node[scale=0.7] at (1.33,1.27) {$Z_1$};
	\node at (0,-1.35) {(e)};
	\end{scope}
	
	\begin{scope}[shift={(15,-9)}]
	\draw [semithick]
	(-0.7,-0.5)--(-0.7,1.1)
	(0.7,-0.5)--(0.7,1.1)
	(-0.7,1.1)--(0.7,1.1)
	(-0.7,1.1)--(-1.4,1.8)
	(0.7,1.1)--(1.4,1.8)
	(1.4,1.8)--(1.4,3.2)
	(2.1,1.8)--(2.1,-0.5)
	(-2.1,1.8)--(-2.1,-0.5)
	(2.1,1.8)--(2.8,2.5)
	(2.8,2.5)--(2.1,3.2)
	(-2.1,1.8)--(-2.8,2.5)
	(-2.8,2.5)--(-2.1,3.2)
	(-2.1,3.2)--(-2.1,5.5)
	(2.1,3.2)--(2.1,5.5)
	;
	\draw [semithick]
	(-0.7,5.5)--(-0.7,3.9)
	(0.7,5.5)--(0.7,3.9)
	(-0.7,3.9)--(0.7,3.9)
	(-0.7,3.9)--(-1.4,3.2)
	(0.7,3.9)--(1.4,3.2)
	(-1.4,3.2)--(-1.4,1.8)
	;
	\node[scale=0.8] at (-2.8,6) [above] {ON$^-$};
	\node[scale=0.8] at (2.8,6) [above] {ON$^-$};
	\draw [dashed, semithick]
	(-2.8,-1)--(-2.8,6)
	(2.8,-1)--(2.8,6);
	\draw [semithick,red]
	(-1.4,3.2)--(-2.1,3.2)
	(1.4,1.8)--(2.1,1.8)
	;
	\draw [semithick,cyan]
	(-1.4,1.8)--(-2.1,1.8)
	(1.4,3.2)--(2.1,3.2)
	;
	\node[scale=0.7] at (1.84,2.88) {$X_1$};
	\node[scale=0.7] at (1.84,2.07) {$X_2$};
	\node[scale=0.7] at (1.32,3.73) {$Y_2$};
	\node[scale=0.7] at (1.33,1.27) {$Z_1$};
	\node at (0,-1.35) {(f)};
	\end{scope}

\end{tikzpicture}
\caption{Weyl symmetry of exchanging $M_2$ and $M_3$ in the ON-plane setup of $(D_4,D_4)$ conformal matter on a circle. }
\label{fig:M2M3inON}
\end{figure}

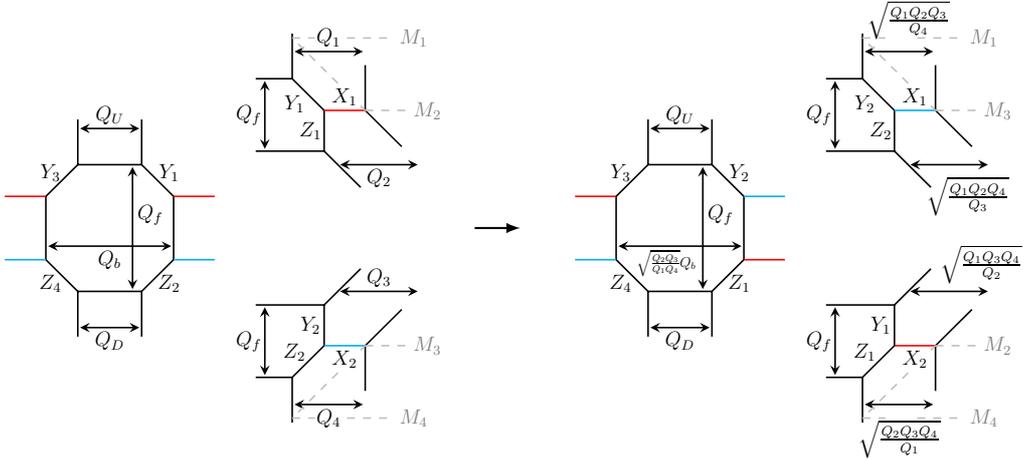
\begin{figure}[htbp]
	\centering
	\begin{tikzpicture}[scale=0.6]
	\draw [semithick]
	(-0.7,0.1)--(-0.7,1.1)
	(0.7,0.1)--(0.7,1.1)
	(-0.7,1.1)--(0.7,1.1)
	(-0.7,1.1)--(-1.4,1.8)
	(0.7,1.1)--(1.4,1.8)
	(1.4,1.8)--(1.4,3.2)
	;
	\draw [ semithick]
	(-0.7,4.9)--(-0.7,3.9)
	(0.7,4.9)--(0.7,3.9)
	(-0.7,3.9)--(0.7,3.9)
	(-0.7,3.9)--(-1.4,3.2)
	(0.7,3.9)--(1.4,3.2)
	(-1.4,3.2)--(-1.4,1.8)
	;
	\draw [semithick,cyan]
	(-1.4,1.8)--(-2.1-0.2,1.8)
	(1.4,1.8)--(2.1+0.2,1.8)
	;
	\draw [semithick,red]
	(-1.4,3.2)--(-2.1-0.2,3.2)
	(1.4,3.2)--(2.1+0.2,3.2)
	;
	
	\draw [ semithick]
	(4,6.8)--(4,5.8)
	(4,5.8)--(3.2,5.8)
	(4,5.8)--(4.7,5.1)
	(4.7,5.1)--(4.7,4.2)
	(4.7,4.2)--(3.2,4.2)
	(5.6,5.1)--(5.6,6.1)
	(5.6,5.1)--(6.4,4.3)
	(4.7,4.2)--(5.5,3.4);
	\draw [semithick,red]
	(4.7,5.1)--(5.6,5.1)
	;
	
	\draw [ semithick]
	(4,5-6.8)--(4,5-5.8)
	(4,5-5.8)--(3.2,5-5.8)
	(4,5-5.8)--(4.7,5-5.1)
	(4.7,5-5.1)--(4.7,5-4.2)
	(4.7,5-4.2)--(3.2,5-4.2)
	(5.6,5-5.1)--(5.6,5-6.1)
	(5.6,5-5.1)--(6.4,5-4.3)
	(4.7,5-4.2)--(5.5,5-3.4);
	\draw [semithick,cyan]
	(4.7,5-5.1)--(5.6,5-5.1)
	;

	\node[scale=0.65] at (0,2.15) [below] {$Q_b$};
	\node[scale=0.65] at (0.45,2.8) [right] {$Q_f$};
	\node[scale=0.65] at (1.3,3.73) {$Y_1$};
	\node[scale=0.65] at (-1.3,3.73) {$Y_3$};
	\node[scale=0.65] at (1.3,1.27) {$Z_2$};
	\node[scale=0.65] at (-1.3,1.27) {$Z_4$};
	\node[scale=0.65] at (0,0) {$Q_D$};
	\node[scale=0.65] at (0,5) {$Q_U$};
	\node[scale=0.65] at (5.15,5-5.4) {$X_2$};
	\node[scale=0.65] at (4.4,5-4.65) {$Y_2$};
	\node[scale=0.65] at (4.05,5-5.25) {$Z_2$};
	\node[scale=0.65] at (5.15,5.4) {$X_1$};
	\node[scale=0.65] at (4.4,4.65) {$Z_1$};
	\node[scale=0.65] at (4.05,5.25) {$Y_1$};
	\node[scale=0.65] at (4.8,6.7) {$Q_1$};
	\node[scale=0.65] at (5.9,3.6) {$Q_2$};
	\node[scale=0.65] at (3.05,5) {$Q_f$};
	\node[scale=0.65] at (4.8,5-6.7) {$Q_4$};
	\node[scale=0.65] at (5.9,5-3.6) {$Q_3$};
	\node[scale=0.65] at (3.05,0) {$Q_f$};
	\draw [<->,semithick] (-1.35,2.1)--(1.35,2.1);
	\draw [<->,semithick] (0.5,1.15)--(0.5,3.85);
	\draw [<->,semithick] (-0.65,0.3)--(0.65,0.3);
	\draw [<->,semithick] (-0.65,5-0.3)--(0.65,5-0.3);
	\draw [<->,semithick] (5.55,6.4)--(4.05,6.4);
	\draw [<->,semithick] (6.45+0.3,4.2-0.3)--(4.75+0.3,4.2-0.3);
	\draw [<->,semithick] (3.4,4.25)--(3.4,5.75);
	\draw [<->,semithick] (5.55,5-6.4)--(4.05,5-6.4);
	\draw [<->,semithick] (6.45+0.3,5-4.2+0.3)--(4.75+0.3,5-4.2+0.3);
	\draw [<->,semithick] (3.4,5-4.25)--(3.4,5-5.75);

	\draw [dashed,lightgray,semithick]
	(5.6,5.1)--(5.5,5.2)
	(5,5.7)--(4,6.7)
	(4,6.7)--(4.5,6.7)
	(5.2,6.7)--(6.2,6.7)
	(5.6,5.1)--(6.5,5.1);
	\draw [dashed,lightgray,semithick]
	(5.6,5-5.1)--(5.5,5-5.2)
	(5,5-5.7)--(4,5-6.7)
	(4,5-6.7)--(4.5,5-6.7)
	(5.2,5-6.7)--(6.2,5-6.7)
	(5.6,5-5.1)--(6.5,5-5.1);
	\node[scale=0.65] at (6.2,6.7) [right,gray] {$M_1$};
	\node[scale=0.65] at (6.5,5.1) [right,gray] {$M_2$};
	\node[scale=0.65] at (6.2,5-6.7) [right,gray] {$M_4$};
	\node[scale=0.65] at (6.5,5-5.1) [right,gray] {$M_3$};

	\begin{scope}[shift={(12.5,0)}]
	\draw [semithick]
	(-0.7,0.1)--(-0.7,1.1)
	(0.7,0.1)--(0.7,1.1)
	(-0.7,1.1)--(0.7,1.1)
	(-0.7,1.1)--(-1.4,1.8)
	(0.7,1.1)--(1.4,1.8)
	(1.4,1.8)--(1.4,3.2)
	;
	\draw [ semithick]
	(-0.7,4.9)--(-0.7,3.9)
	(0.7,4.9)--(0.7,3.9)
	(-0.7,3.9)--(0.7,3.9)
	(-0.7,3.9)--(-1.4,3.2)
	(0.7,3.9)--(1.4,3.2)
	(-1.4,3.2)--(-1.4,1.8)
	;
	\draw [semithick,cyan]
	(-1.4,1.8)--(-2.1-0.2,1.8)
	(1.4,3.2)--(2.1+0.2,3.2)
	;
	\draw [semithick,red]
	(-1.4,3.2)--(-2.1-0.2,3.2)
	(1.4,1.8)--(2.1+0.2,1.8)
	;
	
	\draw [ semithick]
	(4,6.8)--(4,5.8)
	(4,5.8)--(3.2,5.8)
	(4,5.8)--(4.7,5.1)
	(4.7,5.1)--(4.7,4.2)
	(4.7,4.2)--(3.2,4.2)
	(5.6,5.1)--(5.6,6.1)
	(5.6,5.1)--(6.4,4.3)
	(4.7,4.2)--(5.5,3.4);
	\draw [semithick,red]
	(4.7,5-5.1)--(5.6,5-5.1)
	;
	
	\draw [ semithick]
	(4,5-6.8)--(4,5-5.8)
	(4,5-5.8)--(3.2,5-5.8)
	(4,5-5.8)--(4.7,5-5.1)
	(4.7,5-5.1)--(4.7,5-4.2)
	(4.7,5-4.2)--(3.2,5-4.2)
	(5.6,5-5.1)--(5.6,5-6.1)
	(5.6,5-5.1)--(6.4,5-4.3)
	(4.7,5-4.2)--(5.5,5-3.4);
	\draw [semithick,cyan]
	(4.7,5.1)--(5.6,5.1)
	;

	\node[scale=0.45] at (-0.3,2.12) [below] {$\sqrt{\frac{Q_2Q_3}{Q_1Q_4}}Q_b$};
	\node[scale=0.65] at (0.45,2.8) [right] {$Q_f$};
	\node[scale=0.65] at (1.3,3.73) {$Y_2$};
	\node[scale=0.65] at (-1.3,3.73) {$Y_3$};
	\node[scale=0.65] at (1.3,1.27) {$Z_1$};
	\node[scale=0.65] at (-1.3,1.27) {$Z_4$};
	\node[scale=0.65] at (0,0) {$Q_D$};
	\node[scale=0.65] at (0,5) {$Q_U$};
	\node[scale=0.65] at (5.15,5-5.4) {$X_2$};
	\node[scale=0.65] at (4.4,5-4.65) {$Y_1$};
	\node[scale=0.65] at (4.05,5-5.25) {$Z_1$};
	\node[scale=0.65] at (5.15,5.4) {$X_1$};
	\node[scale=0.65] at (4.4,4.65) {$Z_2$};
	\node[scale=0.65] at (4.05,5.25) {$Y_2$};
	\node[scale=0.65] at (5,7.1) {$\sqrt{\frac{Q_1Q_2Q_3}{Q_4}}$};
	\node[scale=0.65] at (6.3,3.2) {$\sqrt{\frac{Q_1Q_2Q_4}{Q_3}}$};
	\node[scale=0.65] at (3.05,5) {$Q_f$};
	\node[scale=0.65] at (4.8,-2.17) {$\sqrt{\frac{Q_2Q_3Q_4}{Q_1}}$};
	\node[scale=0.65] at (6.6,1.7) {$\sqrt{\frac{Q_1Q_3Q_4}{Q_2}}$};
	\node[scale=0.65] at (3.05,0) {$Q_f$};
	\draw [<->,semithick] (-1.35,2.1)--(1.35,2.1);
	\draw [<->,semithick] (0.5,1.15)--(0.5,3.85);
	\draw [<->,semithick] (-0.65,0.3)--(0.65,0.3);
	\draw [<->,semithick] (-0.65,5-0.3)--(0.65,5-0.3);
	\draw [<->,semithick] (5.55,6.4)--(4.05,6.4);
	\draw [<->,semithick] (6.45+0.3,4.2-0.3)--(4.75+0.3,4.2-0.3);
	\draw [<->,semithick] (3.4,4.25)--(3.4,5.75);
	\draw [<->,semithick] (5.55,5-6.4)--(4.05,5-6.4);
	\draw [<->,semithick] (6.45+0.3,5-4.2+0.3)--(4.75+0.3,5-4.2+0.3);
	\draw [<->,semithick] (3.4,5-4.25)--(3.4,5-5.75);
	
	\draw [dashed,lightgray,semithick]
	(5.6,5.1)--(5.5,5.2)
	(5,5.7)--(4,6.7)
	(4,6.7)--(4.5,6.7)
	(5.2,6.7)--(6.2,6.7)
	(5.6,5.1)--(6.5,5.1);
	\draw [dashed,lightgray,semithick]
	(5.6,5-5.1)--(5.5,5-5.2)
	(5,5-5.7)--(4,5-6.7)
	(4,5-6.7)--(4.5,5-6.7)
	(5.2,5-6.7)--(6.2,5-6.7)
	(5.6,5-5.1)--(6.5,5-5.1);
	\node[scale=0.65] at (6.2,6.7) [right,gray] {$M_1$};
	\node[scale=0.65] at (6.5,5.1) [right,gray] {$M_3$};
	\node[scale=0.65] at (6.2,5-6.7) [right,gray] {$M_4$};
	\node[scale=0.65] at (6.5,5-5.1) [right,gray] {$M_2$};
	\end{scope}
	
	\draw[>=latex,->,thick] (8,2.5)--(9,2.5);

	\end{tikzpicture}
\caption{Flop transition related to exchanging $M_2$ and $M_3$ in the quadrivalent gluing brane web of $(D_4,D_4)$ conformal matter on a circle. }
\label{fig:M2M3inQuad}
\end{figure}

The flop transitions $\mathbf{V}_1,\cdots,\mathbf{V}_8$ belong to the $SO(16)$ Weyl group, but they do not generate the complete $SO(16)$ Weyl group. Even if we include the flop transitions $\textbf{V}_{\text{i}},\cdots,\textbf{V}_{\text{vi}}$, we still cannot obtain the complete $SO(16)$ Weyl group, because the Weyl symmetry of exchanging one single mass parameter between different $SU(1)$ subdiagrams is missing. In order to find out the flop transitions that correspond to such Weyl symmetry, let us first switch to the ON-plane realization of $(D_4,D_4)$ conformal matter on a circle which is shown in figure \ref{fig:M2M3inON}(a) with the K\"ahler parameters been labelled in accordance with the ones in figure \ref{fig:D4D4Quad}. We can remove the ON-plane on the right by attaching two D7-branes to the $(1,1)$ and $(1,-1)$ 5-branes on the right as shown in figure \ref{fig:M2M3inON}(b). Then we move the two D7-branes along the two diagonal dashed lines, by Hanany-Witten transition we obtain the diagram in figure \ref{fig:M2M3inON}(c) with four mass parameters been labelled. The Weyl symmetry of exchanging $M_2$ and $M_3$ just corresponds to exchanging the two D5-branes labelled by $M_2$ and $M_3$, so we obtain the diagram in figure \ref{fig:M2M3inON}(d). Due to this exchanging, the K\"ahler parameters $Y_1$ and $Z_2$ are transformed into $Y_2$ and $Z_1$. Then we can obtain figure \ref{fig:M2M3inON}(e) by Hanany-Witten transition and figure \ref{fig:M2M3inON}(f) by removing the two D7-branes and adding the ON-plane on the right. The diagrams in figure \ref{fig:M2M3inON}(a) and \ref{fig:M2M3inON}(f) have corresponding quadrivalent gluing realizations, so we find the flop transition $\mathbf{V}_{\text{e}}$ of exchanging $M_2$ and $M_3$ in quadrivalent gluing as indicated in figure \ref{fig:M2M3inQuad},
\begin{align}
	\mathbf{V}_{\text{e}}=&\big{\{}\resizebox{0.85\hsize}{!}{$Q_1\rightarrow\sqrt{\frac{Q_1Q_2Q_3}{Q_4}},\ Q_2\rightarrow\sqrt{\frac{Q_1Q_2Q_4}{Q_3}},\ Q_3\rightarrow\sqrt{\frac{Q_1Q_3Q_4}{Q_2}},\ Q_4\rightarrow\sqrt{\frac{Q_2Q_3Q_4}{Q_1}},\ Q_0\rightarrow\sqrt{\frac{Q_2Q_3}{Q_1Q_4}}Q_0$}\big{\}},\nn\\
	=&\big{\{}M_2\rightarrow M_3,\ M_3\rightarrow M_2\big{\}}. 
\end{align}
Due to the equivalence of the four $SU(1)$ subdiagrams, any two of the four $SU(1)$ subdiagrams can have the similar flop transition that happens in figure \ref{fig:M2M3inQuad}, which exchanges one single mass parameter. Combining such flop transitions with $\mathbf{V}_1,\cdots,\mathbf{V}_8$, we can obtain the complete $SO(16)$ Weyl group. 

For the flop transition $\mathbf{V}_9$, combining the following transformation which belongs to $SO(16)$ Weyl group:
\begin{align}
	\mathbf{V}_{\text{t}_1}=&\big{\{}M_2\rightarrow M_2^{-1},\ M_3\rightarrow M_3^{-1},\ M_4\rightarrow M_4^{-1},\nn\\
	&\phantom{\big{\{}}M_6\rightarrow M_6^{-1},\ M_7\rightarrow M_7^{-1},\ M_8\rightarrow M_8^{-1}\big{\}},  
\end{align}
we can obtain the Weyl reflection $\mathbf{V}_{\text{s}}=\mathbf{V}_{\text{t}_1}\mathbf{V}_9\mathbf{V}_{\text{t}_1}$:
\begin{align}
	\textbf{V}_{\text{s}}=\big{\{}\resizebox{0.87\hsize}{!}{$M_i\rightarrow M_i\left(\frac{q}{M_1M_2M_3M_4M_5M_6M_7M_8}\right)^{\frac14}\ \forall i\in\{1,\cdots,8\}\ \ ; \ \ 
	A\rightarrow A\left(\frac{q}{M_1M_2M_3M_4M_5M_6M_7M_8}\right)^{\frac14}$}\big{\}}. 
	\label{eq:WeylVs}
\end{align}

For the flop transition $\mathbf{V}_{10}$, we can combine the following transformation of $SO(16)$ Weyl group:
\begin{equation}
	\mathbf{V}_{\text{t}_2}=\big{\{}M_4\rightarrow M_4^{-1},\ M_8\rightarrow M_8^{-1}\big{\}}, 
\end{equation}
to obtain the same Weyl reflection in \eqref{eq:WeylVs}, $\mathbf{V}_{\text{s}}=\mathbf{V}_{\text{t}_2}\mathbf{V}_{10}\mathbf{V}_{\text{t}_2}$. 

After replacing the mass parameter $M_8$ by the new parameter $M_8'\equiv M_8/q$ \cite{Kim:2014dza}, we can obtain the standard affine $E_8$ Weyl reflection basis:
\begin{align}
	&\mathbf{W}_0=\big{\{}M_1\rightarrow M_8\,q,\ M_8\rightarrow M_1\,q^{-1}\big{\}},\nn\\
	&\mathbf{W}_1=\big{\{}M_1\rightarrow M_2,\ M_2\rightarrow M_1\big{\}},\nn\\
	&\mathbf{W}_2=\big{\{}M_2\rightarrow M_3,\ M_3\rightarrow M_2\big{\}},\nn\\
	&\mathbf{W}_3=\big{\{}M_3\rightarrow M_4,\ M_4\rightarrow M_3\big{\}},\nn\\
	&\mathbf{W}_4=\big{\{}M_4\rightarrow M_5,\ M_5\rightarrow M_4\big{\}},\nn\\
	&\mathbf{W}_5=\big{\{}M_5\rightarrow M_6,\ M_6\rightarrow M_5\big{\}},\nn\\
	&\mathbf{W}_6=\big{\{}M_6\rightarrow M_7^{-1},\ M_7\rightarrow M_6^{-1}\big{\}},\nn\\
	&\mathbf{W}_7=\big{\{}\resizebox{0.85\hsize}{!}{$M_i\rightarrow \frac{M_i}{(M_1M_2M_3M_4M_5M_6M_7M_8)^{1/4}}\ \forall i\in \{1,\cdots,8\}\ \ ; \ \ A\rightarrow \frac{A}{(M_1M_2M_3M_4M_5M_6M_7M_8)^{1/4}}$}\big{\}},\nn\\
	&\mathbf{W}_8=\big{\{}M_6\rightarrow M_7,\ M_7\rightarrow M_6\big{\}},
	\label{eq:affineE8Wm}
\end{align}
where for convenience we have dropped the prime on $M_8'$, and we will keep this notation change until the end of this section.

\subsection{Affine $E_8$ invariant Coulomb branch parameter}\label{subsec:inva_Coulomb_rk1}
$\textbf{W}_1,\cdots,\textbf{W}_8$ are the standard $E_8$ Weyl reflection basis, and it is easy\footnote{The Coulomb branch parameter $A$ is $SO(16)$ Weyl invariant, so it is invariant under $\textbf{W}_1,\cdots,\textbf{W}_6,\textbf{W}_8$. $M_8$ is obviously unaffected by $\textbf{W}_1,\cdots,\textbf{W}_6,\textbf{W}_8$. $A$ and $M_8$ are rescaled by the same factor under $\textbf{W}_7$, so $A/M_8$ is invariant under $\textbf{W}_7$. } to see that the following newly defined Coulomb branch parameter:
\begin{equation}
	A'\equiv \frac{A}{M_8}
	\label{eq:E8invArk1}
\end{equation}
is $E_8$ Weyl invariant. 

Under $\textbf{W}_0$, $A'$ transforms in the following way:
\begin{equation}
	A'\to \frac{M_8\,q}{M_1}A'.
\end{equation}

Under $\textbf{W}_0$, the $E_8$ Weyl invariant Jacobi form\footnote{The author thanks Futoshi Yagi for suggesting the Jacobi form $\Theta(q,\boldsymbol{M})$. }
\begin{align}
	\Theta(q,\boldsymbol{M})=\frac12\sum_{l=1}^4\prod_{i=1}^8\theta_l(M_i|q)
	\label{eq:JacobiTheta}
\end{align}
transforms in the following way:
\begin{align}
	\Theta(q,\boldsymbol{M})\rightarrow \Theta(q,\boldsymbol{M})\,\frac{M_1}{M_8\,q}. 
	\label{eq:jacobi_trans}
\end{align}
Thus we can form the following Coulomb branch parameter which is affine $E_8$ Weyl invariant:
\begin{equation}
	\tilde{A}\equiv\Theta(q,\boldsymbol{M})A'=\Theta(q,\boldsymbol{M})\frac{A}{M_8}. 
\end{equation}
We expect the partition function of the theory will have manifest affine $E_8$ symmetry if expanded by the affine $E_8$ Weyl invariant Coulomb branch parameter $\tilde{A}$.

\section{Weyl symmetry in rank 2 $(D_4,D_4)$ conformal matter on a circle}\label{sec:rank2}

\begin{figure}[htbp]
	\centering
	\begin{tikzpicture}[scale=0.8]
	
	\draw [semithick]
	(-0.7,0.1)--(-0.7,1.1)
	(0.7,0.1)--(0.7,1.1)
	(-0.7,1.1)--(0.7,1.1)
	(-0.7,1.1)--(-1.4,1.8)
	(0.7,1.1)--(1.4,1.8)
	(1.4,1.8)--(1.4,3.2)
	;
	\draw [semithick,red]
	(1.4,3.2)--(2.1+0.2,3.2)
	(-1.4,3.2)--(-2.1-0.2,3.2)
	(-1.4,3.2-3.8)--(-2.1-0.2,3.2-3.8)
	(1.4,3.2-3.8)--(2.1+0.2,3.2-3.8)
	;
	\draw [semithick,cyan]
	(-1.4,1.8)--(-2.1-0.2,1.8)
	(1.4,1.8)--(2.1+0.2,1.8)
	(-1.4,1.8-3.8)--(-2.1-0.2,1.8-3.8)
	(1.4,1.8-3.8)--(2.1+0.2,1.8-3.8)
	;
	\draw [semithick]
	(-0.7,4.9)--(-0.7,3.9)
	(0.7,4.9)--(0.7,3.9)
	(-0.7,3.9)--(0.7,3.9)
	(-0.7,3.9)--(-1.4,3.2)
	(0.7,3.9)--(1.4,3.2)
	(-1.4,3.2)--(-1.4,1.8)
	;
	\draw [semithick]
	(-0.7,0.1-3.8)--(-0.7,1.1-3.8)
	(0.7,0.1-3.8)--(0.7,1.1-3.8)
	(-0.7,1.1-3.8)--(0.7,1.1-3.8)
	(-0.7,1.1-3.8)--(-1.4,1.8-3.8)
	(0.7,1.1-3.8)--(1.4,1.8-3.8)

	(1.4,1.8-3.8)--(1.4,3.2-3.8)
	;
	\draw [semithick]
	(-0.7,3.9-3.8)--(0.7,3.9-3.8)
	(-0.7,3.9-3.8)--(-1.4,3.2-3.8)
	(0.7,3.9-3.8)--(1.4,3.2-3.8)
	
	(-1.4,3.2-3.8)--(-1.4,1.8-3.8)
	
	;
	\draw [<->,semithick] (-1.35,2.1-3.8)--(1.35,2.1-3.8);
	\draw [<->,semithick] (0.5,1.15)--(0.5,3.85);
	\draw [<->,semithick] (0.5,1.15-3.8)--(0.5,3.85-3.8);
	\draw [<->,semithick] (-0.65,5-0.3)--(0.65,5-0.3);
	\draw [<->,semithick] (-0.65,-3.5)--(0.65,-3.5);

	\draw [semithick]
	(4+0.5,6.8-2)--(4+0.5,5.8-2)
	(4+0.5,6.8-2)--(4.7+0.5,7.5-2)
	(4.7+0.5,7.5-2)--(4.7+0.5,8.4-2)
	(4.7+0.5,8.4-2)--(3.2+0.5,8.4-2)
	(4.7+0.5,8.4-2)--(5.5+0.5,9.2-2)
	(5.6+0.5,7.5-2)--(6.4+0.5,8.3-2)
	(4+0.5,5.8-2)--(3.2+0.5,5.8-2)
	(3.2+0.5,6.8-2)--(4+0.5,6.8-2)
	(4+0.5,5.8-2)--(4.7+0.5,5.1-2)
	(4.7+0.5,5.1-2)--(4.7+0.5,4.2-2)
	(4.7+0.5,4.2-2)--(3.2+0.5,4.2-2)
	(5.6+0.5,5.1-2)--(5.6+0.5,7.5-2)
	(5.6+0.5,5.1-2)--(6.4+0.5,4.3-2)
	(4.7+0.5,4.2-2)--(5.5+0.5,3.4-2);
	\draw [semithick,red]
	(4.7+0.5,7.5-2)--(5.6+0.5,7.5-2)
	(4.7+0.5,5.1-2)--(5.6+0.5,5.1-2)
	;
	\draw [<->,semithick] (5.55+0.5,6.15-2)--(4.05+0.5,6.15-2);
	\draw [<->,semithick] (6.45+0.3+0.5,4.2-0.3-2)--(4.75+0.3+0.5,4.2-0.3-2);
	\draw [<->,semithick] (6.45+0.3+0.5,8.7-2)--(4.75+0.3+0.5,8.7-2);
	\draw [<->,semithick] (3.4+0.5,4.25-2)--(3.4+0.5,5.75-2);
	\draw [<->,semithick] (3.4+0.5,4.85)--(3.4+0.5,6.35);

	\draw [semithick]
	(4+0.5,1.2-6.8+2)--(4+0.5,1.2-5.8+2)
	(4+0.5,1.2-6.8+2)--(4.7+0.5,1.2-7.5+2)
	(4.7+0.5,1.2-7.5+2)--(4.7+0.5,1.2-8.4+2)
	(4.7+0.5,1.2-8.4+2)--(3.2+0.5,1.2-8.4+2)
	(4.7+0.5,1.2-8.4+2)--(5.5+0.5,1.2-9.2+2)
	(5.6+0.5,1.2-7.5+2)--(6.4+0.5,1.2-8.3+2)
	(4+0.5,1.2-5.8+2)--(3.2+0.5,1.2-5.8+2)
	(3.2+0.5,1.2-6.8+2)--(4+0.5,1.2-6.8+2)
	(4+0.5,1.2-5.8+2)--(4.7+0.5,1.2-5.1+2)
	(4.7+0.5,1.2-5.1+2)--(4.7+0.5,1.2-4.2+2)
	(4.7+0.5,1.2-4.2+2)--(3.2+0.5,1.2-4.2+2)
	(5.6+0.5,1.2-5.1+2)--(5.6+0.5,1.2-7.5+2)
	(5.6+0.5,1.2-5.1+2)--(6.4+0.5,1.2-4.3+2)
	(4.7+0.5,1.2-4.2+2)--(5.5+0.5,1.2-3.4+2);
	\draw [semithick,cyan]
	(4.7+0.5,1.2-7.5+2)--(5.6+0.5,1.2-7.5+2)
	(4.7+0.5,1.2-5.1+2)--(5.6+0.5,1.2-5.1+2)
	;
	\draw [<->,semithick] (5.55+0.5,1.2-6.45+2)--(4.05+0.5,1.2-6.45+2);
	\draw [<->,semithick] (6.45+0.3+0.5,1.2-4.2+0.3+2)--(4.75+0.3+0.5,1.2-4.2+0.3+2);
	\draw [<->,semithick] (6.45+0.3+0.5,1.2-8.7+2)--(4.75+0.3+0.5,1.2-8.7+2);
	\draw [<->,semithick] (3.4+0.5,1.2-4.25+2)--(3.4+0.5,1.2-5.75+2);
	\draw [<->,semithick] (3.4+0.5,1.2-4.25-2.6+2)--(3.4+0.5,1.2-5.75-2.6+2);

	\draw [semithick]
	(-4-0.5,6.8-2)--(-4-0.5,5.8-2)
	(-4-0.5,6.8-2)--(-4.7-0.5,7.5-2)
	(-4.7-0.5,7.5-2)--(-4.7-0.5,8.4-2)
	(-4.7-0.5,8.4-2)--(-3.2-0.5,8.4-2)
	(-4.7-0.5,8.4-2)--(-5.5-0.5,9.2-2)
	(-5.6-0.5,7.5-2)--(-6.4-0.5,8.3-2)
	(-4-0.5,5.8-2)--(-3.2-0.5,5.8-2)
	(-3.2-0.5,6.8-2)--(-4-0.5,6.8-2)
	(-4-0.5,5.8-2)--(-4.7-0.5,5.1-2)
	(-4.7-0.5,5.1-2)--(-4.7-0.5,4.2-2)
	(-4.7-0.5,4.2-2)--(-3.2-0.5,4.2-2)
	(-5.6-0.5,5.1-2)--(-5.6-0.5,7.5-2)
	(-5.6-0.5,5.1-2)--(-6.4-0.5,4.3-2)
	(-4.7-0.5,4.2-2)--(-5.5-0.5,3.4-2);
	\draw [semithick,red]
	(-4.7-0.5,7.5-2)--(-5.6-0.5,7.5-2)
	(-4.7-0.5,5.1-2)--(-5.6-0.5,5.1-2)
	;
	\draw [<->,semithick] (-5.55-0.5,6.15-2)--(-4.05-0.5,6.15-2);
	\draw [<->,semithick] (-6.45-0.3-0.5,4.2-0.3-2)--(-4.75-0.3-0.5,4.2-0.3-2);
	\draw [<->,semithick] (-6.45-0.3-0.5,8.7-2)--(-4.75-0.3-0.5,8.7-2);
	\draw [<->,semithick] (-3.4-0.5,4.25-2)--(-3.4-0.5,5.75-2);
	\draw [<->,semithick] (-3.4-0.5,4.25+2.6-2)--(-3.4-0.5,5.75+2.6-2);

	\draw [semithick]
	(-4-0.5,1.2-6.8+2)--(-4-0.5,1.2-5.8+2)
	(-4-0.5,1.2-6.8+2)--(-4.7-0.5,1.2-7.5+2)
	(-4.7-0.5,1.2-7.5+2)--(-4.7-0.5,1.2-8.4+2)
	(-4.7-0.5,1.2-8.4+2)--(-3.2-0.5,1.2-8.4+2)
	(-4.7-0.5,1.2-8.4+2)--(-5.5-0.5,1.2-9.2+2)
	(-5.6-0.5,1.2-7.5+2)--(-6.4-0.5,1.2-8.3+2)
	(-4-0.5,1.2-5.8+2)--(-3.2-0.5,1.2-5.8+2)
	(-3.2-0.5,1.2-6.8+2)--(-4-0.5,1.2-6.8+2)
	(-4-0.5,1.2-5.8+2)--(-4.7-0.5,1.2-5.1+2)
	(-4.7-0.5,1.2-5.1+2)--(-4.7-0.5,1.2-4.2+2)
	(-4.7-0.5,1.2-4.2+2)--(-3.2-0.5,1.2-4.2+2)
	(-5.6-0.5,1.2-5.1+2)--(-5.6-0.5,1.2-7.5+2)
	(-5.6-0.5,1.2-5.1+2)--(-6.4-0.5,1.2-4.3+2)
	(-4.7-0.5,1.2-4.2+2)--(-5.5-0.5,1.2-3.4+2);
	\draw [semithick,cyan]
	(-4.7-0.5,1.2-7.5+2)--(-5.6-0.5,1.2-7.5+2)
	(-4.7-0.5,1.2-5.1+2)--(-5.6-0.5,1.2-5.1+2)
	;
	\draw [<->,semithick] (-5.55-0.5,1.2-6.45+2)--(-4.05-0.5,1.2-6.45+2);
	\draw [<->,semithick] (-6.45-0.3-0.5,1.2-4.2+0.3+2)--(-4.75-0.3-0.5,1.2-4.2+0.3+2);
	\draw [<->,semithick] (-6.45-0.3-0.5,1.2-8.7+2)--(-4.75-0.3-0.5,1.2-8.7+2);
	\draw [<->,semithick] (-3.4-0.5,1.2-4.25+2)--(-3.4-0.5,1.2-5.75+2);
	\draw [<->,semithick] (-3.4-0.5,1.2-4.25-2.6+2)--(-3.4-0.5,1.2-5.75-2.6+2);

	\node at (0,2.15-3.8) [below] {$Q_b$};
	\node at (0.4,2.5) [right] {$Q_{f_3}$};
	\node at (0.4,2.6-3.8) [right] {$Q_{f_1}$};
	\node at (1.47,3.73) {$Y_{1,2}$};
	\node at (-1.52,3.73) {$Y_{3,2}$};
	\node at (1.47,1.2) {$Z_{2,2}$};
	\node at (-1.33,1.2) {$Z_{4,2}$};
	\node at (0,-3.83) {$Q_D$};
	\node at (0,5) {$Q_U$};
	\node at (1.2,0.6) {$Q_{f_2}$};
	\node at (1.47,3.73-3.8) {$Y_{1,1}$};
	\node at (-1.52,3.73-3.8) {$Y_{3,1}$};
	\node at (1.47,1.2-3.8) {$Z_{2,1}$};
	\node at (-1.33,1.2-3.8) {$Z_{4,1}$};

	\node[scale=0.8] at (5.68,5.75) {$X_{1,2}$};
	\node[scale=0.8] at (5.2,4.95) {$Z_{1,2}$};
	\node at (4.75,5.9) {$Y_{1,2}$};
	\node at (6.48,7.02) {$Q_1$};
	\node at (6.48,1.57) {$Q_2$};
	\node at (3.46,5.6) {$Q_{f_3}$};	
	\node at (3.46,3) {$Q_{f_1}$};	
	\node at (4.07,4.3) {$Q_{f_2}$};
	\node[scale=0.8] at (5.74,8.57-5.75) {$X_{1,1}$};
	\node[scale=0.8] at (5.25,8.5-4.95) {$Y_{1,1}$};
	\node at (4.75,8.6-5.9) {$Z_{1,1}$};
	\node[scale=0.8] at (5.36,4.4) {$P_{1,1}$};
	\node at (6.67,4.3) {$ Q_{1,1}$};

	\node[scale=0.8] at (5.68,5.75-7.4) {$X_{2,2}$};
	\node[scale=0.8] at (5.2,4.95-7.4) {$Z_{2,2}$};
	\node at (4.75,5.9-7.4) {$Y_{2,2}$};
	\node at (6.48,7.02-7.4) {$Q_3$};
	\node at (6.48,1.57-7.4) {$Q_4$};
	\node at (3.46,5.6-7.4) {$Q_{f_3}$};	
	\node at (3.46,3-7.4) {$Q_{f_1}$};	
	\node at (4.07,4.3-7.4) {$Q_{f_2}$};
	\node[scale=0.8] at (5.74,8.57-5.75-7.4) {$X_{2,1}$};
	\node[scale=0.8] at (5.25,8.5-4.95-7.4) {$Y_{2,1}$};
	\node at (4.75,8.6-5.9-7.4) {$Z_{2,1}$};
	\node[scale=0.8] at (5.36,4.4-7.4) {$P_{2,1}$};	
	\node at (6.67,4.3-7.4) {$ Q_{2,1}$};

	\node[scale=0.8] at (-5.68,5.75) {$X_{3,2}$};
	\node[scale=0.8] at (-5.1,4.95) {$Z_{3,2}$};
	\node at (-4.75,5.9) {$Y_{3,2}$};
	\node at (-6.48,7.02) {$Q_5$};
	\node at (-6.48,1.57) {$Q_6$};
	\node at (-3.43,5.6) {$Q_{f_3}$};	
	\node at (-3.43,3) {$Q_{f_1}$};	
	\node at (-4.03,4.3) {$Q_{f_2}$};
	\node[scale=0.8] at (-5.67,8.57-5.75) {$X_{3,1}$};
	\node[scale=0.8] at (-5.25,8.5-4.95) {$Y_{3,1}$};
	\node at (-4.72,8.6-5.9) {$Z_{3,1}$};
	\node[scale=0.8] at (-5.36,4.4) {$P_{3,1}$};
	\node at (-6.67,4.3) {$ Q_{3,1}$};

	\node[scale=0.8] at (-5.68,5.75-7.4) {$X_{4,2}$};
	\node[scale=0.8] at (-5.1,4.95-7.4) {$Z_{4,2}$};
	\node at (-4.75,5.9-7.4) {$Y_{4,2}$};
	\node at (-6.48,7.02-7.4) {$Q_7$};
	\node at (-6.48,1.57-7.4) {$Q_8$};
	\node at (-3.43,5.6-7.4) {$Q_{f_3}$};	
	\node at (-3.43,3-7.4) {$Q_{f_1}$};	
	\node at (-4.03,4.3-7.4) {$Q_{f_2}$};
	\node[scale=0.8] at (-5.67,8.57-5.75-7.4) {$X_{4,1}$};
	\node[scale=0.8] at (-5.25,8.5-4.95-7.4) {$Y_{4,1}$};
	\node at (-4.72,8.6-5.9-7.4) {$Z_{4,1}$};
	\node[scale=0.8] at (-5.36,4.4-7.4) {$P_{4,1}$};	
	\node at (-6.67,4.3-7.4) {$ Q_{4,1}$};

	\end{tikzpicture}
\caption{Rank 2 $(D_4,D_4)$ conformal matter on a circle in terms of brane web with quadrivalent gluing. }
\label{fig:D4D4R2}
\end{figure}
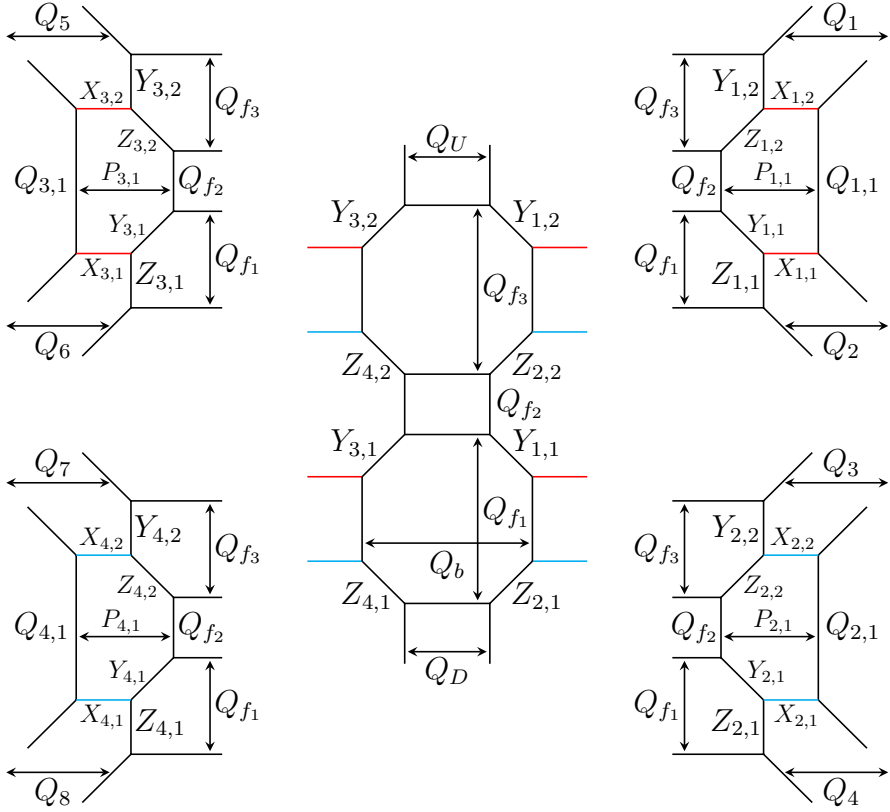

We extend the discussion of Weyl symmetry of $(D_4,D_4)$ conformal matter on a circle to the rank 2 case whose brane web construction by quadrivalent gluing is given in \cite{Hayashi:2021pcj}. We depict the brane web in figure \ref{fig:D4D4R2} which corresponds to the following affine $D_4$ quiver:
\begin{align*}
	\begin{tikzpicture}
	\node at (0,0) {$SU(4)$};
	\node at (2.15,0.9){$SU(2)$};
	\node at (2.15,-0.9){$SU(2)$};
	\node at (-2.15,0.9){$SU(2)$};
	\node at (-2.15,-0.9){$SU(2)$};
	\node at (3,-1){.};
	\draw[semithick]
	(0.6,0.2)--(1.5,0.8)
	(0.6,-0.2)--(1.5,-0.8)
	(-0.6,0.2)--(-1.5,0.8)
	(-0.6,-0.2)--(-1.5,-0.8)
	;
	\end{tikzpicture}
\end{align*}
The rank 2 theory also has nine independent K\"ahler parameters $Q_1,\cdots,Q_8,Q_b$ for the global symmetry.
In the rank 1 case, we can regard the theory as 5d $SU(2)+8\mathbf{F}$, so we have eight mass parameters, but in the rank 2 affine $D_4$ quiver gauge theory description, we do not have manifest eight flavors in the quadrivalently glued brane web, but still we can use the same physical-K\"ahler parameter relationships in equation \eqref{eq:MtoQ} of the rank 1 case to define the mass parameters for the rank 2 case which characterize the global symmetry of the rank 2 theory:
\begin{align}
	&M_1=\sqrt{Q_1Q_2},\quad M_2=\sqrt{\frac{Q_2}{Q_1}},\quad M_3=\sqrt{\frac{Q_4}{Q_3}},\quad M_4=\sqrt{\frac{1}{Q_3 Q_4}},\nn\\
	&M_5=\sqrt{Q_5Q_6},\quad M_6=\sqrt{\frac{Q_6}{Q_5}},\quad M_7=\sqrt{\frac{Q_8}{Q_7}},\quad M_8=\sqrt{\frac{1}{Q_7 Q_8}}. 
	\label{eq:MtoQ_r2}
\end{align}
Due to the increase of the rank, the rank 2 theory has seven Coulomb branch parameters, we define the three Coulomb branch parameters of the $SU(4)$ gauge node as the following:
\begin{equation}
	A_1\equiv\sqrt{Q_{f_1}},\quad A_2\equiv\sqrt{Q_{f_2}},\quad A_3\equiv\sqrt{Q_{f_3}}. 
\end{equation}
The Coulomb branch parameters of the four $SU(2)$ gauge nodes are $Q_{1,1},\cdots,Q_{4,1}$. In total the rank 2 theory has sixteen independent K\"ahler parameters $Q_1,\cdots,Q_8,Q_b,Q_{f_1}\allowbreak,Q_{f_2},Q_{f_3},Q_{1,1},\cdots,Q_{4,1}$. 

Mimicking the rank 1 case, we can find that for $I\in\{1,2,3,4\}$, $i\in\{1,2\}$, 
\begin{equation}
	X_{I,i}=\sqrt{\frac{P_{I,i}P_{I,i-1}}{Q_{f_{2i-1}}}},\quad Y_{I,i}=\sqrt{\frac{P_{I,i}Q_{f_{2i-1}}}{P_{I,i-1}}},\quad Z_{I,i}=\sqrt{\frac{P_{I,i-1}Q_{f_{2i-1}}}{P_{I,i}}},
\end{equation}
in which
\begin{alignat}{4}
    P_{1,0} &\equiv Q_2, \quad & P_{2,0} &\equiv Q_4, \quad & P_{3,0} &\equiv Q_6, \quad & P_{4,0} &\equiv Q_8, \\
    P_{1,2} &\equiv Q_1, \quad & P_{2,2} &\equiv Q_3, \quad & P_{3,2} &\equiv Q_5, \quad & P_{4,2} &\equiv Q_7.
\end{alignat}
The K\"ahler parameters $Q_U$ and $Q_D$ are
\begin{equation}
	Q_U=Q_0\sqrt{\frac{Q_2Q_6}{Q_1Q_5}},\quad Q_D=Q_0\sqrt{\frac{Q_3Q_7}{Q_4Q_8}},
\end{equation}
where we define $Q_0$ as the following to replace $Q_b$ as the independent K\"ahler parameter:
\begin{equation}
	Q_0\equiv\frac{Q_b}{Q_{f_1}}\sqrt{\frac{P_{2,1}P_{4,1}}{Q_3Q_7}}. 
\end{equation}

The period $q$ is
\begin{align}	q&=(\sqrt{Q_UQ_D})^2\sqrt{Q_1Q_2}\sqrt{Q_3Q_4}\sqrt{Q_5Q_6}\sqrt{Q_7Q_8}\nn\\
	&=Q_2Q_3Q_6Q_7Q_0^2. \label{eq:periodofR1}
\end{align}

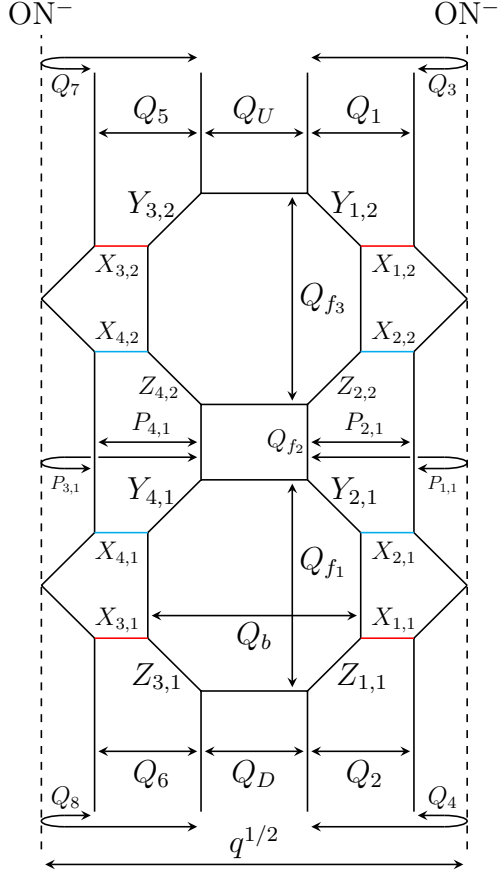
\begin{figure}[htbp]
	\centering
	\begin{tikzpicture}
	\draw [semithick]
	(-0.7,0.1)--(-0.7,1.1)
	(0.7,0.1)--(0.7,1.1)
	(-0.7,1.1)--(0.7,1.1)
	(-0.7,1.1)--(-1.4,1.8)
	(0.7,1.1)--(1.4,1.8)
	(1.4,1.8)--(1.4,3.2)
	(2.1,1.8)--(2.1,-0.6)
	(-2.1,1.8)--(-2.1,-0.6)
	(2.1,1.8)--(2.8,2.5)
	(2.8,2.5)--(2.1,3.2)
	(-2.1,1.8)--(-2.8,2.5)
	(-2.8,2.5)--(-2.1,3.2)
	(-2.1,3.2)--(-2.1,5.5)
	(2.1,3.2)--(2.1,5.5)
	;
	\draw [semithick]
	(-0.7,5.5)--(-0.7,3.9)
	(0.7,5.5)--(0.7,3.9)
	(-0.7,3.9)--(0.7,3.9)
	(-0.7,3.9)--(-1.4,3.2)
	(0.7,3.9)--(1.4,3.2)
	(-1.4,3.2)--(-1.4,1.8)
	;
	\node at (0.45,2.5) [right] {$Q_{f_3}$};
	\node[scale=0.8] at (0.42,0.6) {$Q_{f_2}$};
	\node at (-2.8,6) [above] {ON$^-$};
	\node at (2.8,6) [above] {ON$^-$};
	\draw [<->,semithick] (0.5,1.15)--(0.5,3.85);
	\draw [dashed, semithick]
	(-2.8,-4.8)--(-2.8,6)
	(2.8,-4.8)--(2.8,6);
	\draw [<->,semithick] (-2.75,-5)--(2.75,-5);
	\node at (0,-4.7) {$q^{1/2}$};
	\draw [semithick,red]
	(-1.4,3.2)--(-2.1,3.2)
	(1.4,3.2)--(2.1,3.2)
	;
	\draw [semithick,cyan]
	(-1.4,1.8)--(-2.1,1.8)
	(1.4,1.8)--(2.1,1.8)
	;
	\node at (1.32,3.73) {$Y_{1,2}$};
	\node at (-1.35,3.73) {$Y_{3,2}$};
	\node[scale=0.8] at (1.35,1.3) {$Z_{2,2}$};
	\node[scale=0.8] at (-1.26,1.3) {$Z_{4,2}$};
	\node at (0,5) {$Q_U$};
	\node[scale=0.8] at (1.84,2.92) {$X_{1,2}$};
	\node[scale=0.8] at (1.84,2.04) {$X_{2,2}$};
	\node[scale=0.8] at (-1.8,2.92) {$X_{3,2}$};
	\node[scale=0.8] at (-1.8,2.04) {$X_{4,2}$};
	\node at (-1.35,5) {$Q_5$};
	\node at (1.45,5) {$Q_1$};
	\node[scale=0.8] at (1.45,0.83) {$P_{2,1}$};
	\node[scale=0.8] at (-1.35,0.83) {$P_{4,1}$};
	\node[scale=0.6] at (2.48,0.03) {$P_{1,1}$};
	\node[scale=0.6] at (-2.48,0.03) {$P_{3,1}$};
	\node[scale=0.8] at (2.48,5.33) {$Q_3$};
	\node[scale=0.8] at (-2.48,5.33) {$Q_7$};
	\node[scale=0.8] at (-2.48,-4.12) {$Q_8$};
	\node[scale=0.8] at (2.48,-4.12) {$Q_4$};
	\draw [<->,semithick] (-0.65,5-0.3)--(0.65,5-0.3);
	\draw [<->,semithick] (-2.05,0.6)--(-0.75,0.6);
	\draw [<->,semithick] (2.05,0.6)--(0.75,0.6);
	\draw [<->,semithick] (-2.05,5-0.3)--(-0.75,5-0.3);
	\draw [<->,semithick] (2.05,5-0.3)--(0.75,5-0.3);
	
	\draw [->,semithick] (2.5,6-0.3)--(0.75,6-0.3);
	\draw [->,semithick] (2.5,5.85-0.3)--(2.15,5.85-0.3);
	\draw [semithick] (2.5,5.85-0.3) arc(-90:90:0.3cm and 0.075cm);
	\draw [->,semithick] (-2.5,6-0.3)--(-0.75,6-0.3);
	\draw [->,semithick] (-2.5,5.85-0.3)--(-2.15,5.85-0.3);
	\draw [semithick] (-2.5,5.85-0.3) arc(270:90:0.3cm and 0.075cm);
	
	\draw [->,semithick] (2.05,6-0.3-5.3)--(0.75,6-0.3-5.3);
	\draw [semithick] (2.5,6-0.3-5.3)--(2.15,6-0.3-5.3);
	\draw [->,semithick] (2.5,5.85-0.3-5.3)--(2.15,5.85-0.3-5.3);
	\draw [semithick] (2.5,5.85-0.3-5.3) arc(-90:90:0.3cm and 0.075cm);
	
	\draw [->,semithick] (-2.05,6-0.3-5.3)--(-0.75,6-0.3-5.3);
	\draw [semithick] (-2.5,6-0.3-5.3)--(-2.15,6-0.3-5.3);
	\draw [->,semithick] (-2.5,5.85-0.3-5.3)--(-2.15,5.85-0.3-5.3);
	\draw [semithick] (-2.5,5.85-0.3-5.3) arc(270:90:0.3cm and 0.075cm);

	\begin{scope}[shift={(0,-3.8)}]
	\draw [semithick]
	(-0.7,-0.5)--(-0.7,1.1)
	(0.7,-0.5)--(0.7,1.1)
	(-0.7,1.1)--(0.7,1.1)
	(-0.7,1.1)--(-1.4,1.8)
	(0.7,1.1)--(1.4,1.8)
	(1.4,1.8)--(1.4,3.2)
	(2.1,1.8)--(2.1,-0.5)
	(-2.1,1.8)--(-2.1,-0.5)
	(2.1,1.8)--(2.8,2.5)
	(2.8,2.5)--(2.1,3.2)
	(-2.1,1.8)--(-2.8,2.5)
	(-2.8,2.5)--(-2.1,3.2)
	;
	\draw [semithick]
	(-0.7,3.9)--(0.7,3.9)
	(-0.7,3.9)--(-1.4,3.2)
	(0.7,3.9)--(1.4,3.2)
	(-1.4,3.2)--(-1.4,1.8)
	;
	\node at (0,2.15) [below] {$Q_b$};
	\node at (0.45,2.8) [right] {$Q_{f_1}$};
	\draw [<->,semithick] (-1.35,2.1)--(1.35,2.1);
	\draw [<->,semithick] (0.5,1.15)--(0.5,3.85);
	\draw [semithick,red]
	(-1.4,1.8)--(-2.1,1.8)
	(1.4,1.8)--(2.1,1.8)
	;
	\draw [semithick,cyan]
	(-1.4,3.2)--(-2.1,3.2)
	(1.4,3.2)--(2.1,3.2)
	;
	\node at (1.32,3.73) {$Y_{2,1}$};
	\node at (-1.35,3.73) {$Y_{4,1}$};
	\node at (1.42,1.27) {$Z_{1,1}$};
	\node at (-1.28,1.27) {$Z_{3,1}$};
	\node at (0,0) {$Q_D$};
	
	\node[scale=0.8] at (1.84,2.92) {$X_{2,1}$};
	\node[scale=0.8] at (1.84,2.04) {$X_{1,1}$};
	\node[scale=0.8] at (-1.8,2.92) {$X_{4,1}$};
	\node[scale=0.8] at (-1.8,2.04) {$X_{3,1}$};

	\node at (1.45,0) {$Q_2$};
	\node at (-1.35,0) {$Q_6$};
	\draw [<->,semithick] (-0.65,0.3)--(0.65,0.3);
	\draw [<->,semithick] (-2.05,0.3)--(-0.75,0.3);
	\draw [<->,semithick] (2.05,0.3)--(0.75,0.3);
	\draw [->,semithick] (2.5,-0.7)--(0.75,-0.7);
	\draw [->,semithick] (2.5,-0.55)--(2.15,-0.55);
	\draw [semithick] (2.5,-0.7) arc(-90:90:0.3cm and 0.075cm);
	\draw [->,semithick] (-2.5,-0.7)--(-0.75,-0.7);
	\draw [->,semithick] (-2.5,-0.55)--(-2.15,-0.55);
	\draw [semithick] (-2.5,-0.7) arc(270:90:0.3cm and 0.075cm);
	\end{scope}

	\end{tikzpicture}
	\caption{Rank 2 $(D_4,D_4)$ conformal matter on a circle in the ON-plane setup. }
	\label{fig:D4D4R2_ON}	
\end{figure}

The rank 2 theory on a circle can also be realized by 5-brane web with two ON-planes as depicted in figure \ref{fig:D4D4R2_ON} in which we have labelled all the parameters that are in one-to-one correspondence with the quadrivalent gluing in figure \ref{fig:D4D4R2}. The equivalence of this two diagrams can be proven by using the topological vertex formalism with ON-planes \cite{Kim:2022dbr}.

\subsection{Flop transitions in the rank 2 quadrivalent gluing web}
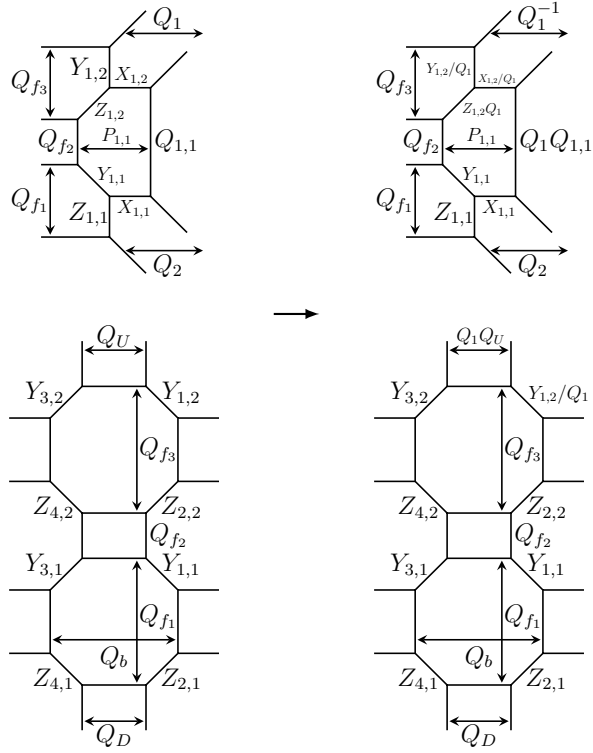
\begin{figure}[htbp]
	\centering
	\begin{tikzpicture}[scale=0.6]
	
	\begin{scope}[shift={(-0.5,0)}]
	\draw [semithick]
	(4+0.5,6.8-2)--(4+0.5,5.8-2)
	(4+0.5,6.8-2)--(4.7+0.5,7.5-2)
	(4.7+0.5,7.5-2)--(4.7+0.5,8.4-2)
	(4.7+0.5,8.4-2)--(3.2+0.5,8.4-2)
	(4.7+0.5,8.4-2)--(5.5+0.5,9.2-2)
	(4.7+0.5,7.5-2)--(5.6+0.5,7.5-2)
	(5.6+0.5,7.5-2)--(6.4+0.5,8.3-2)
	(4+0.5,5.8-2)--(3.2+0.5,5.8-2)
	(3.2+0.5,6.8-2)--(4+0.5,6.8-2)
	(4+0.5,5.8-2)--(4.7+0.5,5.1-2)
	(4.7+0.5,5.1-2)--(4.7+0.5,4.2-2)
	(4.7+0.5,4.2-2)--(3.2+0.5,4.2-2)
	(4.7+0.5,5.1-2)--(5.6+0.5,5.1-2)
	(5.6+0.5,5.1-2)--(5.6+0.5,7.5-2)
	(5.6+0.5,5.1-2)--(6.4+0.5,4.3-2)
	(4.7+0.5,4.2-2)--(5.5+0.5,3.4-2);
	\draw [<->,semithick] (5.55+0.5,6.15-2)--(4.05+0.5,6.15-2);
	\draw [<->,semithick] (6.45+0.3+0.5,4.2-0.3-2)--(4.75+0.3+0.5,4.2-0.3-2);
	\draw [<->,semithick] (6.45+0.3+0.5,8.7-2)--(4.75+0.3+0.5,8.7-2);
	\draw [<->,semithick] (3.4+0.5,4.25-2)--(3.4+0.5,5.75-2);
	\draw [<->,semithick] (3.4+0.5,4.85)--(3.4+0.5,6.35);

	\node[scale=0.6] at (5.68,5.75) {$X_{1,2}$};
	\node[scale=0.6] at (5.2,4.95) {$Z_{1,2}$};
	\node[scale=0.8] at (4.75,5.9) {$Y_{1,2}$};
	\node[scale=0.8] at (6.48,7.02) {$Q_1$};
	\node[scale=0.8] at (6.48,1.57) {$Q_2$};
	\node[scale=0.8] at (3.46,5.6) {$Q_{f_3}$};	
	\node[scale=0.8] at (3.46,3) {$Q_{f_1}$};	
	\node[scale=0.8] at (4.05,4.3) {$Q_{f_2}$};
	\node[scale=0.6] at (5.74,8.57-5.75) {$X_{1,1}$};
	\node[scale=0.6] at (5.25,8.5-4.95) {$Y_{1,1}$};
	\node[scale=0.8] at (4.75,8.6-5.9) {$Z_{1,1}$};
	\node[scale=0.6] at (5.36,4.4) {$P_{1,1}$};
	\node[scale=0.8] at (6.67,4.3) {$ Q_{1,1}$};
	\end{scope}
	
	\draw[>=latex,->,thick] (8.3,0.5)--(9.3,0.5);
	
	\begin{scope}[shift={(7.5,0)}]
	\draw [semithick]
	(4+0.5,6.8-2)--(4+0.5,5.8-2)
	(4+0.5,6.8-2)--(4.7+0.5,7.5-2)
	(4.7+0.5,7.5-2)--(4.7+0.5,8.4-2)
	(4.7+0.5,8.4-2)--(3.2+0.5,8.4-2)
	(4.7+0.5,8.4-2)--(5.5+0.5,9.2-2)
	(4.7+0.5,7.5-2)--(5.6+0.5,7.5-2)
	(5.6+0.5,7.5-2)--(6.4+0.5,8.3-2)
	(4+0.5,5.8-2)--(3.2+0.5,5.8-2)
	(3.2+0.5,6.8-2)--(4+0.5,6.8-2)
	(4+0.5,5.8-2)--(4.7+0.5,5.1-2)
	(4.7+0.5,5.1-2)--(4.7+0.5,4.2-2)
	(4.7+0.5,4.2-2)--(3.2+0.5,4.2-2)
	(4.7+0.5,5.1-2)--(5.6+0.5,5.1-2)
	(5.6+0.5,5.1-2)--(5.6+0.5,7.5-2)
	(5.6+0.5,5.1-2)--(6.4+0.5,4.3-2)
	(4.7+0.5,4.2-2)--(5.5+0.5,3.4-2);
	\draw [<->,semithick] (5.55+0.5,6.15-2)--(4.05+0.5,6.15-2);
	\draw [<->,semithick] (6.45+0.3+0.5,4.2-0.3-2)--(4.75+0.3+0.5,4.2-0.3-2);
	\draw [<->,semithick] (6.45+0.3+0.5,8.7-2)--(4.75+0.3+0.5,8.7-2);
	\draw [<->,semithick] (3.4+0.5,4.25-2)--(3.4+0.5,5.75-2);
	\draw [<->,semithick] (3.4+0.5,4.85)--(3.4+0.5,6.35);

	\node[scale=0.35] at (5.7,5.69) {$X_{1,2}/Q_1$};
	\node[scale=0.45] at (5.37,5.01) {$Z_{1,2}Q_1$};
	\node[scale=0.45] at (4.65,5.9) {$Y_{1,2}/Q_1$};
	\node[scale=0.8] at (6.62,7.08) {$Q_1^{-1}$};
	\node[scale=0.8] at (6.48,1.57) {$Q_2$};
	\node[scale=0.8] at (3.46,5.6) {$Q_{f_3}$};	
	\node[scale=0.8] at (3.46,3) {$Q_{f_1}$};	
	\node[scale=0.8] at (4.05,4.3) {$Q_{f_2}$};
	\node[scale=0.6] at (5.74,8.57-5.75) {$X_{1,1}$};
	\node[scale=0.6] at (5.25,8.5-4.95) {$Y_{1,1}$};
	\node[scale=0.8] at (4.75,8.6-5.9) {$Z_{1,1}$};
	\node[scale=0.6] at (5.36,4.4) {$P_{1,1}$};
	\node[scale=0.8] at (7,4.3) {$Q_1 Q_{1,1}$};
	\end{scope}

	\begin{scope}[shift={(5.3-0.5,-5)}]
	\draw [semithick]
	(-0.7,0.1)--(-0.7,1.1)
	(0.7,0.1)--(0.7,1.1)
	(-0.7,1.1)--(0.7,1.1)
	(-0.7,1.1)--(-1.4,1.8)
	(0.7,1.1)--(1.4,1.8)
	(-1.4,1.8)--(-2.1-0.2,1.8)
	(1.4,1.8)--(2.1+0.2,1.8)
	(1.4,1.8)--(1.4,3.2);
	\draw [semithick]
	(-0.7,4.9)--(-0.7,3.9)
	(0.7,4.9)--(0.7,3.9)
	(-0.7,3.9)--(0.7,3.9)
	(-0.7,3.9)--(-1.4,3.2)
	(0.7,3.9)--(1.4,3.2)
	(-1.4,3.2)--(-2.1-0.2,3.2)
	(-1.4,3.2)--(-1.4,1.8)
	(1.4,3.2)--(2.1+0.2,3.2);
	\draw [semithick]
	(-0.7,0.1-3.8)--(-0.7,1.1-3.8)
	(0.7,0.1-3.8)--(0.7,1.1-3.8)
	(-0.7,1.1-3.8)--(0.7,1.1-3.8)
	(-0.7,1.1-3.8)--(-1.4,1.8-3.8)
	(0.7,1.1-3.8)--(1.4,1.8-3.8)
	(-1.4,1.8-3.8)--(-2.1-0.2,1.8-3.8)
	(1.4,1.8-3.8)--(2.1+0.2,1.8-3.8)
	(1.4,1.8-3.8)--(1.4,3.2-3.8);
	\draw [semithick]
	(-0.7,3.9-3.8)--(0.7,3.9-3.8)
	(-0.7,3.9-3.8)--(-1.4,3.2-3.8)
	(0.7,3.9-3.8)--(1.4,3.2-3.8)
	(-1.4,3.2-3.8)--(-2.1-0.2,3.2-3.8)
	(-1.4,3.2-3.8)--(-1.4,1.8-3.8)
	(1.4,3.2-3.8)--(2.1+0.2,3.2-3.8);
	\draw [<->,semithick] (-1.35,2.1-3.8)--(1.35,2.1-3.8);
	\draw [<->,semithick] (0.5,1.15)--(0.5,3.85);
	\draw [<->,semithick] (0.5,1.15-3.8)--(0.5,3.85-3.8);
	\draw [<->,semithick] (-0.65,5-0.3)--(0.65,5-0.3);
	\draw [<->,semithick] (-0.65,-3.5)--(0.65,-3.5);

	\node[scale=0.8] at (0,2.15-3.8) [below] {$Q_b$};
	\node[scale=0.8] at (0.36,2.5) [right] {$Q_{f_3}$};
	\node[scale=0.8] at (0.36,2.6-3.8) [right] {$Q_{f_1}$};
	\node[scale=0.8] at (1.47,3.73) {$Y_{1,2}$};
	\node[scale=0.8] at (-1.52,3.73) {$Y_{3,2}$};
	\node[scale=0.8] at (1.47,1.2) {$Z_{2,2}$};
	\node[scale=0.8] at (-1.33,1.2) {$Z_{4,2}$};
	\node[scale=0.8] at (0,-3.83) {$Q_D$};
	\node[scale=0.8] at (0,5) {$Q_U$};
	\node[scale=0.8] at (1.2,0.6) {$Q_{f_2}$};
	\node[scale=0.8] at (1.47,3.73-3.8) {$Y_{1,1}$};
	\node[scale=0.8] at (-1.52,3.73-3.8) {$Y_{3,1}$};
	\node[scale=0.8] at (1.47,1.2-3.8) {$Z_{2,1}$};
	\node[scale=0.8] at (-1.33,1.2-3.8) {$Z_{4,1}$};	
	\end{scope}

	\begin{scope}[shift={(5.3+7.5,-5)}]
	\draw [semithick]
	(-0.7,0.1)--(-0.7,1.1)
	(0.7,0.1)--(0.7,1.1)
	(-0.7,1.1)--(0.7,1.1)
	(-0.7,1.1)--(-1.4,1.8)
	(0.7,1.1)--(1.4,1.8)
	(-1.4,1.8)--(-2.1-0.2,1.8)
	(1.4,1.8)--(2.1+0.2,1.8)
	(1.4,1.8)--(1.4,3.2);
	\draw [semithick]
	(-0.7,4.9)--(-0.7,3.9)
	(0.7,4.9)--(0.7,3.9)
	(-0.7,3.9)--(0.7,3.9)
	(-0.7,3.9)--(-1.4,3.2)
	(0.7,3.9)--(1.4,3.2)
	(-1.4,3.2)--(-2.1-0.2,3.2)
	(-1.4,3.2)--(-1.4,1.8)
	(1.4,3.2)--(2.1+0.2,3.2);
	\draw [semithick]
	(-0.7,0.1-3.8)--(-0.7,1.1-3.8)
	(0.7,0.1-3.8)--(0.7,1.1-3.8)
	(-0.7,1.1-3.8)--(0.7,1.1-3.8)
	(-0.7,1.1-3.8)--(-1.4,1.8-3.8)
	(0.7,1.1-3.8)--(1.4,1.8-3.8)
	(-1.4,1.8-3.8)--(-2.1-0.2,1.8-3.8)
	(1.4,1.8-3.8)--(2.1+0.2,1.8-3.8)
	(1.4,1.8-3.8)--(1.4,3.2-3.8);
	\draw [semithick]
	(-0.7,3.9-3.8)--(0.7,3.9-3.8)
	(-0.7,3.9-3.8)--(-1.4,3.2-3.8)
	(0.7,3.9-3.8)--(1.4,3.2-3.8)
	(-1.4,3.2-3.8)--(-2.1-0.2,3.2-3.8)
	(-1.4,3.2-3.8)--(-1.4,1.8-3.8)
	(1.4,3.2-3.8)--(2.1+0.2,3.2-3.8);
	\draw [<->,semithick] (-1.35,2.1-3.8)--(1.35,2.1-3.8);
	\draw [<->,semithick] (0.5,1.15)--(0.5,3.85);
	\draw [<->,semithick] (0.5,1.15-3.8)--(0.5,3.85-3.8);
	\draw [<->,semithick] (-0.65,5-0.3)--(0.65,5-0.3);
	\draw [<->,semithick] (-0.65,-3.5)--(0.65,-3.5);

	\node[scale=0.8] at (0,2.15-3.8) [below] {$Q_b$};
	\node[scale=0.8] at (0.36,2.5) [right] {$Q_{f_3}$};
	\node[scale=0.8] at (0.36,2.6-3.8) [right] {$Q_{f_1}$};
	\node[scale=0.6] at (1.75,3.73) {$Y_{1,2}/Q_1$};
	\node[scale=0.8] at (-1.52,3.73) {$Y_{3,2}$};
	\node[scale=0.8] at (1.47,1.2) {$Z_{2,2}$};
	\node[scale=0.8] at (-1.33,1.2) {$Z_{4,2}$};
	\node[scale=0.8] at (0,-3.83) {$Q_D$};
	\node[scale=0.6] at (0.04,5) {$Q_1Q_U$};
	\node[scale=0.8] at (1.2,0.6) {$Q_{f_2}$};
	\node[scale=0.8] at (1.47,3.73-3.8) {$Y_{1,1}$};
	\node[scale=0.8] at (-1.52,3.73-3.8) {$Y_{3,1}$};
	\node[scale=0.8] at (1.47,1.2-3.8) {$Z_{2,1}$};
	\node[scale=0.8] at (-1.33,1.2-3.8) {$Z_{4,1}$};	
	\end{scope}

	\end{tikzpicture}
	\caption{Flop transition related to $Q_1$ in rank 2 $(D_4,D_4)$ conformal matter on a circle. }
	\label{fig:R2_Q1_flop}
\end{figure}

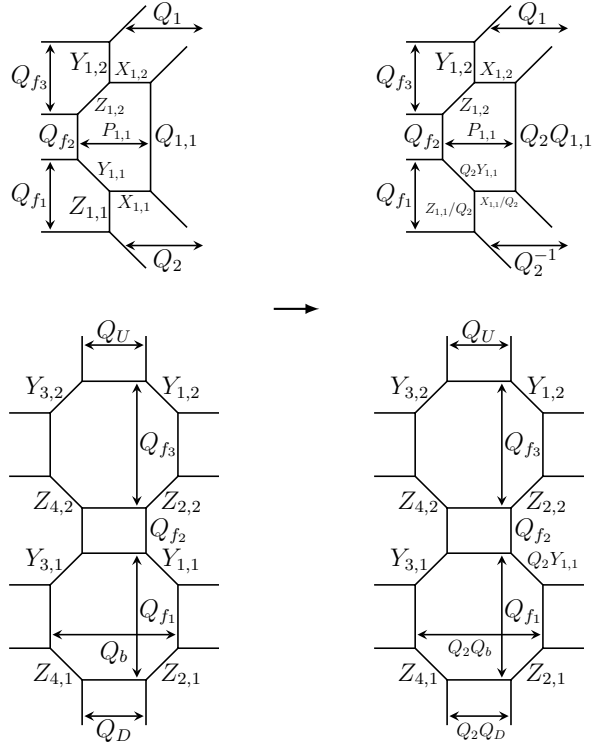
\begin{figure}[htbp]
	\centering
	\begin{tikzpicture}[scale=0.6]
	
	\begin{scope}[shift={(-0.5,0)}]
	\draw [semithick]
	(4+0.5,6.8-2)--(4+0.5,5.8-2)
	(4+0.5,6.8-2)--(4.7+0.5,7.5-2)
	(4.7+0.5,7.5-2)--(4.7+0.5,8.4-2)
	(4.7+0.5,8.4-2)--(3.2+0.5,8.4-2)
	(4.7+0.5,8.4-2)--(5.5+0.5,9.2-2)
	(4.7+0.5,7.5-2)--(5.6+0.5,7.5-2)
	(5.6+0.5,7.5-2)--(6.4+0.5,8.3-2)
	(4+0.5,5.8-2)--(3.2+0.5,5.8-2)
	(3.2+0.5,6.8-2)--(4+0.5,6.8-2)
	(4+0.5,5.8-2)--(4.7+0.5,5.1-2)
	(4.7+0.5,5.1-2)--(4.7+0.5,4.2-2)
	(4.7+0.5,4.2-2)--(3.2+0.5,4.2-2)
	(4.7+0.5,5.1-2)--(5.6+0.5,5.1-2)
	(5.6+0.5,5.1-2)--(5.6+0.5,7.5-2)
	(5.6+0.5,5.1-2)--(6.4+0.5,4.3-2)
	(4.7+0.5,4.2-2)--(5.5+0.5,3.4-2);
	\draw [<->,semithick] (5.55+0.5,6.15-2)--(4.05+0.5,6.15-2);
	\draw [<->,semithick] (6.45+0.3+0.5,4.2-0.3-2)--(4.75+0.3+0.5,4.2-0.3-2);
	\draw [<->,semithick] (6.45+0.3+0.5,8.7-2)--(4.75+0.3+0.5,8.7-2);
	\draw [<->,semithick] (3.4+0.5,4.25-2)--(3.4+0.5,5.75-2);
	\draw [<->,semithick] (3.4+0.5,4.85)--(3.4+0.5,6.35);

	\node[scale=0.6] at (5.68,5.75) {$X_{1,2}$};
	\node[scale=0.6] at (5.2,4.95) {$Z_{1,2}$};
	\node[scale=0.8] at (4.75,5.9) {$Y_{1,2}$};
	\node[scale=0.8] at (6.48,7.02) {$Q_1$};
	\node[scale=0.8] at (6.48,1.57) {$Q_2$};
	\node[scale=0.8] at (3.46,5.6) {$Q_{f_3}$};	
	\node[scale=0.8] at (3.46,3) {$Q_{f_1}$};	
	\node[scale=0.8] at (4.07,4.3) {$Q_{f_2}$};
	\node[scale=0.6] at (5.74,8.57-5.75) {$X_{1,1}$};
	\node[scale=0.6] at (5.25,8.5-4.95) {$Y_{1,1}$};
	\node[scale=0.8] at (4.75,8.6-5.9) {$Z_{1,1}$};
	\node[scale=0.6] at (5.36,4.4) {$P_{1,1}$};
	\node[scale=0.8] at (6.67,4.3) {$ Q_{1,1}$};
	\end{scope}
	
	\draw[>=latex,->,thick] (8.3,0.5)--(9.3,0.5);
	
	\begin{scope}[shift={(7.5,0)}]
	\draw [semithick]
	(4+0.5,6.8-2)--(4+0.5,5.8-2)
	(4+0.5,6.8-2)--(4.7+0.5,7.5-2)
	(4.7+0.5,7.5-2)--(4.7+0.5,8.4-2)
	(4.7+0.5,8.4-2)--(3.2+0.5,8.4-2)
	(4.7+0.5,8.4-2)--(5.5+0.5,9.2-2)
	(4.7+0.5,7.5-2)--(5.6+0.5,7.5-2)
	(5.6+0.5,7.5-2)--(6.4+0.5,8.3-2)
	(4+0.5,5.8-2)--(3.2+0.5,5.8-2)
	(3.2+0.5,6.8-2)--(4+0.5,6.8-2)
	(4+0.5,5.8-2)--(4.7+0.5,5.1-2)
	(4.7+0.5,5.1-2)--(4.7+0.5,4.2-2)
	(4.7+0.5,4.2-2)--(3.2+0.5,4.2-2)
	(4.7+0.5,5.1-2)--(5.6+0.5,5.1-2)
	(5.6+0.5,5.1-2)--(5.6+0.5,7.5-2)
	(5.6+0.5,5.1-2)--(6.4+0.5,4.3-2)
	(4.7+0.5,4.2-2)--(5.5+0.5,3.4-2);
	\draw [<->,semithick] (5.55+0.5,6.15-2)--(4.05+0.5,6.15-2);
	\draw [<->,semithick] (6.45+0.3+0.5,4.2-0.3-2)--(4.75+0.3+0.5,4.2-0.3-2);
	\draw [<->,semithick] (6.45+0.3+0.5,8.7-2)--(4.75+0.3+0.5,8.7-2);
	\draw [<->,semithick] (3.4+0.5,4.25-2)--(3.4+0.5,5.75-2);
	\draw [<->,semithick] (3.4+0.5,4.85)--(3.4+0.5,6.35);

	\node[scale=0.6] at (5.68,5.75) {$X_{1,2}$};
	\node[scale=0.6] at (5.2,4.95) {$Z_{1,2}$};
	\node[scale=0.8] at (4.75,5.9) {$Y_{1,2}$};
	\node[scale=0.8] at (6.48,7.02) {$Q_1$};
	\node[scale=0.8] at (6.55,1.52) {$Q_2^{-1}$};
	\node[scale=0.8] at (3.46,5.6) {$Q_{f_3}$};	
	\node[scale=0.8] at (3.46,3) {$Q_{f_1}$};	
	\node[scale=0.8] at (4.07,4.3) {$Q_{f_2}$};
	\node[scale=0.35] at (5.74,8.6-5.75) {$X_{1,1}/Q_2$};
	\node[scale=0.45] at (5.3,8.5-4.95) {$Q_2Y_{1,1}$};
	\node[scale=0.45] at (4.64,8.6-5.9) {$Z_{1,1}/Q_2$};
	\node[scale=0.6] at (5.36,4.4) {$P_{1,1}$};
	\node[scale=0.8] at (7,4.3) {$Q_2 Q_{1,1}$};
	\end{scope}

	\begin{scope}[shift={(5.3-0.5,-5)}]
	\draw [semithick]
	(-0.7,0.1)--(-0.7,1.1)
	(0.7,0.1)--(0.7,1.1)
	(-0.7,1.1)--(0.7,1.1)
	(-0.7,1.1)--(-1.4,1.8)
	(0.7,1.1)--(1.4,1.8)
	(-1.4,1.8)--(-2.1-0.2,1.8)
	(1.4,1.8)--(2.1+0.2,1.8)
	(1.4,1.8)--(1.4,3.2);
	\draw [semithick]
	(-0.7,4.9)--(-0.7,3.9)
	(0.7,4.9)--(0.7,3.9)
	(-0.7,3.9)--(0.7,3.9)
	(-0.7,3.9)--(-1.4,3.2)
	(0.7,3.9)--(1.4,3.2)
	(-1.4,3.2)--(-2.1-0.2,3.2)
	(-1.4,3.2)--(-1.4,1.8)
	(1.4,3.2)--(2.1+0.2,3.2);
	\draw [semithick]
	(-0.7,0.1-3.8)--(-0.7,1.1-3.8)
	(0.7,0.1-3.8)--(0.7,1.1-3.8)
	(-0.7,1.1-3.8)--(0.7,1.1-3.8)
	(-0.7,1.1-3.8)--(-1.4,1.8-3.8)
	(0.7,1.1-3.8)--(1.4,1.8-3.8)
	(-1.4,1.8-3.8)--(-2.1-0.2,1.8-3.8)
	(1.4,1.8-3.8)--(2.1+0.2,1.8-3.8)
	(1.4,1.8-3.8)--(1.4,3.2-3.8);
	\draw [semithick]
	(-0.7,3.9-3.8)--(0.7,3.9-3.8)
	(-0.7,3.9-3.8)--(-1.4,3.2-3.8)
	(0.7,3.9-3.8)--(1.4,3.2-3.8)
	(-1.4,3.2-3.8)--(-2.1-0.2,3.2-3.8)
	(-1.4,3.2-3.8)--(-1.4,1.8-3.8)
	(1.4,3.2-3.8)--(2.1+0.2,3.2-3.8);
	\draw [<->,semithick] (-1.35,2.1-3.8)--(1.35,2.1-3.8);
	\draw [<->,semithick] (0.5,1.15)--(0.5,3.85);
	\draw [<->,semithick] (0.5,1.15-3.8)--(0.5,3.85-3.8);
	\draw [<->,semithick] (-0.65,5-0.3)--(0.65,5-0.3);
	\draw [<->,semithick] (-0.65,-3.5)--(0.65,-3.5);

	\node[scale=0.8] at (0,2.15-3.8) [below] {$Q_b$};
	\node[scale=0.8] at (0.36,2.5) [right] {$Q_{f_3}$};
	\node[scale=0.8] at (0.36,2.6-3.8) [right] {$Q_{f_1}$};
	\node[scale=0.8] at (1.47,3.73) {$Y_{1,2}$};
	\node[scale=0.8] at (-1.52,3.73) {$Y_{3,2}$};
	\node[scale=0.8] at (1.47,1.2) {$Z_{2,2}$};
	\node[scale=0.8] at (-1.33,1.2) {$Z_{4,2}$};
	\node[scale=0.8] at (0,-3.83) {$Q_D$};
	\node[scale=0.8] at (0,5) {$Q_U$};
	\node[scale=0.8] at (1.2,0.6) {$Q_{f_2}$};
	\node[scale=0.8] at (1.47,3.73-3.8) {$Y_{1,1}$};
	\node[scale=0.8] at (-1.52,3.73-3.8) {$Y_{3,1}$};
	\node[scale=0.8] at (1.47,1.2-3.8) {$Z_{2,1}$};
	\node[scale=0.8] at (-1.33,1.2-3.8) {$Z_{4,1}$};	
	\end{scope}

	\begin{scope}[shift={(5.3+7.5,-5)}]
	\draw [semithick]
	(-0.7,0.1)--(-0.7,1.1)
	(0.7,0.1)--(0.7,1.1)
	(-0.7,1.1)--(0.7,1.1)
	(-0.7,1.1)--(-1.4,1.8)
	(0.7,1.1)--(1.4,1.8)
	(-1.4,1.8)--(-2.1-0.2,1.8)
	(1.4,1.8)--(2.1+0.2,1.8)
	(1.4,1.8)--(1.4,3.2);
	\draw [semithick]
	(-0.7,4.9)--(-0.7,3.9)
	(0.7,4.9)--(0.7,3.9)
	(-0.7,3.9)--(0.7,3.9)
	(-0.7,3.9)--(-1.4,3.2)
	(0.7,3.9)--(1.4,3.2)
	(-1.4,3.2)--(-2.1-0.2,3.2)
	(-1.4,3.2)--(-1.4,1.8)
	(1.4,3.2)--(2.1+0.2,3.2);
	\draw [semithick]
	(-0.7,0.1-3.8)--(-0.7,1.1-3.8)
	(0.7,0.1-3.8)--(0.7,1.1-3.8)
	(-0.7,1.1-3.8)--(0.7,1.1-3.8)
	(-0.7,1.1-3.8)--(-1.4,1.8-3.8)
	(0.7,1.1-3.8)--(1.4,1.8-3.8)
	(-1.4,1.8-3.8)--(-2.1-0.2,1.8-3.8)
	(1.4,1.8-3.8)--(2.1+0.2,1.8-3.8)
	(1.4,1.8-3.8)--(1.4,3.2-3.8);
	\draw [semithick]
	(-0.7,3.9-3.8)--(0.7,3.9-3.8)
	(-0.7,3.9-3.8)--(-1.4,3.2-3.8)
	(0.7,3.9-3.8)--(1.4,3.2-3.8)
	(-1.4,3.2-3.8)--(-2.1-0.2,3.2-3.8)
	(-1.4,3.2-3.8)--(-1.4,1.8-3.8)
	(1.4,3.2-3.8)--(2.1+0.2,3.2-3.8);
	\draw [<->,semithick] (-1.35,2.1-3.8)--(1.35,2.1-3.8);
	\draw [<->,semithick] (0.5,1.15)--(0.5,3.85);
	\draw [<->,semithick] (0.5,1.15-3.8)--(0.5,3.85-3.8);
	\draw [<->,semithick] (-0.65,5-0.3)--(0.65,5-0.3);
	\draw [<->,semithick] (-0.65,-3.5)--(0.65,-3.5);

	\node[scale=0.6] at (-0.2,2.13-3.8) [below] {$Q_2Q_b$};
	\node[scale=0.8] at (0.36,2.5) [right] {$Q_{f_3}$};
	\node[scale=0.8] at (0.36,2.6-3.8) [right] {$Q_{f_1}$};
	\node[scale=0.8] at (1.47,3.73) {$Y_{1,2}$};
	\node[scale=0.8] at (-1.52,3.73) {$Y_{3,2}$};
	\node[scale=0.8] at (1.47,1.2) {$Z_{2,2}$};
	\node[scale=0.8] at (-1.33,1.2) {$Z_{4,2}$};
	\node[scale=0.6] at (0.05,-3.83) {$Q_2Q_D$};
	\node[scale=0.8] at (0,5) {$Q_U$};
	\node[scale=0.8] at (1.2,0.6) {$Q_{f_2}$};
	\node[scale=0.6] at (1.62,3.73-3.8) {$Q_2Y_{1,1}$};
	\node[scale=0.8] at (-1.52,3.73-3.8) {$Y_{3,1}$};
	\node[scale=0.8] at (1.47,1.2-3.8) {$Z_{2,1}$};
	\node[scale=0.8] at (-1.33,1.2-3.8) {$Z_{4,1}$};	
	\end{scope}

	\end{tikzpicture}
	\caption{Flop transition related to $Q_2$ in rank 2 $(D_4,D_4)$ conformal matter on a circle. }
	\label{fig:R2_Q2_flop}
\end{figure}

\begin{figure}[htbp]
	\centering
	\begin{tikzpicture}[scale=0.8]
	
	\draw [semithick]
	(-0.7,0.1)--(-0.7,1.1)
	(0.7,0.1)--(0.7,1.1)
	(-0.7,1.1)--(0.7,1.1)
	(-0.7,1.1)--(-1.4,1.8)
	(0.7,1.1)--(1.4,1.8)
	(1.4,1.8)--(1.4,3.2)
	;
	\draw [semithick,red]
	(1.4,3.2)--(2.1+0.2,3.2)
	(-1.4,3.2)--(-2.1-0.2,3.2)
	(-1.4,3.2-3.8)--(-2.1-0.2,3.2-3.8)
	(1.4,3.2-3.8)--(2.1+0.2,3.2-3.8)
	;
	\draw [semithick,cyan]
	(-1.4,1.8)--(-2.1-0.2,1.8)
	(1.4,1.8)--(2.1+0.2,1.8)
	(-1.4,1.8-3.8)--(-2.1-0.2,1.8-3.8)
	(1.4,1.8-3.8)--(2.1+0.2,1.8-3.8)
	;
	\draw [semithick]
	(-0.7,4.9)--(-0.7,3.9)
	(0.7,4.9)--(0.7,3.9)
	(-0.7,3.9)--(0.7,3.9)
	(-0.7,3.9)--(-1.4,3.2)
	(0.7,3.9)--(1.4,3.2)
	(-1.4,3.2)--(-1.4,1.8)
	;
	\draw [semithick]
	(-0.7,0.1-3.8)--(-0.7,1.1-3.8)
	(0.7,0.1-3.8)--(0.7,1.1-3.8)
	(-0.7,1.1-3.8)--(0.7,1.1-3.8)
	(-0.7,1.1-3.8)--(-1.4,1.8-3.8)
	(0.7,1.1-3.8)--(1.4,1.8-3.8)

	(1.4,1.8-3.8)--(1.4,3.2-3.8)
	;
	\draw [semithick]
	(-0.7,3.9-3.8)--(0.7,3.9-3.8)
	(-0.7,3.9-3.8)--(-1.4,3.2-3.8)
	(0.7,3.9-3.8)--(1.4,3.2-3.8)
	
	(-1.4,3.2-3.8)--(-1.4,1.8-3.8)
	
	;
	\draw [<->,semithick] (-1.35,2.1-3.8)--(1.35,2.1-3.8);
	\draw [<->,semithick] (0.5,1.15)--(0.5,3.85);
	\draw [<->,semithick] (0.5,1.15-3.8)--(0.5,3.85-3.8);
	\draw [<->,semithick] (-0.65,5-0.3)--(0.65,5-0.3);
	\draw [<->,semithick] (-0.65,-3.5)--(0.65,-3.5);

	\draw [semithick]
	(4+0.5,6.8-2)--(4+0.5,5.8-2)
	(4+0.5,6.8-2)--(4.7+0.5,7.5-2)
	(4.7+0.5,7.5-2)--(4.7+0.5,8.4-2)
	(4.7+0.5,8.4-2)--(3.2+0.5,8.4-2)
	(4.7+0.5,8.4-2)--(5.5+0.5,9.2-2)
	(5.6+0.5,7.5-2)--(6.4+0.5,8.3-2)
	(4+0.5,5.8-2)--(3.2+0.5,5.8-2)
	(3.2+0.5,6.8-2)--(4+0.5,6.8-2)
	(4+0.5,5.8-2)--(4.7+0.5,5.1-2)
	(4.7+0.5,5.1-2)--(4.7+0.5,4.2-2)
	(4.7+0.5,4.2-2)--(3.2+0.5,4.2-2)
	(5.6+0.5,5.1-2)--(5.6+0.5,7.5-2)
	(5.6+0.5,5.1-2)--(6.4+0.5,4.3-2)
	(4.7+0.5,4.2-2)--(5.5+0.5,3.4-2);
	\draw [semithick,red]
	(4.7+0.5,7.5-2)--(5.6+0.5,7.5-2)
	(4.7+0.5,5.1-2)--(5.6+0.5,5.1-2)
	;
	\draw [<->,semithick] (5.55+0.5,6.15-2)--(4.05+0.5,6.15-2);
	\draw [<->,semithick] (6.45+0.3+0.5,4.2-0.3-2)--(4.75+0.3+0.5,4.2-0.3-2);
	\draw [<->,semithick] (6.45+0.3+0.5,8.7-2)--(4.75+0.3+0.5,8.7-2);
	\draw [<->,semithick] (3.4+0.5,4.25-2)--(3.4+0.5,5.75-2);
	\draw [<->,semithick] (3.4+0.5,4.85)--(3.4+0.5,6.35);

	\draw [semithick]
	(4+0.5,1.2-6.8+2)--(4+0.5,1.2-5.8+2)
	(4+0.5,1.2-6.8+2)--(4.7+0.5,1.2-7.5+2)
	(4.7+0.5,1.2-7.5+2)--(4.7+0.5,1.2-8.4+2)
	(4.7+0.5,1.2-8.4+2)--(3.2+0.5,1.2-8.4+2)
	(4.7+0.5,1.2-8.4+2)--(5.5+0.5,1.2-9.2+2)
	(5.6+0.5,1.2-7.5+2)--(6.4+0.5,1.2-8.3+2)
	(4+0.5,1.2-5.8+2)--(3.2+0.5,1.2-5.8+2)
	(3.2+0.5,1.2-6.8+2)--(4+0.5,1.2-6.8+2)
	(4+0.5,1.2-5.8+2)--(4.7+0.5,1.2-5.1+2)
	(4.7+0.5,1.2-5.1+2)--(4.7+0.5,1.2-4.2+2)
	(4.7+0.5,1.2-4.2+2)--(3.2+0.5,1.2-4.2+2)
	(5.6+0.5,1.2-5.1+2)--(5.6+0.5,1.2-7.5+2)
	(5.6+0.5,1.2-5.1+2)--(6.4+0.5,1.2-4.3+2)
	(4.7+0.5,1.2-4.2+2)--(5.5+0.5,1.2-3.4+2);
	\draw [semithick,cyan]
	(4.7+0.5,1.2-7.5+2)--(5.6+0.5,1.2-7.5+2)
	(4.7+0.5,1.2-5.1+2)--(5.6+0.5,1.2-5.1+2)
	;
	\draw [<->,semithick] (5.55+0.5,1.2-6.45+2)--(4.05+0.5,1.2-6.45+2);
	\draw [<->,semithick] (6.45+0.3+0.5,1.2-4.2+0.3+2)--(4.75+0.3+0.5,1.2-4.2+0.3+2);
	\draw [<->,semithick] (6.45+0.3+0.5,1.2-8.7+2)--(4.75+0.3+0.5,1.2-8.7+2);
	\draw [<->,semithick] (3.4+0.5,1.2-4.25+2)--(3.4+0.5,1.2-5.75+2);
	\draw [<->,semithick] (3.4+0.5,1.2-4.25-2.6+2)--(3.4+0.5,1.2-5.75-2.6+2);

	\draw [semithick]
	(-4-0.5,6.8-2)--(-4-0.5,5.8-2)
	(-4-0.5,6.8-2)--(-4.7-0.5,7.5-2)
	(-4.7-0.5,7.5-2)--(-4.7-0.5,8.4-2)
	(-4.7-0.5,8.4-2)--(-3.2-0.5,8.4-2)
	(-4.7-0.5,8.4-2)--(-5.5-0.5,9.2-2)
	(-5.6-0.5,7.5-2)--(-6.4-0.5,8.3-2)
	(-4-0.5,5.8-2)--(-3.2-0.5,5.8-2)
	(-3.2-0.5,6.8-2)--(-4-0.5,6.8-2)
	(-4-0.5,5.8-2)--(-4.7-0.5,5.1-2)
	(-4.7-0.5,5.1-2)--(-4.7-0.5,4.2-2)
	(-4.7-0.5,4.2-2)--(-3.2-0.5,4.2-2)
	(-5.6-0.5,5.1-2)--(-5.6-0.5,7.5-2)
	(-5.6-0.5,5.1-2)--(-6.4-0.5,4.3-2)
	(-4.7-0.5,4.2-2)--(-5.5-0.5,3.4-2);
	\draw [semithick,red]
	(-4.7-0.5,7.5-2)--(-5.6-0.5,7.5-2)
	(-4.7-0.5,5.1-2)--(-5.6-0.5,5.1-2)
	;
	\draw [<->,semithick] (-5.55-0.5,6.15-2)--(-4.05-0.5,6.15-2);
	\draw [<->,semithick] (-6.45-0.3-0.5,4.2-0.3-2)--(-4.75-0.3-0.5,4.2-0.3-2);
	\draw [<->,semithick] (-6.45-0.3-0.5,8.7-2)--(-4.75-0.3-0.5,8.7-2);
	\draw [<->,semithick] (-3.4-0.5,4.25-2)--(-3.4-0.5,5.75-2);
	\draw [<->,semithick] (-3.4-0.5,4.25+2.6-2)--(-3.4-0.5,5.75+2.6-2);

	\draw [semithick]
	(-4-0.5,1.2-6.8+2)--(-4-0.5,1.2-5.8+2)
	(-4-0.5,1.2-6.8+2)--(-4.7-0.5,1.2-7.5+2)
	(-4.7-0.5,1.2-7.5+2)--(-4.7-0.5,1.2-8.4+2)
	(-4.7-0.5,1.2-8.4+2)--(-3.2-0.5,1.2-8.4+2)
	(-4.7-0.5,1.2-8.4+2)--(-5.5-0.5,1.2-9.2+2)
	(-5.6-0.5,1.2-7.5+2)--(-6.4-0.5,1.2-8.3+2)
	(-4-0.5,1.2-5.8+2)--(-3.2-0.5,1.2-5.8+2)
	(-3.2-0.5,1.2-6.8+2)--(-4-0.5,1.2-6.8+2)
	(-4-0.5,1.2-5.8+2)--(-4.7-0.5,1.2-5.1+2)
	(-4.7-0.5,1.2-5.1+2)--(-4.7-0.5,1.2-4.2+2)
	(-4.7-0.5,1.2-4.2+2)--(-3.2-0.5,1.2-4.2+2)
	(-5.6-0.5,1.2-5.1+2)--(-5.6-0.5,1.2-7.5+2)
	(-5.6-0.5,1.2-5.1+2)--(-6.4-0.5,1.2-4.3+2)
	(-4.7-0.5,1.2-4.2+2)--(-5.5-0.5,1.2-3.4+2);
	\draw [semithick,cyan]
	(-4.7-0.5,1.2-7.5+2)--(-5.6-0.5,1.2-7.5+2)
	(-4.7-0.5,1.2-5.1+2)--(-5.6-0.5,1.2-5.1+2)
	;
	\draw [<->,semithick] (-5.55-0.5,1.2-6.45+2)--(-4.05-0.5,1.2-6.45+2);
	\draw [<->,semithick] (-6.45-0.3-0.5,1.2-4.2+0.3+2)--(-4.75-0.3-0.5,1.2-4.2+0.3+2);
	\draw [<->,semithick] (-6.45-0.3-0.5,1.2-8.7+2)--(-4.75-0.3-0.5,1.2-8.7+2);
	\draw [<->,semithick] (-3.4-0.5,1.2-4.25+2)--(-3.4-0.5,1.2-5.75+2);
	\draw [<->,semithick] (-3.4-0.5,1.2-4.25-2.6+2)--(-3.4-0.5,1.2-5.75-2.6+2);

	\node at (0,2.15-3.8) [below] {$Q_b$};
	\node[scale=0.53] at (0.45,2.5) [right] {$Q_UQ_{f_3}$};
	\node at (0.4,2.6-3.8) [right] {$Q_{f_1}$};
	\node[scale=0.8] at (1.66,3.73) {$Q_UY_{1,2}$};
	\node[scale=0.8] at (-1.69,3.73) {$Q_UY_{3,2}$};
	\node at (1.47,1.2) {$Z_{2,2}$};
	\node at (-1.33,1.2) {$Z_{4,2}$};
	\node at (0,-3.83) {$Q_D$};
	\node at (0.1,5.1) {$Q_U^{-1}$};
	\node at (1.2,0.6) {$Q_{f_2}$};
	\node at (1.47,3.73-3.8) {$Y_{1,1}$};
	\node at (-1.52,3.73-3.8) {$Y_{3,1}$};
	\node at (1.47,1.2-3.8) {$Z_{2,1}$};
	\node at (-1.33,1.2-3.8) {$Z_{4,1}$};

	\node[scale=0.8] at (5.68,5.75) {$X_{1,2}$};
	\node[scale=0.8] at (5.2,4.95) {$Z_{1,2}$};
	\node[scale=0.6] at (4.68,5.9) {$Q_UY_{1,2}$};
	\node[scale=0.8] at (6.58,7.02) {$Q_UQ_1$};
	\node at (6.48,1.57) {$Q_2$};
	\node at (3.1,5.6) {$Q_UQ_{f_3}$};	
	\node at (3.46,3) {$Q_{f_1}$};	
	\node at (4.07,4.3) {$Q_{f_2}$};
	\node[scale=0.8] at (5.74,8.57-5.75) {$X_{1,1}$};
	\node[scale=0.8] at (5.25,8.5-4.95) {$Y_{1,1}$};
	\node at (4.75,8.6-5.9) {$Z_{1,1}$};
	\node[scale=0.8] at (5.36,4.4) {$P_{1,1}$};
	\node at (6.67,4.3) {$ Q_{1,1}$};

	\node[scale=0.8] at (5.68,5.75-7.4) {$X_{2,2}$};
	\node[scale=0.8] at (5.2,4.95-7.4) {$Z_{2,2}$};
	\node[scale=0.6] at (4.68,5.9-7.4) {$Q_UY_{2,2}$};
	\node[scale=0.8] at (6.58,7.02-7.4) {$Q_UQ_3$};
	\node at (6.48,1.57-7.4) {$Q_4$};
	\node[scale=0.8] at (3.25,5.6-7.4) {$Q_UQ_{f_3}$};	
	\node at (3.46,3-7.4) {$Q_{f_1}$};	
	\node at (4.07,4.3-7.4) {$Q_{f_2}$};
	\node[scale=0.8] at (5.74,8.57-5.75-7.4) {$X_{2,1}$};
	\node[scale=0.8] at (5.25,8.5-4.95-7.4) {$Y_{2,1}$};
	\node at (4.75,8.6-5.9-7.4) {$Z_{2,1}$};
	\node[scale=0.8] at (5.36,4.4-7.4) {$P_{2,1}$};	
	\node at (6.67,4.3-7.4) {$ Q_{2,1}$};

	\node[scale=0.8] at (-5.68,5.75) {$X_{3,2}$};
	\node[scale=0.8] at (-5.1,4.95) {$Z_{3,2}$};
	\node[scale=0.6] at (-4.67,5.9) {$Q_UY_{3,2}$};
	\node[scale=0.8] at (-6.52,7.02) {$Q_UQ_5$};
	\node at (-6.48,1.57) {$Q_6$};
	\node at (-3.05,5.6) {$Q_UQ_{f_3}$};	
	\node at (-3.43,3) {$Q_{f_1}$};	
	\node at (-4.03,4.3) {$Q_{f_2}$};
	\node[scale=0.8] at (-5.67,8.57-5.75) {$X_{3,1}$};
	\node[scale=0.8] at (-5.25,8.5-4.95) {$Y_{3,1}$};
	\node at (-4.72,8.6-5.9) {$Z_{3,1}$};
	\node[scale=0.8] at (-5.36,4.4) {$P_{3,1}$};
	\node at (-6.67,4.3) {$ Q_{3,1}$};

	\node[scale=0.8] at (-5.68,5.75-7.4) {$X_{4,2}$};
	\node[scale=0.8] at (-5.1,4.95-7.4) {$Z_{4,2}$};
	\node[scale=0.6] at (-4.67,5.9-7.4) {$Q_UY_{4,2}$};
	\node[scale=0.8] at (-6.52,7.02-7.4) {$Q_UQ_7$};
	\node at (-6.48,1.57-7.4) {$Q_8$};
	\node[scale=0.8] at (-3.22,5.6-7.4) {$Q_UQ_{f_3}$};	
	\node at (-3.43,3-7.4) {$Q_{f_1}$};	
	\node at (-4.03,4.3-7.4) {$Q_{f_2}$};
	\node[scale=0.8] at (-5.67,8.57-5.75-7.4) {$X_{4,1}$};
	\node[scale=0.8] at (-5.25,8.5-4.95-7.4) {$Y_{4,1}$};
	\node at (-4.72,8.6-5.9-7.4) {$Z_{4,1}$};
	\node[scale=0.8] at (-5.36,4.4-7.4) {$P_{4,1}$};	
	\node at (-6.67,4.3-7.4) {$ Q_{4,1}$};

	\end{tikzpicture}
\caption{Flop transition related to $Q_U$ in rank 2 $(D_4,D_4)$ conformal matter on a circle. }
\label{fig:R2_QU_flop}
\end{figure}
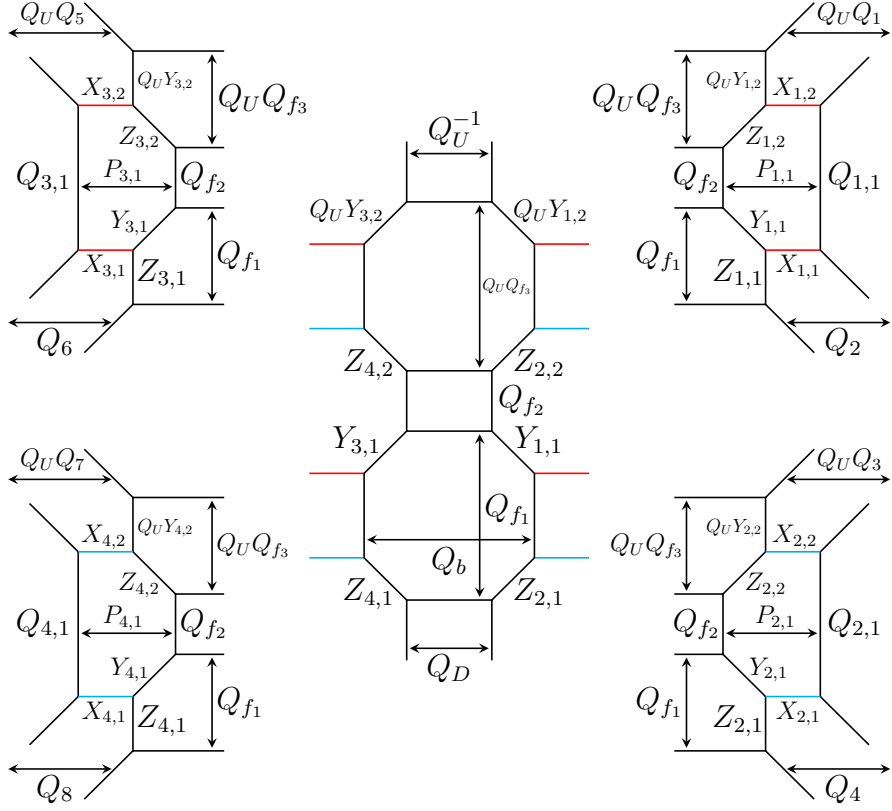

From the experience of rank 1 case, 
we expect that the exchanging of external parallel branes still give rise to the affine $E_8$ symmetry, so we still consider the flop transitions related to these exchangings in the rank 2 case.
For K\"ahler parameter $Q_1$, $Q_2$ and $Q_U$, the corresponding flop transitions are depicted in figure \ref{fig:R2_Q1_flop}, \ref{fig:R2_Q2_flop} and \ref{fig:R2_QU_flop} respectively which are similar to the rank 1 case, other flop transitions related to the exchanging of external parallel branes can also be derived in the similar way.

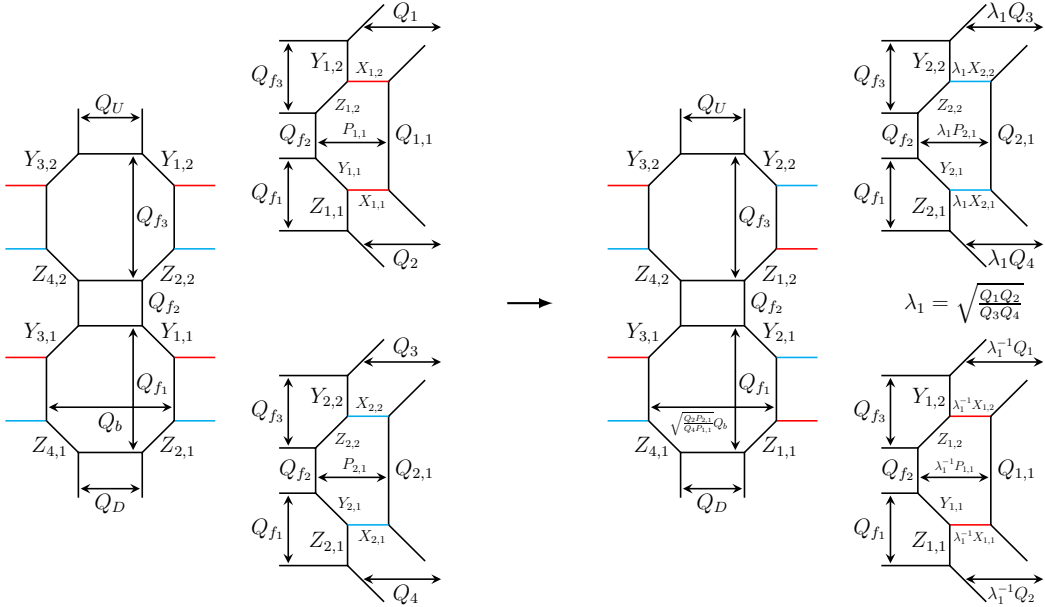
\begin{figure}[htbp]
	\centering
	\begin{tikzpicture}[scale=0.6]
	\draw [semithick]
	(-0.7,0.1)--(-0.7,1.1)
	(0.7,0.1)--(0.7,1.1)
	(-0.7,1.1)--(0.7,1.1)
	(-0.7,1.1)--(-1.4,1.8)
	(0.7,1.1)--(1.4,1.8)
	(1.4,1.8)--(1.4,3.2)
	;
	\draw [semithick,red]
	(1.4,3.2)--(2.1+0.2,3.2)
	(-1.4,3.2)--(-2.1-0.2,3.2)
	(-1.4,3.2-3.8)--(-2.1-0.2,3.2-3.8)
	(1.4,3.2-3.8)--(2.1+0.2,3.2-3.8)
	;
	\draw [semithick,cyan]
	(-1.4,1.8)--(-2.1-0.2,1.8)
	(1.4,1.8)--(2.1+0.2,1.8)
	(-1.4,1.8-3.8)--(-2.1-0.2,1.8-3.8)
	(1.4,1.8-3.8)--(2.1+0.2,1.8-3.8)
	;
	\draw [semithick]
	(-0.7,4.9)--(-0.7,3.9)
	(0.7,4.9)--(0.7,3.9)
	(-0.7,3.9)--(0.7,3.9)
	(-0.7,3.9)--(-1.4,3.2)
	(0.7,3.9)--(1.4,3.2)
	(-1.4,3.2)--(-1.4,1.8)
	;
	\draw [semithick]
	(-0.7,0.1-3.8)--(-0.7,1.1-3.8)
	(0.7,0.1-3.8)--(0.7,1.1-3.8)
	(-0.7,1.1-3.8)--(0.7,1.1-3.8)
	(-0.7,1.1-3.8)--(-1.4,1.8-3.8)
	(0.7,1.1-3.8)--(1.4,1.8-3.8)
	(1.4,1.8-3.8)--(1.4,3.2-3.8)
	;
	\draw [semithick]
	(-0.7,3.9-3.8)--(0.7,3.9-3.8)
	(-0.7,3.9-3.8)--(-1.4,3.2-3.8)
	(0.7,3.9-3.8)--(1.4,3.2-3.8)
	(-1.4,3.2-3.8)--(-1.4,1.8-3.8)
	;
	\draw [<->,semithick] (-1.35,2.1-3.8)--(1.35,2.1-3.8);
	\draw [<->,semithick] (0.5,1.15)--(0.5,3.85);
	\draw [<->,semithick] (0.5,1.15-3.8)--(0.5,3.85-3.8);
	\draw [<->,semithick] (-0.65,5-0.3)--(0.65,5-0.3);
	\draw [<->,semithick] (-0.65,-3.5)--(0.65,-3.5);
	
	\draw [semithick]
	(4+0.5,6.8-2)--(4+0.5,5.8-2)
	(4+0.5,6.8-2)--(4.7+0.5,7.5-2)
	(4.7+0.5,7.5-2)--(4.7+0.5,8.4-2)
	(4.7+0.5,8.4-2)--(3.2+0.5,8.4-2)
	(4.7+0.5,8.4-2)--(5.5+0.5,9.2-2)
	(5.6+0.5,7.5-2)--(6.4+0.5,8.3-2)
	(4+0.5,5.8-2)--(3.2+0.5,5.8-2)
	(3.2+0.5,6.8-2)--(4+0.5,6.8-2)
	(4+0.5,5.8-2)--(4.7+0.5,5.1-2)
	(4.7+0.5,5.1-2)--(4.7+0.5,4.2-2)
	(4.7+0.5,4.2-2)--(3.2+0.5,4.2-2)
	(5.6+0.5,5.1-2)--(5.6+0.5,7.5-2)
	(5.6+0.5,5.1-2)--(6.4+0.5,4.3-2)
	(4.7+0.5,4.2-2)--(5.5+0.5,3.4-2);
	\draw [semithick,red]
	(4.7+0.5,7.5-2)--(5.6+0.5,7.5-2)
	(4.7+0.5,5.1-2)--(5.6+0.5,5.1-2)
	;
	\draw [<->,semithick] (5.55+0.5,6.15-2)--(4.05+0.5,6.15-2);
	\draw [<->,semithick] (6.45+0.3+0.5,4.2-0.3-2)--(4.75+0.3+0.5,4.2-0.3-2);
	\draw [<->,semithick] (6.45+0.3+0.5,8.7-2)--(4.75+0.3+0.5,8.7-2);
	\draw [<->,semithick] (3.4+0.5,4.25-2)--(3.4+0.5,5.75-2);
	\draw [<->,semithick] (3.4+0.5,4.85)--(3.4+0.5,6.35);
	
	\draw [semithick]
	(4+0.5,1.2-6.8+2)--(4+0.5,1.2-5.8+2)
	(4+0.5,1.2-6.8+2)--(4.7+0.5,1.2-7.5+2)
	(4.7+0.5,1.2-7.5+2)--(4.7+0.5,1.2-8.4+2)
	(4.7+0.5,1.2-8.4+2)--(3.2+0.5,1.2-8.4+2)
	(4.7+0.5,1.2-8.4+2)--(5.5+0.5,1.2-9.2+2)
	(5.6+0.5,1.2-7.5+2)--(6.4+0.5,1.2-8.3+2)
	(4+0.5,1.2-5.8+2)--(3.2+0.5,1.2-5.8+2)
	(3.2+0.5,1.2-6.8+2)--(4+0.5,1.2-6.8+2)
	(4+0.5,1.2-5.8+2)--(4.7+0.5,1.2-5.1+2)
	(4.7+0.5,1.2-5.1+2)--(4.7+0.5,1.2-4.2+2)
	(4.7+0.5,1.2-4.2+2)--(3.2+0.5,1.2-4.2+2)
	(5.6+0.5,1.2-5.1+2)--(5.6+0.5,1.2-7.5+2)
	(5.6+0.5,1.2-5.1+2)--(6.4+0.5,1.2-4.3+2)
	(4.7+0.5,1.2-4.2+2)--(5.5+0.5,1.2-3.4+2);
	\draw [semithick,cyan]
	(4.7+0.5,1.2-7.5+2)--(5.6+0.5,1.2-7.5+2)
	(4.7+0.5,1.2-5.1+2)--(5.6+0.5,1.2-5.1+2)
	;
	\draw [<->,semithick] (5.55+0.5,1.2-6.45+2)--(4.05+0.5,1.2-6.45+2);
	\draw [<->,semithick] (6.45+0.3+0.5,1.2-4.2+0.3+2)--(4.75+0.3+0.5,1.2-4.2+0.3+2);
	\draw [<->,semithick] (6.45+0.3+0.5,1.2-8.7+2)--(4.75+0.3+0.5,1.2-8.7+2);
	\draw [<->,semithick] (3.4+0.5,1.2-4.25+2)--(3.4+0.5,1.2-5.75+2);
	\draw [<->,semithick] (3.4+0.5,1.2-4.25-2.6+2)--(3.4+0.5,1.2-5.75-2.6+2);

	\node[scale=0.7] at (0,2.15-3.8) [below] {$Q_b$};
	\node[scale=0.7] at (0.4,2.5) [right] {$Q_{f_3}$};
	\node[scale=0.7] at (0.4,2.6-3.8) [right] {$Q_{f_1}$};
	\node[scale=0.7] at (1.47,3.73) {$Y_{1,2}$};
	\node[scale=0.7] at (-1.52,3.73) {$Y_{3,2}$};
	\node[scale=0.7] at (1.47,1.2) {$Z_{2,2}$};
	\node[scale=0.7] at (-1.33,1.2) {$Z_{4,2}$};
	\node[scale=0.7] at (0,-3.83) {$Q_D$};
	\node[scale=0.7] at (0,5) {$Q_U$};
	\node[scale=0.7] at (1.2,0.6) {$Q_{f_2}$};
	\node[scale=0.7] at (1.47,3.73-3.8) {$Y_{1,1}$};
	\node[scale=0.7] at (-1.52,3.73-3.8) {$Y_{3,1}$};
	\node[scale=0.7] at (1.47,1.2-3.8) {$Z_{2,1}$};
	\node[scale=0.7] at (-1.33,1.2-3.8) {$Z_{4,1}$};
		
	\node[scale=0.5] at (5.68,5.75) {$X_{1,2}$};
	\node[scale=0.5] at (5.2,4.95) {$Z_{1,2}$};
	\node[scale=0.7] at (4.75,5.9) {$Y_{1,2}$};
	\node[scale=0.7] at (6.48,7.02) {$Q_1$};
	\node[scale=0.7] at (6.48,1.57) {$Q_2$};
	\node[scale=0.7] at (3.46,5.6) {$Q_{f_3}$};	
	\node[scale=0.7] at (3.46,3) {$Q_{f_1}$};	
	\node[scale=0.7] at (4.07,4.3) {$Q_{f_2}$};
	\node[scale=0.5] at (5.74,8.57-5.75) {$X_{1,1}$};
	\node[scale=0.5] at (5.25,8.5-4.95) {$Y_{1,1}$};
	\node[scale=0.7] at (4.75,8.6-5.9) {$Z_{1,1}$};
	\node[scale=0.5] at (5.36,4.4) {$P_{1,1}$};
	\node[scale=0.7] at (6.67,4.3) {$ Q_{1,1}$};
		
	\node[scale=0.5] at (5.68,5.75-7.4) {$X_{2,2}$};
	\node[scale=0.5] at (5.2,4.95-7.4) {$Z_{2,2}$};
	\node[scale=0.7] at (4.75,5.9-7.4) {$Y_{2,2}$};
	\node[scale=0.7] at (6.48,7.02-7.4) {$Q_3$};
	\node[scale=0.7] at (6.48,1.57-7.4) {$Q_4$};
	\node[scale=0.7] at (3.46,5.6-7.4) {$Q_{f_3}$};	
	\node[scale=0.7] at (3.46,3-7.4) {$Q_{f_1}$};	
	\node[scale=0.7] at (4.07,4.3-7.4) {$Q_{f_2}$};
	\node[scale=0.5] at (5.74,8.57-5.75-7.4) {$X_{2,1}$};
	\node[scale=0.5] at (5.25,8.5-4.95-7.4) {$Y_{2,1}$};
	\node[scale=0.7] at (4.75,8.6-5.9-7.4) {$Z_{2,1}$};
	\node[scale=0.5] at (5.36,4.4-7.4) {$P_{2,1}$};	
	\node[scale=0.7] at (6.67,4.3-7.4) {$ Q_{2,1}$};
	
	\draw[>=latex,->,thick] (8.7,0.6)--(9.7,0.6);
	
	\begin{scope}[shift={(13.2,0)}]
	\draw [semithick]
	(-0.7,0.1)--(-0.7,1.1)
	(0.7,0.1)--(0.7,1.1)
	(-0.7,1.1)--(0.7,1.1)
	(-0.7,1.1)--(-1.4,1.8)
	(0.7,1.1)--(1.4,1.8)
	(1.4,1.8)--(1.4,3.2)
	;
	\draw [semithick,red]
	(1.4,1.8)--(2.1+0.2,1.8)
	(-1.4,3.2)--(-2.1-0.2,3.2)
	(-1.4,3.2-3.8)--(-2.1-0.2,3.2-3.8)
	(1.4,1.8-3.8)--(2.1+0.2,1.8-3.8)
	;
	\draw [semithick,cyan]
	(1.4,3.2)--(2.1+0.2,3.2)
	(-1.4,1.8)--(-2.1-0.2,1.8)
	(1.4,3.2-3.8)--(2.1+0.2,3.2-3.8)
	(-1.4,1.8-3.8)--(-2.1-0.2,1.8-3.8)
	;
	\draw [semithick]
	(-0.7,4.9)--(-0.7,3.9)
	(0.7,4.9)--(0.7,3.9)
	(-0.7,3.9)--(0.7,3.9)
	(-0.7,3.9)--(-1.4,3.2)
	(0.7,3.9)--(1.4,3.2)
	(-1.4,3.2)--(-1.4,1.8)
	;
	\draw [semithick]
	(-0.7,0.1-3.8)--(-0.7,1.1-3.8)
	(0.7,0.1-3.8)--(0.7,1.1-3.8)
	(-0.7,1.1-3.8)--(0.7,1.1-3.8)
	(-0.7,1.1-3.8)--(-1.4,1.8-3.8)
	(0.7,1.1-3.8)--(1.4,1.8-3.8)
	(1.4,1.8-3.8)--(1.4,3.2-3.8)
	;
	\draw [semithick]
	(-0.7,3.9-3.8)--(0.7,3.9-3.8)
	(-0.7,3.9-3.8)--(-1.4,3.2-3.8)
	(0.7,3.9-3.8)--(1.4,3.2-3.8)
	(-1.4,3.2-3.8)--(-1.4,1.8-3.8)
	;
	\draw [<->,semithick] (-1.35,2.1-3.8)--(1.35,2.1-3.8);
	\draw [<->,semithick] (0.5,1.15)--(0.5,3.85);
	\draw [<->,semithick] (0.5,1.15-3.8)--(0.5,3.85-3.8);
	\draw [<->,semithick] (-0.65,5-0.3)--(0.65,5-0.3);
	\draw [<->,semithick] (-0.65,-3.5)--(0.65,-3.5);
	
	\draw [semithick]
	(4+0.5,6.8-2)--(4+0.5,5.8-2)
	(4+0.5,6.8-2)--(4.7+0.5,7.5-2)
	(4.7+0.5,7.5-2)--(4.7+0.5,8.4-2)
	(4.7+0.5,8.4-2)--(3.2+0.5,8.4-2)
	(4.7+0.5,8.4-2)--(5.5+0.5,9.2-2)
	(5.6+0.5,7.5-2)--(6.4+0.5,8.3-2)
	(4+0.5,5.8-2)--(3.2+0.5,5.8-2)
	(3.2+0.5,6.8-2)--(4+0.5,6.8-2)
	(4+0.5,5.8-2)--(4.7+0.5,5.1-2)
	(4.7+0.5,5.1-2)--(4.7+0.5,4.2-2)
	(4.7+0.5,4.2-2)--(3.2+0.5,4.2-2)
	(5.6+0.5,5.1-2)--(5.6+0.5,7.5-2)
	(5.6+0.5,5.1-2)--(6.4+0.5,4.3-2)
	(4.7+0.5,4.2-2)--(5.5+0.5,3.4-2);
	\draw [semithick,red]
	(4.7+0.5,1.2-7.5+2)--(5.6+0.5,1.2-7.5+2)
	(4.7+0.5,1.2-5.1+2)--(5.6+0.5,1.2-5.1+2)
	;
	\draw [<->,semithick] (5.55+0.5,6.15-2)--(4.05+0.5,6.15-2);
	\draw [<->,semithick] (6.45+0.3+0.5,4.2-0.3-2)--(4.75+0.3+0.5,4.2-0.3-2);
	\draw [<->,semithick] (6.45+0.3+0.5,8.7-2)--(4.75+0.3+0.5,8.7-2);
	\draw [<->,semithick] (3.4+0.5,4.25-2)--(3.4+0.5,5.75-2);
	\draw [<->,semithick] (3.4+0.5,4.85)--(3.4+0.5,6.35);
	
	\draw [semithick]
	(4+0.5,1.2-6.8+2)--(4+0.5,1.2-5.8+2)
	(4+0.5,1.2-6.8+2)--(4.7+0.5,1.2-7.5+2)
	(4.7+0.5,1.2-7.5+2)--(4.7+0.5,1.2-8.4+2)
	(4.7+0.5,1.2-8.4+2)--(3.2+0.5,1.2-8.4+2)
	(4.7+0.5,1.2-8.4+2)--(5.5+0.5,1.2-9.2+2)
	(5.6+0.5,1.2-7.5+2)--(6.4+0.5,1.2-8.3+2)
	(4+0.5,1.2-5.8+2)--(3.2+0.5,1.2-5.8+2)
	(3.2+0.5,1.2-6.8+2)--(4+0.5,1.2-6.8+2)
	(4+0.5,1.2-5.8+2)--(4.7+0.5,1.2-5.1+2)
	(4.7+0.5,1.2-5.1+2)--(4.7+0.5,1.2-4.2+2)
	(4.7+0.5,1.2-4.2+2)--(3.2+0.5,1.2-4.2+2)
	(5.6+0.5,1.2-5.1+2)--(5.6+0.5,1.2-7.5+2)
	(5.6+0.5,1.2-5.1+2)--(6.4+0.5,1.2-4.3+2)
	(4.7+0.5,1.2-4.2+2)--(5.5+0.5,1.2-3.4+2);
	\draw [semithick,cyan]
	(4.7+0.5,7.5-2)--(5.6+0.5,7.5-2)
	(4.7+0.5,5.1-2)--(5.6+0.5,5.1-2)
	;
	\draw [<->,semithick] (5.55+0.5,1.2-6.45+2)--(4.05+0.5,1.2-6.45+2);
	\draw [<->,semithick] (6.45+0.3+0.5,1.2-4.2+0.3+2)--(4.75+0.3+0.5,1.2-4.2+0.3+2);
	\draw [<->,semithick] (6.45+0.3+0.5,1.2-8.7+2)--(4.75+0.3+0.5,1.2-8.7+2);
	\draw [<->,semithick] (3.4+0.5,1.2-4.25+2)--(3.4+0.5,1.2-5.75+2);
	\draw [<->,semithick] (3.4+0.5,1.2-4.25-2.6+2)--(3.4+0.5,1.2-5.75-2.6+2);

	\node[scale=0.4] at (-0.3,2.1-3.8) [below] {$\sqrt{\frac{Q_2P_{2,1}}{Q_4P_{1,1}}}Q_b$};
	\node[scale=0.7] at (0.4,2.5) [right] {$Q_{f_3}$};
	\node[scale=0.7] at (0.4,2.6-3.8) [right] {$Q_{f_1}$};
	\node[scale=0.7] at (1.47,3.73) {$Y_{2,2}$};
	\node[scale=0.7] at (-1.52,3.73) {$Y_{3,2}$};
	\node[scale=0.7] at (1.47,1.2) {$Z_{1,2}$};
	\node[scale=0.7] at (-1.33,1.2) {$Z_{4,2}$};
	\node[scale=0.7] at (0,-3.83) {$Q_D$};
	\node[scale=0.7] at (0,5) {$Q_U$};
	\node[scale=0.7] at (1.2,0.6) {$Q_{f_2}$};
	\node[scale=0.7] at (1.47,3.73-3.8) {$Y_{2,1}$};
	\node[scale=0.7] at (-1.52,3.73-3.8) {$Y_{3,1}$};
	\node[scale=0.7] at (1.47,1.2-3.8) {$Z_{1,1}$};
	\node[scale=0.7] at (-1.33,1.2-3.8) {$Z_{4,1}$};
		
	\node[scale=0.5] at (5.72,5.75) {$\lambda_1 X_{2,2}$};
	\node[scale=0.5] at (5.2,4.95) {$Z_{2,2}$};
	\node[scale=0.7] at (4.75,5.9) {$Y_{2,2}$};
	\node[scale=0.7] at (6.53,7.02) {$\lambda_1 Q_3$};
	\node[scale=0.7] at (6.55,1.57) {$\lambda_1 Q_4$};
	\node[scale=0.7] at (3.46,5.6) {$Q_{f_3}$};	
	\node[scale=0.7] at (3.46,3) {$Q_{f_1}$};	
	\node[scale=0.7] at (4.07,4.3) {$Q_{f_2}$};
	\node[scale=0.5] at (5.74,8.57-5.75) {$\lambda_1 X_{2,1}$};
	\node[scale=0.5] at (5.25,8.5-4.95) {$Y_{2,1}$};
	\node[scale=0.7] at (4.75,8.6-5.9) {$Z_{2,1}$};
	\node[scale=0.5] at (5.36,4.4) {$\lambda_1 P_{2,1}$};
	\node[scale=0.7] at (6.67,4.3) {$ Q_{2,1}$};
		
	\node[scale=0.4] at (5.72,5.75-7.4) {$\lambda_1^{-1} X_{1,2}$};
	\node[scale=0.5] at (5.2,4.95-7.4) {$Z_{1,2}$};
	\node[scale=0.7] at (4.75,5.9-7.4) {$Y_{1,2}$};
	\node[scale=0.55] at (6.53,7.02-7.4) {$\lambda_1^{-1} Q_1$};
	\node[scale=0.55] at (6.55,1.57-7.4) {$\lambda_1^{-1} Q_2$};
	\node[scale=0.7] at (3.46,5.6-7.4) {$Q_{f_3}$};	
	\node[scale=0.7] at (3.46,3-7.4) {$Q_{f_1}$};	
	\node[scale=0.7] at (4.07,4.3-7.4) {$Q_{f_2}$};
	\node[scale=0.4] at (5.74,8.57-5.75-7.4) {$\lambda_1^{-1} X_{1,1}$};
	\node[scale=0.5] at (5.25,8.5-4.95-7.4) {$Y_{1,1}$};
	\node[scale=0.7] at (4.75,8.6-5.9-7.4) {$Z_{1,1}$};
	\node[scale=0.4] at (5.31,4.4-7.4) {$\lambda_1^{-1} P_{1,1}$};	
	\node[scale=0.7] at (6.67,4.3-7.4) {$ Q_{1,1}$};
	
	\node[scale=0.7] at (5.5,0.6) {$\lambda_1=\sqrt{\frac{Q_1Q_2}{Q_3Q_4}}$};
	\end{scope}

	\end{tikzpicture}
\caption{Flop transition related to exchanging $M_2$ and $M_3$ in the rank 2 $(D_4,D_4)$ conformal matter on a circle. }
\label{fig:M2M3inR2}
\end{figure}

\begin{figure}[htbp]
	\centering
	\begin{tikzpicture}[scale=0.6]
	\draw [semithick]
	(-0.7,0.1)--(-0.7,1.1)
	(0.7,0.1)--(0.7,1.1)
	(-0.7,1.1)--(0.7,1.1)
	(-0.7,1.1)--(-1.4,1.8)
	(0.7,1.1)--(1.4,1.8)
	(1.4,1.8)--(1.4,3.2)
	;
	\draw [semithick,red]
	(1.4,3.2)--(2.1+0.2,3.2)
	(-1.4,3.2)--(-2.1-0.2,3.2)
	(-1.4,3.2-3.8)--(-2.1-0.2,3.2-3.8)
	(1.4,3.2-3.8)--(2.1+0.2,3.2-3.8)
	;
	\draw [semithick,cyan]
	(-1.4,1.8)--(-2.1-0.2,1.8)
	(1.4,1.8)--(2.1+0.2,1.8)
	(-1.4,1.8-3.8)--(-2.1-0.2,1.8-3.8)
	(1.4,1.8-3.8)--(2.1+0.2,1.8-3.8)
	;
	\draw [semithick]
	(-0.7,4.9)--(-0.7,3.9)
	(0.7,4.9)--(0.7,3.9)
	(-0.7,3.9)--(0.7,3.9)
	(-0.7,3.9)--(-1.4,3.2)
	(0.7,3.9)--(1.4,3.2)
	(-1.4,3.2)--(-1.4,1.8)
	;
	\draw [semithick]
	(-0.7,0.1-3.8)--(-0.7,1.1-3.8)
	(0.7,0.1-3.8)--(0.7,1.1-3.8)
	(-0.7,1.1-3.8)--(0.7,1.1-3.8)
	(-0.7,1.1-3.8)--(-1.4,1.8-3.8)
	(0.7,1.1-3.8)--(1.4,1.8-3.8)
	(1.4,1.8-3.8)--(1.4,3.2-3.8)
	;
	\draw [semithick]
	(-0.7,3.9-3.8)--(0.7,3.9-3.8)
	(-0.7,3.9-3.8)--(-1.4,3.2-3.8)
	(0.7,3.9-3.8)--(1.4,3.2-3.8)
	(-1.4,3.2-3.8)--(-1.4,1.8-3.8)
	;
	\draw [<->,semithick] (-1.35,2.1-3.8)--(1.35,2.1-3.8);
	\draw [<->,semithick] (0.5,1.15)--(0.5,3.85);
	\draw [<->,semithick] (0.5,1.15-3.8)--(0.5,3.85-3.8);
	\draw [<->,semithick] (-0.65,5-0.3)--(0.65,5-0.3);
	\draw [<->,semithick] (-0.65,-3.5)--(0.65,-3.5);
	
	\draw [semithick]
	(4+0.5,6.8-2)--(4+0.5,5.8-2)
	(4+0.5,6.8-2)--(4.7+0.5,7.5-2)
	(4.7+0.5,7.5-2)--(4.7+0.5,8.4-2)
	(4.7+0.5,8.4-2)--(3.2+0.5,8.4-2)
	(4.7+0.5,8.4-2)--(5.5+0.5,9.2-2)
	(5.6+0.5,7.5-2)--(6.4+0.5,8.3-2)
	(4+0.5,5.8-2)--(3.2+0.5,5.8-2)
	(3.2+0.5,6.8-2)--(4+0.5,6.8-2)
	(4+0.5,5.8-2)--(4.7+0.5,5.1-2)
	(4.7+0.5,5.1-2)--(4.7+0.5,4.2-2)
	(4.7+0.5,4.2-2)--(3.2+0.5,4.2-2)
	(5.6+0.5,5.1-2)--(5.6+0.5,7.5-2)
	(5.6+0.5,5.1-2)--(6.4+0.5,4.3-2)
	(4.7+0.5,4.2-2)--(5.5+0.5,3.4-2);
	\draw [semithick,red]
	(4.7+0.5,7.5-2)--(5.6+0.5,7.5-2)
	(4.7+0.5,5.1-2)--(5.6+0.5,5.1-2)
	;
	\draw [<->,semithick] (5.55+0.5,6.15-2)--(4.05+0.5,6.15-2);
	\draw [<->,semithick] (6.45+0.3+0.5,4.2-0.3-2)--(4.75+0.3+0.5,4.2-0.3-2);
	\draw [<->,semithick] (6.45+0.3+0.5,8.7-2)--(4.75+0.3+0.5,8.7-2);
	\draw [<->,semithick] (3.4+0.5,4.25-2)--(3.4+0.5,5.75-2);
	\draw [<->,semithick] (3.4+0.5,4.85)--(3.4+0.5,6.35);
	
	\draw [semithick]
	(4+0.5,1.2-6.8+2)--(4+0.5,1.2-5.8+2)
	(4+0.5,1.2-6.8+2)--(4.7+0.5,1.2-7.5+2)
	(4.7+0.5,1.2-7.5+2)--(4.7+0.5,1.2-8.4+2)
	(4.7+0.5,1.2-8.4+2)--(3.2+0.5,1.2-8.4+2)
	(4.7+0.5,1.2-8.4+2)--(5.5+0.5,1.2-9.2+2)
	(5.6+0.5,1.2-7.5+2)--(6.4+0.5,1.2-8.3+2)
	(4+0.5,1.2-5.8+2)--(3.2+0.5,1.2-5.8+2)
	(3.2+0.5,1.2-6.8+2)--(4+0.5,1.2-6.8+2)
	(4+0.5,1.2-5.8+2)--(4.7+0.5,1.2-5.1+2)
	(4.7+0.5,1.2-5.1+2)--(4.7+0.5,1.2-4.2+2)
	(4.7+0.5,1.2-4.2+2)--(3.2+0.5,1.2-4.2+2)
	(5.6+0.5,1.2-5.1+2)--(5.6+0.5,1.2-7.5+2)
	(5.6+0.5,1.2-5.1+2)--(6.4+0.5,1.2-4.3+2)
	(4.7+0.5,1.2-4.2+2)--(5.5+0.5,1.2-3.4+2);
	\draw [semithick,cyan]
	(4.7+0.5,1.2-7.5+2)--(5.6+0.5,1.2-7.5+2)
	(4.7+0.5,1.2-5.1+2)--(5.6+0.5,1.2-5.1+2)
	;
	\draw [<->,semithick] (5.55+0.5,1.2-6.45+2)--(4.05+0.5,1.2-6.45+2);
	\draw [<->,semithick] (6.45+0.3+0.5,1.2-4.2+0.3+2)--(4.75+0.3+0.5,1.2-4.2+0.3+2);
	\draw [<->,semithick] (6.45+0.3+0.5,1.2-8.7+2)--(4.75+0.3+0.5,1.2-8.7+2);
	\draw [<->,semithick] (3.4+0.5,1.2-4.25+2)--(3.4+0.5,1.2-5.75+2);
	\draw [<->,semithick] (3.4+0.5,1.2-4.25-2.6+2)--(3.4+0.5,1.2-5.75-2.6+2);

	\node[scale=0.7] at (0,2.15-3.8) [below] {$Q_b$};
	\node[scale=0.7] at (0.4,2.5) [right] {$Q_{f_3}$};
	\node[scale=0.7] at (0.4,2.6-3.8) [right] {$Q_{f_1}$};
	\node[scale=0.7] at (1.47,3.73) {$Y_{1,2}$};
	\node[scale=0.7] at (-1.52,3.73) {$Y_{3,2}$};
	\node[scale=0.7] at (1.47,1.2) {$Z_{2,2}$};
	\node[scale=0.7] at (-1.33,1.2) {$Z_{4,2}$};
	\node[scale=0.7] at (0,-3.83) {$Q_D$};
	\node[scale=0.7] at (0,5) {$Q_U$};
	\node[scale=0.7] at (1.2,0.6) {$Q_{f_2}$};
	\node[scale=0.7] at (1.47,3.73-3.8) {$Y_{1,1}$};
	\node[scale=0.7] at (-1.52,3.73-3.8) {$Y_{3,1}$};
	\node[scale=0.7] at (1.47,1.2-3.8) {$Z_{2,1}$};
	\node[scale=0.7] at (-1.33,1.2-3.8) {$Z_{4,1}$};
		
	\node[scale=0.5] at (5.68,5.75) {$X_{1,2}$};
	\node[scale=0.5] at (5.2,4.95) {$Z_{1,2}$};
	\node[scale=0.7] at (4.75,5.9) {$Y_{1,2}$};
	\node[scale=0.7] at (6.48,7.02) {$Q_1$};
	\node[scale=0.7] at (6.48,1.57) {$Q_2$};
	\node[scale=0.7] at (3.46,5.6) {$Q_{f_3}$};	
	\node[scale=0.7] at (3.46,3) {$Q_{f_1}$};	
	\node[scale=0.7] at (4.07,4.3) {$Q_{f_2}$};
	\node[scale=0.5] at (5.74,8.57-5.75) {$X_{1,1}$};
	\node[scale=0.5] at (5.25,8.5-4.95) {$Y_{1,1}$};
	\node[scale=0.7] at (4.75,8.6-5.9) {$Z_{1,1}$};
	\node[scale=0.5] at (5.36,4.4) {$P_{1,1}$};
	\node[scale=0.7] at (6.67,4.3) {$ Q_{1,1}$};
		
	\node[scale=0.5] at (5.68,5.75-7.4) {$X_{2,2}$};
	\node[scale=0.5] at (5.2,4.95-7.4) {$Z_{2,2}$};
	\node[scale=0.7] at (4.75,5.9-7.4) {$Y_{2,2}$};
	\node[scale=0.7] at (6.48,7.02-7.4) {$Q_3$};
	\node[scale=0.7] at (6.48,1.57-7.4) {$Q_4$};
	\node[scale=0.7] at (3.46,5.6-7.4) {$Q_{f_3}$};	
	\node[scale=0.7] at (3.46,3-7.4) {$Q_{f_1}$};	
	\node[scale=0.7] at (4.07,4.3-7.4) {$Q_{f_2}$};
	\node[scale=0.5] at (5.74,8.57-5.75-7.4) {$X_{2,1}$};
	\node[scale=0.5] at (5.25,8.5-4.95-7.4) {$Y_{2,1}$};
	\node[scale=0.7] at (4.75,8.6-5.9-7.4) {$Z_{2,1}$};
	\node[scale=0.5] at (5.36,4.4-7.4) {$P_{2,1}$};	
	\node[scale=0.7] at (6.67,4.3-7.4) {$ Q_{2,1}$};
	
	\draw[>=latex,->,thick] (8.7,0.6)--(9.7,0.6);
	
	\begin{scope}[shift={(13.2,0)}]
	\draw [semithick]
	(-0.7,0.1)--(-0.7,1.1)
	(0.7,0.1)--(0.7,1.1)
	(-0.7,1.1)--(0.7,1.1)
	(-0.7,1.1)--(-1.4,1.8)
	(0.7,1.1)--(1.4,1.8)
	(1.4,1.8)--(1.4,3.2)
	;
	\draw [semithick,red]
	(1.4,3.2)--(2.1+0.2,3.2)
	(-1.4,3.2)--(-2.1-0.2,3.2)
	(-1.4,3.2-3.8)--(-2.1-0.2,3.2-3.8)
	(1.4,3.2-3.8)--(2.1+0.2,3.2-3.8)
	;
	\draw [semithick,cyan]
	(-1.4,1.8)--(-2.1-0.2,1.8)
	(1.4,1.8)--(2.1+0.2,1.8)
	(-1.4,1.8-3.8)--(-2.1-0.2,1.8-3.8)
	(1.4,1.8-3.8)--(2.1+0.2,1.8-3.8)
	;
	\draw [semithick]
	(-0.7,4.9)--(-0.7,3.9)
	(0.7,4.9)--(0.7,3.9)
	(-0.7,3.9)--(0.7,3.9)
	(-0.7,3.9)--(-1.4,3.2)
	(0.7,3.9)--(1.4,3.2)
	(-1.4,3.2)--(-1.4,1.8)
	;
	\draw [semithick]
	(-0.7,0.1-3.8)--(-0.7,1.1-3.8)
	(0.7,0.1-3.8)--(0.7,1.1-3.8)
	(-0.7,1.1-3.8)--(0.7,1.1-3.8)
	(-0.7,1.1-3.8)--(-1.4,1.8-3.8)
	(0.7,1.1-3.8)--(1.4,1.8-3.8)
	(1.4,1.8-3.8)--(1.4,3.2-3.8)
	;
	\draw [semithick]
	(-0.7,3.9-3.8)--(0.7,3.9-3.8)
	(-0.7,3.9-3.8)--(-1.4,3.2-3.8)
	(0.7,3.9-3.8)--(1.4,3.2-3.8)
	(-1.4,3.2-3.8)--(-1.4,1.8-3.8)
	;
	\draw [<->,semithick] (-1.35,2.1-3.8)--(1.35,2.1-3.8);
	\draw [<->,semithick] (0.5,1.15)--(0.5,3.85);
	\draw [<->,semithick] (0.5,1.15-3.8)--(0.5,3.85-3.8);
	\draw [<->,semithick] (-0.65,5-0.3)--(0.65,5-0.3);
	\draw [<->,semithick] (-0.65,-3.5)--(0.65,-3.5);
	
	\draw [semithick]
	(4+0.5,6.8-2)--(4+0.5,5.8-2)
	(4+0.5,6.8-2)--(4.7+0.5,7.5-2)
	(4.7+0.5,7.5-2)--(4.7+0.5,8.4-2)
	(4.7+0.5,8.4-2)--(3.2+0.5,8.4-2)
	(4.7+0.5,8.4-2)--(5.5+0.5,9.2-2)
	(5.6+0.5,7.5-2)--(6.4+0.5,8.3-2)
	(4+0.5,5.8-2)--(3.2+0.5,5.8-2)
	(3.2+0.5,6.8-2)--(4+0.5,6.8-2)
	(4+0.5,5.8-2)--(4.7+0.5,5.1-2)
	(4.7+0.5,5.1-2)--(4.7+0.5,4.2-2)
	(4.7+0.5,4.2-2)--(3.2+0.5,4.2-2)
	(5.6+0.5,5.1-2)--(5.6+0.5,7.5-2)
	(5.6+0.5,5.1-2)--(6.4+0.5,4.3-2)
	(4.7+0.5,4.2-2)--(5.5+0.5,3.4-2);
	\draw [semithick,cyan]
	(4.7+0.5,1.2-7.5+2)--(5.6+0.5,1.2-7.5+2)
	(4.7+0.5,1.2-5.1+2)--(5.6+0.5,1.2-5.1+2)
	;
	\draw [<->,semithick] (5.55+0.5,6.15-2)--(4.05+0.5,6.15-2);
	\draw [<->,semithick] (6.45+0.3+0.5,4.2-0.3-2)--(4.75+0.3+0.5,4.2-0.3-2);
	\draw [<->,semithick] (6.45+0.3+0.5,8.7-2)--(4.75+0.3+0.5,8.7-2);
	\draw [<->,semithick] (3.4+0.5,4.25-2)--(3.4+0.5,5.75-2);
	\draw [<->,semithick] (3.4+0.5,4.85)--(3.4+0.5,6.35);
	
	\draw [semithick]
	(4+0.5,1.2-6.8+2)--(4+0.5,1.2-5.8+2)
	(4+0.5,1.2-6.8+2)--(4.7+0.5,1.2-7.5+2)
	(4.7+0.5,1.2-7.5+2)--(4.7+0.5,1.2-8.4+2)
	(4.7+0.5,1.2-8.4+2)--(3.2+0.5,1.2-8.4+2)
	(4.7+0.5,1.2-8.4+2)--(5.5+0.5,1.2-9.2+2)
	(5.6+0.5,1.2-7.5+2)--(6.4+0.5,1.2-8.3+2)
	(4+0.5,1.2-5.8+2)--(3.2+0.5,1.2-5.8+2)
	(3.2+0.5,1.2-6.8+2)--(4+0.5,1.2-6.8+2)
	(4+0.5,1.2-5.8+2)--(4.7+0.5,1.2-5.1+2)
	(4.7+0.5,1.2-5.1+2)--(4.7+0.5,1.2-4.2+2)
	(4.7+0.5,1.2-4.2+2)--(3.2+0.5,1.2-4.2+2)
	(5.6+0.5,1.2-5.1+2)--(5.6+0.5,1.2-7.5+2)
	(5.6+0.5,1.2-5.1+2)--(6.4+0.5,1.2-4.3+2)
	(4.7+0.5,1.2-4.2+2)--(5.5+0.5,1.2-3.4+2);
	\draw [semithick,red]
	(4.7+0.5,7.5-2)--(5.6+0.5,7.5-2)
	(4.7+0.5,5.1-2)--(5.6+0.5,5.1-2)
	;
	\draw [<->,semithick] (5.55+0.5,1.2-6.45+2)--(4.05+0.5,1.2-6.45+2);
	\draw [<->,semithick] (6.45+0.3+0.5,1.2-4.2+0.3+2)--(4.75+0.3+0.5,1.2-4.2+0.3+2);
	\draw [<->,semithick] (6.45+0.3+0.5,1.2-8.7+2)--(4.75+0.3+0.5,1.2-8.7+2);
	\draw [<->,semithick] (3.4+0.5,1.2-4.25+2)--(3.4+0.5,1.2-5.75+2);
	\draw [<->,semithick] (3.4+0.5,1.2-4.25-2.6+2)--(3.4+0.5,1.2-5.75-2.6+2);

	\node[scale=0.7] at (0,2.15-3.8) [below] {$Q_b$};
	\node[scale=0.7] at (0.4,2.5) [right] {$Q_{f_3}$};
	\node[scale=0.7] at (0.4,2.6-3.8) [right] {$Q_{f_1}$};
	\node[scale=0.7] at (1.47,3.73) {$Y_{1,2}$};
	\node[scale=0.7] at (-1.52,3.73) {$Y_{3,2}$};
	\node[scale=0.7] at (1.47,1.2) {$Z_{2,2}$};
	\node[scale=0.7] at (-1.33,1.2) {$Z_{4,2}$};
	\node[scale=0.7] at (0,-3.83) {$Q_D$};
	\node[scale=0.7] at (0,5) {$Q_U$};
	\node[scale=0.7] at (1.2,0.6) {$Q_{f_2}$};
	\node[scale=0.7] at (1.47,3.73-3.8) {$Y_{1,1}$};
	\node[scale=0.7] at (-1.52,3.73-3.8) {$Y_{3,1}$};
	\node[scale=0.7] at (1.47,1.2-3.8) {$Z_{2,1}$};
	\node[scale=0.7] at (-1.33,1.2-3.8) {$Z_{4,1}$};
		
	\node[scale=0.5] at (5.72,5.75-7.4) {$\lambda_1 X_{2,2}$};
	\node[scale=0.5] at (5.2,4.95-7.4) {$Z_{2,2}$};
	\node[scale=0.7] at (4.75,5.9-7.4) {$Y_{2,2}$};
	\node[scale=0.7] at (6.53,7.02-7.4) {$\lambda_1 Q_3$};
	\node[scale=0.7] at (6.55,1.57-7.4) {$\lambda_1 Q_4$};
	\node[scale=0.7] at (3.46,5.6-7.4) {$Q_{f_3}$};	
	\node[scale=0.7] at (3.46,3-7.4) {$Q_{f_1}$};	
	\node[scale=0.7] at (4.07,4.3-7.4) {$Q_{f_2}$};
	\node[scale=0.5] at (5.74,8.57-5.75-7.4) {$\lambda_1 X_{2,1}$};
	\node[scale=0.5] at (5.25,8.5-4.95-7.4) {$Y_{2,1}$};
	\node[scale=0.7] at (4.75,8.6-5.9-7.4) {$Z_{2,1}$};
	\node[scale=0.5] at (5.36,4.4-7.4) {$\lambda_1 P_{2,1}$};
	\node[scale=0.7] at (6.67,4.3-7.4) {$ Q_{2,1}$};
		
	\node[scale=0.4] at (5.72,5.75) {$\lambda_1^{-1} X_{1,2}$};
	\node[scale=0.5] at (5.2,4.95) {$Z_{1,2}$};
	\node[scale=0.7] at (4.75,5.9) {$Y_{1,2}$};
	\node[scale=0.55] at (6.53,7.02) {$\lambda_1^{-1} Q_1$};
	\node[scale=0.55] at (6.55,1.57) {$\lambda_1^{-1} Q_2$};
	\node[scale=0.7] at (3.46,5.6) {$Q_{f_3}$};	
	\node[scale=0.7] at (3.46,3) {$Q_{f_1}$};	
	\node[scale=0.7] at (4.07,4.3) {$Q_{f_2}$};
	\node[scale=0.4] at (5.74,8.57-5.75) {$\lambda_1^{-1} X_{1,1}$};
	\node[scale=0.5] at (5.25,8.5-4.95) {$Y_{1,1}$};
	\node[scale=0.7] at (4.75,8.6-5.9) {$Z_{1,1}$};
	\node[scale=0.4] at (5.31,4.4) {$\lambda_1^{-1} P_{1,1}$};	
	\node[scale=0.7] at (6.67,4.3) {$ Q_{1,1}$};
	
	\node[scale=0.7] at (5.5,0.6) {$\lambda_1=\sqrt{\frac{Q_1Q_2}{Q_3Q_4}}$};
	\end{scope}

	\end{tikzpicture}
\caption{Flop transition related to exchanging $M_1$ and $M_4^{-1}$ in the rank 2 $(D_4,D_4)$ conformal matter on a circle. }
\label{fig:M1M4ReinR2}
\end{figure}
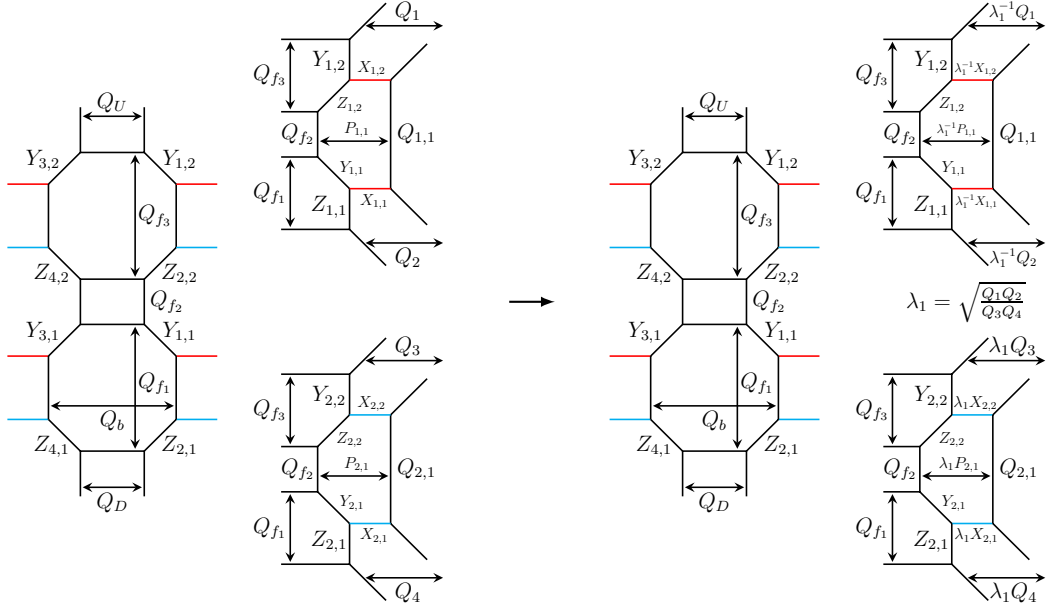

We summarize the 10 flop transitions related to $Q_1,\cdots,Q_8,Q_U,Q_D$ as the following
\begin{align}
	&\mathbf{V}_1=\big{\{}Q_1\rightarrow Q_1^{-1}\ \ ; \ \  Q_{1,1}\rightarrow Q_1 Q_{1,1}\big{\}},\nn\\
	&\mathbf{V}_2=\big{\{}Q_2\rightarrow Q_2^{-1},\ Q_0\rightarrow Q_2Q_0\ \ ; \ \  Q_{1,1}\rightarrow Q_2 Q_{1,1}\big{\}},\nn\\
	&\mathbf{V}_3=\big{\{}Q_3\rightarrow Q_3^{-1},\ Q_0\rightarrow Q_3Q_0\ \ ; \ \  Q_{2,1}\rightarrow Q_3 Q_{2,1}\big{\}},\nn\\
	&\mathbf{V}_4=\big{\{}Q_4\rightarrow Q_4^{-1}\ \ ; \ \  Q_{2,1}\rightarrow Q_4 Q_{2,1}\big{\}},\nn\\
	&\mathbf{V}_5=\big{\{}Q_5\rightarrow Q_5^{-1}\ \ ; \ \  Q_{3,1}\rightarrow Q_5 Q_{3,1}\big{\}},\nn\\
	&\mathbf{V}_6=\big{\{}Q_6\rightarrow Q_6^{-1},\ Q_0\rightarrow Q_6Q_0\ \ ; \ \  Q_{3,1}\rightarrow Q_6 Q_{3,1}\big{\}},\nn\\
	&\mathbf{V}_7=\big{\{}Q_7\rightarrow Q_7^{-1},\ Q_0\rightarrow Q_7Q_0\ \ ; \ \  Q_{4,1}\rightarrow Q_7 Q_{4,1}\big{\}},\nn\\
	&\mathbf{V}_8=\big{\{}Q_8\rightarrow Q_8^{-1}\ \ ; \ \  Q_{4,1}\rightarrow Q_8 Q_{4,1}\big{\}},\nn\\
	&\mathbf{V}_9=\big{\{}\resizebox{0.65\hsize}{!}{$Q_1\rightarrow \sqrt{\frac{Q_1Q_2Q_6}{Q_5}}Q_0,\ Q_3\rightarrow \sqrt{\frac{Q_2Q_6}{Q_1Q_5}}Q_0 Q_3,\ Q_5\rightarrow \sqrt{\frac{Q_2Q_5Q_6}{Q_1}}Q_0,$}\nn\\
	&\qquad\quad\ \resizebox{0.6\hsize}{!}{$Q_7\rightarrow \sqrt{\frac{Q_2Q_6}{Q_1Q_5}}Q_0 Q_7,\ Q_0\rightarrow\sqrt{\frac{Q_1Q_5}{Q_2Q_6}}\ \ ; \ \  Q_{f_3}\rightarrow \sqrt{\frac{Q_2Q_6}{Q_1Q_5}}Q_0 Q_{f_3}$}\big{\}},\nn\\
	&\mathbf{V}_{10}=\big{\{}\resizebox{0.65\hsize}{!}{$Q_2\rightarrow \sqrt{\frac{Q_3Q_7}{Q_4Q_8}}Q_0 Q_2,\ Q_4\rightarrow \sqrt{\frac{Q_3Q_4Q_7}{Q_8}}Q_0,\ Q_6\rightarrow \sqrt{\frac{Q_3Q_7}{Q_4Q_8}}Q_0 Q_6,$}\nn\\
	&\qquad\quad\ \ \resizebox{0.6\hsize}{!}{$Q_8\rightarrow \sqrt{\frac{Q_3Q_7Q_8}{Q_4}}Q_0,\ Q_0\rightarrow\sqrt{\frac{Q_4Q_8}{Q_3Q_7}}\ \ ; \ \  Q_{f_1}\rightarrow \sqrt{\frac{Q_3Q_7}{Q_4Q_8}}Q_0 Q_{f_1}$}\big{\}}. 
	\label{eq:D4D4R2flops}
\end{align}

In terms of physical parameters, the 10 flop transitions are in the following form:
\begin{align}
	&\mathbf{V}_1=\big{\{}M_1\rightarrow M_2,\ M_2\rightarrow M_1\ \ ; \ \ Q_{1,1}\to \frac{M_1 Q_{1,1}}{M_2}\big{\}},\nn\\
	&\mathbf{V}_2=\big{\{}M_1\rightarrow M_2^{-1},\ M_2\rightarrow M_1^{-1}\ \ ; \ \ Q_{1,1}\to M_1 M_2 Q_{1,1}\big{\}},\nn\\
	&\mathbf{V}_3=\big{\{}M_3\rightarrow M_4^{-1},\ M_4\rightarrow M_3^{-1}\ \ ; \ \ Q_{2,1}\to \frac{Q_{2,1}}{M_3 M_4}\big{\}},\nn\\
	&\mathbf{V}_4=\big{\{}M_3\rightarrow M_4,\ M_4\rightarrow M_3\ \ ; \ \ Q_{2,1}\to \frac{M_3 Q_{2,1}}{M_4}\big{\}},\nn\\
	&\mathbf{V}_5=\big{\{}M_5\rightarrow M_6,\ M_6\rightarrow M_5\ \ ; \ \ Q_{3,1}\to \frac{M_5 Q_{3,1}}{M_6}\big{\}},\nn\\
	&\mathbf{V}_6=\big{\{}M_5\rightarrow M_6^{-1},\ M_6\rightarrow M_5^{-1}\ \ ; \ \ Q_{3,1}\to M_5 M_6 Q_{3,1}\big{\}},\nn\\
	&\mathbf{V}_7=\big{\{}M_7\rightarrow M_8^{-1},\ M_8\rightarrow M_7^{-1}\ \ ; \ \ Q_{4,1}\to \frac{Q_{4,1}}{M_7 M_8}\big{\}},\nn\\
	&\mathbf{V}_8=\big{\{}M_7\rightarrow M_8,\ M_8\rightarrow M_7\ \ ; \ \ Q_{4,1}\to \frac{M_7 Q_{4,1}}{M_8}\big{\}},\nn\\
	&\mathbf{V}_9=\big{\{}\resizebox{0.85\hsize}{!}{$M_1\rightarrow\left(\frac{M_1^3M_2M_3M_4M_6M_7M_8q}{M_5}\right)^{\frac14},\ M_2\rightarrow\left(\frac{M_1M_2^3M_5}{M_3M_4M_6M_7M_8q}\right)^{\frac14},\ M_3\rightarrow\left(\frac{M_1M_3^3M_5}{M_2M_4M_6M_7M_8q}\right)^{\frac14},$}\nn\\
	&\qquad\quad\,\resizebox{0.85\hsize}{!}{$M_4\rightarrow\left(\frac{M_1M_4^3M_5}{M_2M_3M_6M_7M_8q}\right)^{\frac14},\ M_5\rightarrow\left(\frac{M_2M_3M_4M_5^3M_6M_7M_8q}{M_1}\right)^{\frac14},\ M_6\rightarrow\left(\frac{M_1M_5M_6^3}{M_2M_3M_4M_7M_8q}\right)^{\frac14},$}\nn\\
	&\qquad\quad\,\resizebox{0.84\hsize}{!}{$M_7\rightarrow\left(\frac{M_1M_5M_7^3}{M_2M_3M_4M_6M_8q}\right)^{\frac14},\ M_8\rightarrow\left(\frac{M_1M_5M_8^3}{M_2M_3M_4M_6M_7q}\right)^{\frac14}\ \ ; \ \ A_3\rightarrow A_3\left(\frac{M_2M_3M_4M_6M_7M_8q}{M_1M_5}\right)^{\frac14}$}\big{\}},\nn\\
	&\mathbf{V}_{10}=\big{\{}\resizebox{0.76\hsize}{!}{$M_1\rightarrow\left(\frac{M_1^3M_4M_8q}{M_2M_3M_5M_6M_7}\right)^{\frac14},\ M_2\rightarrow\left(\frac{M_2^3M_4M_8q}{M_1M_3M_5M_6M_7}\right)^{\frac14},\ M_3\rightarrow\left(\frac{M_3^3M_4M_8q}{M_1M_2M_5M_6M_7}\right)^{\frac14},$}\nn\\
	&\qquad\quad\ \,\resizebox{0.82\hsize}{!}{$M_4\rightarrow\left(\frac{M_1M_2M_3M_4^3M_5M_6M_7}{M_8q}\right)^{\frac14},\ M_5\rightarrow\left(\frac{M_4M_5^3M_8q}{M_1M_2M_3M_6M_7}\right)^{\frac14},\ M_6\rightarrow\left(\frac{M_4M_6^3M_8q}{M_1M_2M_3M_5M_7}\right)^{\frac14},$}\nn\\
	&\qquad\quad\ \,\resizebox{0.84\hsize}{!}{$M_7\rightarrow\left(\frac{M_4M_7^3M_8q}{M_1M_2M_3M_5M_6}\right)^{\frac14},\ M_8\rightarrow\left(\frac{M_1M_2M_3M_5M_6M_7M_8^3}{M_4q}\right)^{\frac14}\ \ ; \ \ A_1\rightarrow A_1\left(\frac{M_4M_8q}{M_1M_2M_3M_5M_6M_7}\right)^{\frac14}$}\big{\}}.
	\label{eq:D4D4quadflopsMR2}
\end{align}

\subsection{Hidden flop transitions for the $SO(16)$ symmetry}
For the mass parameters, the transformations in equation \eqref{eq:D4D4quadflopsMR2} are the same as the ones in equation \eqref{eq:D4D4quadflopsM}, so the transformations $\mathbf{V}_1$ to $\mathbf{V}_8$ still belong to $SO(16)$ Weyl group. As discussed in the rank 1 case, these transformations do not generate the complete $SO(16)$ Weyl group, so again we need to figure out the transformations that exchange one single mass parameter between different $SU(2)$ subdiagrams.

As an example, we try to find out the transformation that exchanges $M_2$ and $M_3$ in the rank 2 case whose corresponding rank 1 transformation is in figure \ref{fig:M2M3inQuad}. We find the consistent flop transition which is depicted in figure \ref{fig:M2M3inR2}, where
\begin{equation}
	\lambda_1=\sqrt{\frac{Q_1Q_2}{Q_3Q_4}}. 
\end{equation}
The corresponding Weyl reflection is 
\begin{align}
	\mathbf{V}_{\text{e}_1}=&\Big{\{}\resizebox{0.8\hsize}{!}{$Q_1\rightarrow\sqrt{\frac{Q_1Q_2Q_3}{Q_4}},\ Q_2\rightarrow\sqrt{\frac{Q_1Q_2Q_4}{Q_3}},\ Q_3\rightarrow\sqrt{\frac{Q_1Q_3Q_4}{Q_2}},\ Q_4\rightarrow\sqrt{\frac{Q_2Q_3Q_4}{Q_1}}, $}\nn\\
	&\phantom{\Big{\{}}Q_0\rightarrow\sqrt{\frac{Q_2Q_3}{Q_1Q_4}}Q_0\ \  ; \ \  Q_{1,1}\rightarrow Q_{2,1},\  Q_{2,1}\rightarrow Q_{1,1}\Big{\}}\nn\\
	=&\Big{\{}M_2\rightarrow M_3,\ M_3\rightarrow M_2\ \ ; \ \  Q_{1,1}\rightarrow Q_{2,1},\  Q_{2,1}\rightarrow Q_{1,1}\Big{\}}.  
\end{align}

By using the permutation symmetry between the $SU(2)$ subdiagrams, we can obtain the flop transition depicted in figure \ref{fig:M1M4ReinR2}, in which the corresponding Weyl reflection is 
\begin{align}
	\mathbf{V}_{\text{I}}=&\Big{\{}\resizebox{0.8\hsize}{!}{$Q_1\rightarrow\sqrt{\frac{Q_1Q_3Q_4}{Q_2}},\ Q_2\rightarrow\sqrt{\frac{Q_2Q_3Q_4}{Q_1}},\ Q_3\rightarrow\sqrt{\frac{Q_1Q_2Q_3}{Q_4}},\ Q_4\rightarrow\sqrt{\frac{Q_1Q_2Q_4}{Q_3}}$}\Big{\}}\nn\\
	=&\big{\{}M_1\rightarrow M_4^{-1},\ M_4\rightarrow M_1^{-1}\big{\}}. 
\end{align}
From figure \ref{fig:M1M4ReinR2} we can see that the two Coulomb branch parameters $ Q_{1,1}$ and $ Q_{2,1}$ are not affected by the exchanging of $M_1$ and $M_4^{-1}$, while the two Coulomb branch parameters also exchange with each other by the exchanging of $M_2$ and $M_3$ in figure \ref{fig:M2M3inR2}. By combining $\mathbf{V}_{\text{I}}$ with the flop transitions $\mathbf{V}_1$ and $\mathbf{V}_3$ in \eqref{eq:D4D4R2flops}, we can obtain the following flop transition which also exchanges $M_2$ and $M_3$ but does not exchange the Coulomb branch parameters $ Q_{1,1}$ and $ Q_{2,1}$, $\mathbf{V}_{\text{e}_2}=\mathbf{V}_1\mathbf{V}_{\text{I}}\mathbf{V}_3\mathbf{V}_{\text{I}}\mathbf{V}_1$:
\begin{align}
	\mathbf{V}_{\text{e}_2}=&\Big{\{}\resizebox{0.8\hsize}{!}{$Q_1\rightarrow\sqrt{\frac{Q_1Q_2Q_3}{Q_4}},\ Q_2\rightarrow\sqrt{\frac{Q_1Q_2Q_4}{Q_3}},\ Q_3\rightarrow\sqrt{\frac{Q_1Q_3Q_4}{Q_2}},\ Q_4\rightarrow\sqrt{\frac{Q_2Q_3Q_4}{Q_1}}, $}\nn\\
	&\phantom{\Big{\{}}Q_0\rightarrow\sqrt{\frac{Q_2Q_3}{Q_1Q_4}}Q_0\ \  ; \ \  Q_{1,1}\rightarrow\sqrt{\frac{Q_1Q_4}{Q_2Q_3}} Q_{1,1},\  Q_{2,1}\rightarrow\sqrt{\frac{Q_2Q_3}{Q_1Q_4}} Q_{2,1}\Big{\}}\nn\\
	=&\Big{\{}M_2\rightarrow M_3,\ M_3\rightarrow M_2\ \ ; \ \  Q_{1,1}\rightarrow\frac{M_3}{M_2} Q_{1,1},\  Q_{2,1}\rightarrow\frac{M_2}{M_3} Q_{2,1}\Big{\}}.  
\end{align}
The flop transition $\mathbf{V}_{\text{e}_2}$ is better than $\mathbf{V}_{\text{e}_1}$ in the sense that it is more hopeful to find out affine $E_8$ invariant Coulomb branch parameters by including $\mathbf{V}_{\text{e}_2}$ into the $SO(16)$ Weyl group rather than including $\mathbf{V}_{\text{e}_1}$. The appearance of multiple different flop transitions that correspond to the same element of the Weyl group is a characteristic of the rank 2 theory, Coulomb branch parameters transform differently in these flop transitions. In order to find out affine $E_8$ invariant Coulomb branch parameters in the rank 2 theory, we will only consider the flop transitions that only transform Coulomb branch parameters at most by a factor of mass parameters. The flop transitions $\mathbf{V}_{1},\cdots,\mathbf{V}_{10}$ belong to such category as well as $\mathbf{V}_{\text{I}}$. The flop transition $\mathbf{V}_{\text{I}}$ is the transition between the first and second $SU(2)$ subdiagrams, we can similarly obtain other five flop transitions which are the similar transitions between different $SU(2)$ subdiagrams:
\begin{alignat}{2}
	\mathbf{V}_{\text{II}}&=\big{\{}M_4\to M_5^{-1},\ M_5\to M_4^{-1}\big{\}},\quad &
	\mathbf{V}_{\text{III}}&=\big{\{}M_5\to M_8^{-1},\ M_8\to M_5^{-1}\big{\}},\nn\\
	\mathbf{V}_{\text{IV}}&=\big{\{}M_1\to M_8^{-1},\ M_8\to M_1^{-1}\big{\}},\quad &
	\mathbf{V}_{\text{V}}&=\big{\{}M_1\to M_5,\ M_5\to M_1\big{\}},\nn\\
	\mathbf{V}_{\text{VI}}&=\big{\{}\ M_4\to M_8,\ M_8\to M_4\big{\}}, 
\end{alignat}
in which $\mathbf{V}_{\text{IV}},\mathbf{V}_{\text{V}},\mathbf{V}_{\text{VI}}$ can be obtained by $\mathbf{V}_{\text{I}},\mathbf{V}_{\text{II}},\mathbf{V}_{\text{III}}$. 

\subsection{Different formations of the affine $E_8$ symmetry}
Due to the Coulomb branch parameter transformations in $\mathbf{V}_{1},\cdots,\mathbf{V}_{10}$, in the rank 2 theory, these 10 flop transitions become independent of each other. To form the $SO(16)$ symmetry, we need to pick one pair of flop transitions that correspond to one of the four $SU(2)$ subdiagrams, then pick one flop transition from each of the remaining three $SU(2)$ subdiagrams, and combing $\mathbf{V}_{\text{I}},\cdots,\mathbf{V}_{\text{VI}}$ we can generate the full $SO(16)$ Weyl group. We can further generate the affine $E_8$ Weyl group by picking $\mathbf{V}_{9}$ or $\mathbf{V}_{10}$. In total, we have 64 different choices of $\mathbf{V}_{i_1},\mathbf{V}_{i_2},\mathbf{V}_{i_3},\mathbf{V}_{i_4},\mathbf{V}_{i_5},\mathbf{V}_{i_6}$ to form the affine $E_8$ Weyl group, we list the 64 sets of $\{i_1,i_2,i_3,i_4,i_5,i_6\}$ in the following:
\begin{align}
	&\{1,2,3,5,7,9\},\{1,2,3,5,8,9\},\{1,2,3,6,7,9\},\{1,2,3,6,8,9\},\{1,2,4,5,7,9\},\nn\\
	&\{1,2,4,5,8,9\},\{1,2,4,6,7,9\},\{1,2,4,6,8,9\},\{3,4,1,5,7,9\},\{3,4,1,5,8,9\},\nn\\
	&\{3,4,1,6,7,9\},\{3,4,1,6,8,9\},\{3,4,2,5,7,9\},\{3,4,2,5,8,9\},\{3,4,2,6,7,9\},\nn\\
	&\{3,4,2,6,8,9\},\{5,6,1,3,7,9\},\{5,6,1,3,8,9\},\{5,6,1,4,7,9\},\{5,6,1,4,8,9\},\nn\\
	&\{5,6,2,3,7,9\},\{5,6,2,3,8,9\},\{5,6,2,4,7,9\},\{5,6,2,4,8,9\},\{7,8,1,3,5,9\},\nn\\
	&\{7,8,1,3,6,9\},\{7,8,1,4,5,9\},\{7,8,1,4,6,9\},\{7,8,2,3,5,9\},\{7,8,2,3,6,9\},\nn\\
	&\{7,8,2,4,5,9\},\{7,8,2,4,6,9\},\{1,2,3,5,7,10\},\{1,2,3,5,8,10\},\{1,2,3,6,7,10\},\nn\\
	&\{1,2,3,6,8,10\},\{1,2,4,5,7,10\},\{1,2,4,5,8,10\},\{1,2,4,6,7,10\},\{1,2,4,6,8,10\},\nn\\
	&\{3,4,1,5,7,10\},\{3,4,1,5,8,10\},\{3,4,1,6,7,10\},\{3,4,1,6,8,10\},\{3,4,2,5,7,10\},\nn\\
	&\{3,4,2,5,8,10\},\{3,4,2,6,7,10\},\{3,4,2,6,8,10\},\{5,6,1,3,7,10\},\{5,6,1,3,8,10\},\nn\\
	&\{5,6,1,4,7,10\},\{5,6,1,4,8,10\},\{5,6,2,3,7,10\},\{5,6,2,3,8,10\},\{5,6,2,4,7,10\},\nn\\
	&\{5,6,2,4,8,10\},\{7,8,1,3,5,10\},\{7,8,1,3,6,10\},\{7,8,1,4,5,10\},\{7,8,1,4,6,10\},\nn\\
	&\{7,8,2,3,5,10\},\{7,8,2,3,6,10\},\{7,8,2,4,5,10\},\{7,8,2,4,6,10\}. 
	\label{eq:64choicesR2}
\end{align}

As an example, we pick $\mathbf{V}_1,\mathbf{V}_2,\mathbf{V}_3,\mathbf{V}_6,\mathbf{V}_8,\mathbf{V}_9$ to illustrate how to use them to obtain the standard affine $E_8$ Weyl reflections. 
We use the same notation as in equation \eqref{eq:affineE8Wy} to denote the standard affine $E_8$ reflections but also including the transformations on the Coulomb branch parameters. Then we can obtain the following flop transitions:
\begin{align}
	\mathbf{W}_1&=\mathbf{V}_1\nn\\
	&=\big{\{}M_1\rightarrow M_2,\ M_2\rightarrow M_1\ \ ; \ \ Q_{1,1}\to \frac{M_1 Q_{1,1}}{M_2}\big{\}},\label{eq:W1rk2}\\
	\mathbf{W}_2&=\mathbf{W}_1\mathbf{V}_{\text{I}}\mathbf{V}_3\mathbf{V}_{\text{I}}\mathbf{W}_1\nn\\
	&=\big\{M_2\to M_3,\ M_3\to M_2\ \ ; \ \  Q_{1,1}\to \frac{M_3 Q_{1,1}}{M_2},\ Q_{2,1}\to \frac{M_2 Q_{2,1}}{M_3}\big\},\\
	\mathbf{W}_3&=\mathbf{W}_2\mathbf{V}_2\mathbf{V}_{\text{I}}\mathbf{V}_2\mathbf{W}_2\nn\\
	&=\big\{M_3\to M_4,\ M_4\to M_3\ \ ; \ \  Q_{1,1}\to \frac{M_3^2 Q_{1,1}}{M_4^2},\ Q_{2,1}\to \frac{M_4 Q_{2,1}}{M_3}\big\},\\
	\mathbf{W}_4&=\mathbf{W}_3\mathbf{W}_2\mathbf{W}_1\mathbf{W}_2\mathbf{W}_3\mathbf{V}_{\text{V}}\mathbf{W}_3\mathbf{W}_2\mathbf{W}_1\mathbf{W}_2\mathbf{W}_3\nn\\
	&=\big\{M_4\to M_5,\ M_5\to M_4\ \ ; \ \  Q_{1,1}\to \frac{M_5^2 Q_{1,1}}{M_4^2}\big\},\\
	\mathbf{W}_5&=\mathbf{W}_4\mathbf{V}_{\text{II}}\mathbf{V}_6\mathbf{V}_{\text{II}}\mathbf{W}_4\nn\\
	&=\big\{M_5\to M_6,\ M_6\to M_5\ \ ; \ \  Q_{1,1}\to \frac{M_5^2 Q_{1,1}}{M_6^2},\ Q_{3,1}\to \frac{M_6 Q_{3,1}}{M_5}\big\},\\
	\mathbf{W}_6&=\mathbf{W}_5\mathbf{V}_{\text{III}}\mathbf{V}_8\mathbf{V}_{\text{III}}\mathbf{W}_5\nn\\
	&=\big\{\resizebox{0.87\hsize}{!}{$M_6\to \frac{1}{M_7},\ M_7\to \frac{1}{M_6}\ \ ; \ \  Q_{1,1}\to \frac{Q_{1,1}}{M_6^2 M_7^2},\ Q_{3,1}\to M_6 M_7 Q_{3,1},\ Q_{4,1}\to M_6 M_7 Q_{4,1}$}\big\},\\
	\mathbf{W}_8&=\mathbf{W}_5\mathbf{W}_4\mathbf{V}_{\text{VI}}\mathbf{V}_8\mathbf{V}_{\text{VI}}\mathbf{W}_4\mathbf{W}_5\nn\\
	&=\big\{M_6\to M_7,\ M_7\to M_6\ \ ; \ \ Q_{3,1}\to \frac{M_6 Q_{3,1}}{M_7},\ Q_{4,1}\to \frac{M_7 Q_{4,1}}{M_6}\big\},\label{eq:W8rk2}\\
	\mathbf{V}_0&=\mathbf{W}_1\mathbf{W}_2\mathbf{W}_3\mathbf{W}_2\mathbf{W}_1\mathbf{V}_{\text{VI}}\mathbf{W}_1\mathbf{W}_2\mathbf{W}_3\mathbf{W}_2\mathbf{W}_1\nn\\
	&=\big\{M_1\to M_8,\ M_8\to M_1\ \ ; \ \ Q_{1,1}\to \frac{M_1^2 Q_{1,1}}{M_8^2}\big\}. 
\end{align}
$\mathbf{V}_0,\mathbf{W}_1,\cdots,\mathbf{W}_6,\mathbf{W}_8$ form the standard $SO(16)$ Weyl reflections, we can use them to generate the following flop transition:
\begin{align}
	\mathbf{V}_{\text{t}_1}=&\mathbf{W}_6\mathbf{W}_8\mathbf{W}_1\mathbf{V}_{\text{V}}\mathbf{W}_5\mathbf{V}_{\text{V}}\mathbf{W}_1\mathbf{W}_3\mathbf{W}_4\mathbf{W}_5\mathbf{W}_8\mathbf{W}_5\mathbf{W}_4\mathbf{W}_3\mathbf{W}_6\mathbf{W}_8\mathbf{W}_3\mathbf{W}_4\mathbf{W}_5\mathbf{W}_8\mathbf{W}_5\nn\\
	&\mathbf{W}_4\mathbf{W}_3\mathbf{W}_1\mathbf{V}_{\text{V}}\mathbf{W}_5\mathbf{V}_{\text{V}}\mathbf{W}_1\mathbf{W}_3\mathbf{V}_0\mathbf{W}_1\mathbf{V}_0\mathbf{W}_1\mathbf{V}_{\text{V}}\mathbf{W}_5\mathbf{V}_{\text{V}}\mathbf{W}_1\mathbf{W}_3\mathbf{W}_4\mathbf{W}_5\mathbf{W}_8\mathbf{W}_5\mathbf{W}_4\nn\\
	&\mathbf{W}_3\mathbf{W}_6\mathbf{W}_8\mathbf{W}_3\mathbf{W}_4\mathbf{W}_5\mathbf{W}_8\mathbf{W}_5\mathbf{W}_4\mathbf{W}_3\mathbf{W}_1\mathbf{V}_{\text{V}}\mathbf{W}_5\mathbf{V}_{\text{V}}\mathbf{W}_1\mathbf{V}_0\mathbf{W}_1\mathbf{V}_0\mathbf{W}_3\nn\\
	=&\big\{M_2\to \frac{1}{M_2},\ M_3\to \frac{1}{M_3},\ M_4\to \frac{1}{M_4},\ M_6\to \frac{1}{M_6},\ M_7\to \frac{1}{M_7},\ M_8\to \frac{1}{M_8}\ \ ; \ \ \nn\\
	&\phantom{\big\{}Q_{1,1}\to \frac{M_3^2 Q_{1,1}}{M_4^2 M_6^2 M_7^2 M_8^2},\ Q_{2,1}\to \frac{Q_{2,1}}{M_3^2},\ Q_{3,1}\to M_6^2 Q_{3,1},\ Q_{4,1}\to M_7^2 Q_{4,1}\big\}.  
\end{align}
Then we can transform $\mathbf{V}_9$ into the following form by $\mathbf{V}_{\text{t}_1}$:
\begin{align}
	\mathbf{V}_{\text{s}_1}=&\mathbf{V}_{\text{t}_1}\mathbf{V}_9\mathbf{V}_{\text{t}_1}\nn\\
	=&\big{\{}\resizebox{0.87\hsize}{!}{$M_i\rightarrow M_i\left(\frac{q}{M_1M_2M_3M_4M_5M_6M_7M_8}\right)^{\frac14}\ \forall i\in\{1,\cdots,8\}\ \ ; \ \ 
	A_3\rightarrow \left(\frac{q}{M_1M_2M_3M_4M_5M_6M_7M_8}\right)^{\frac14}A_3,\ $}\nn\\
	&\phantom{\big{\{}}\resizebox{0.84\hsize}{!}{$Q_{1,1}\to \left(\frac{q}{M_1M_2M_3M_4M_5M_6M_7M_8}\right)^{\frac32}Q_{1,1},\ Q_{2,1}\to \left(\frac{q}{M_1M_2M_3M_4M_5M_6M_7M_8}\right)^{\frac12}Q_{2,1},\ $}\nn\\
	&\phantom{\big{\{}}\resizebox{0.825\hsize}{!}{$Q_{3,1}\to \left(\frac{M_1M_2M_3M_4M_5M_6M_7M_8}{q}\right)^{\frac12}Q_{3,1},\ Q_{4,1}\to \left(\frac{M_1M_2M_3M_4M_5M_6M_7M_8}{q}\right)^{\frac12}Q_{4,1}$}\big{\}}.
\end{align}
Replacing the mass parameter $M_8$ by the new parameter $M_8'\equiv M_8/q$, the flop transitions $\mathbf{V}_0$ and $\mathbf{V}_{\text{s}_1}$ become the following forms respectively\footnote{As the flop transitions $\mathbf{W}_1,\cdots,\mathbf{W}_6,\mathbf{W}_8$ do not involve $M_8$, their forms are unchanged. }:
\begin{align}
	\mathbf{W}_0=&\big\{M_1\to q\,M_8,\ M_8\to q^{-1}M_1\ \ ; \ \ Q_{1,1}\to \frac{M_1^2 Q_{1,1}}{q^2 M_8^2}\big\}, \label{eq:W0rk2}\\
	\mathbf{W}_7=&\big{\{}\resizebox{0.87\hsize}{!}{$M_i\rightarrow \frac{M_i}{(M_1M_2M_3M_4M_5M_6M_7M_8)^{1/4}}\ \forall i\in\{1,\cdots,8\}\ \ ; \ \ 
	A_3\rightarrow \frac{A_3}{(M_1M_2M_3M_4M_5M_6M_7M_8)^{1/4}},\ $}\nn\\
	&\phantom{\big{\{}}\resizebox{0.73\hsize}{!}{$Q_{1,1}\to \frac{Q_{1,1}}{(M_1M_2M_3M_4M_5M_6M_7M_8)^{3/2}},\ Q_{2,1}\to \frac{Q_{2,1}}{(M_1M_2M_3M_4M_5M_6M_7M_8)^{1/2}},\ $}\nn\\
	&\phantom{\big{\{}}\resizebox{0.88\hsize}{!}{$Q_{3,1}\to \left(M_1M_2M_3M_4M_5M_6M_7M_8\right)^{\frac12}Q_{3,1},\ Q_{4,1}\to \left(M_1M_2M_3M_4M_5M_6M_7M_8\right)^{\frac12}Q_{4,1}$}\big{\}},\label{eq:W7rk2}
\end{align}
where for convenience we have dropped the prime on $M_8'$, and we will keep this notation change until the end of this section. 

Thus we have obtained the standard affine $E_8$ Weyl reflections $\mathbf{W}_0,\cdots,\mathbf{W}_8$ from the choice $\mathbf{V}_1,\mathbf{V}_2,\mathbf{V}_3,\mathbf{V}_6,\mathbf{V}_8,\mathbf{V}_9$. 

In order to find out affine $E_8$ invariant Coulomb branch parameters, we first try to find out the $E_8$ invariant Coulomb branch parameters just like what we have done in section \ref{subsec:inva_Coulomb_rk1}. For the Coulomb branch parameter $Q_{1,1}$, we use the following ansatz to represent the corresponding $E_8$ invariant Coulomb branch parameter $Q_{1,1}'$:
\begin{equation}
	Q_{1,1}'=M_1^{\alpha_1}M_2^{\alpha_2}M_3^{\alpha_3}M_4^{\alpha_4}M_5^{\alpha_5}M_6^{\alpha_6}M_7^{\alpha_7}M_8^{\alpha_8}Q_{1,1}. 
\end{equation}
Requiring $Q_{1,1}'$ to be invariant under the $E_8$ Weyl reflections $\mathbf{W}_1,\cdots,\mathbf{W}_8$ in equation \eqref{eq:W1rk2}-\eqref{eq:W8rk2} and equation \eqref{eq:W7rk2}, we can determine these $\alpha_i$ and we find that 
\begin{equation}
	Q_{1,1}'=\frac{M_1 M_3 M_5 Q_{1,1}}{M_4 M_6 M_7 M_8^6}.
\end{equation}
Similarly we can obtain the $E_8$ invariant Coulomb branch parameters $Q_{2,1}',Q_{3,1}',Q_{4,1}'$: 
\begin{equation}
	Q_{2,1}'=\frac{Q_{2,1}}{M_3 M_8},\quad Q_{3,1}'=M_6 M_8 Q_{3,1},\quad Q_{4,1}'=M_7 M_8 Q_{4,1}. 
\end{equation}
The Coulomb branch parameter $A_3$ transforms in equation \eqref{eq:W7rk2} in the same way as the Coulomb branch parameter $A$ in $\mathbf{W}_7$ of equation \eqref{eq:affineE8Wm}, so the corresponding $E_8$ invariant Coulomb branch parameter $A_3'$ is
\begin{equation}
	A_3'=\frac{A_3}{M_8}. 
\end{equation}
Under $\mathbf{W}_0$ in equation \eqref{eq:W0rk2}, the $E_8$ invariant Coulomb branch parameters transform in the following way:
\begin{align}
&Q_{1,1}'\to \left(\frac{M_8\,q}{M_1}\right)^5 Q_{1,1}',\quad Q_{2,1}'\to \frac{M_8\,q}{M_1}Q_{2,1}',\quad Q_{3,1}'\to \left(\frac{M_8\,q}{M_1}\right)^{-1}Q_{3,1}',\nn\\
&Q_{4,1}'\to \left(\frac{M_8\,q}{M_1}\right)^{-1}Q_{4,1}',\quad A_3'\to \frac{M_8\,q}{M_1}A_3'. 
\end{align}
Then by equation \eqref{eq:jacobi_trans}, we can define the following affine $E_8$ invariant Coulomb branch parameters:
\begin{align}
&\tilde{Q}_{1,1}\equiv\Theta(q,\boldsymbol{M})^5\frac{M_1 M_3 M_5 Q_{1,1}}{M_4 M_6 M_7 M_8^6},\quad \tilde{Q}_{2,1}\equiv\Theta(q,\boldsymbol{M})\frac{Q_{2,1}}{M_3 M_8},\nn\\
&\tilde{Q}_{3,1}\equiv \Theta(q,\boldsymbol{M})^{-1}M_6 M_8 Q_{3,1},\quad \tilde{Q}_{4,1}\equiv \Theta(q,\boldsymbol{M})^{-1}M_7 M_8 Q_{4,1},\nn\\
&\tilde{A}_3\equiv\Theta(q,\boldsymbol{M})\frac{A_3}{M_8}.
\end{align}
The Coulomb branch parameters $A_1$, $A_2$ are unaffected by the affine $E_8$ Weyl reflections, so they are automatically affine $E_8$ invariant. 

The remaining 63 different choices to form the standard affine $E_8$ Weyl reflections and consequently the corresponding affine $E_8$ invariant Coulomb branch parameters can also be similarly computed. The author has made a Mathematica code to compute all the 64 different choices which can generate the standard affine $E_8$ Weyl reflections as well as the affine $E_8$ invariant Coulomb branch parameters, the code can be found in \cite{WeylSymmetryD4D4}. We list the results of the first three choices of equation \eqref{eq:64choicesR2} in appendix \ref{app:rank2}. 

\section{Weyl symmetry in rank $N\geq 3$ $(D_4,D_4)$ conformal matter on a circle}\label{sec:rankN}
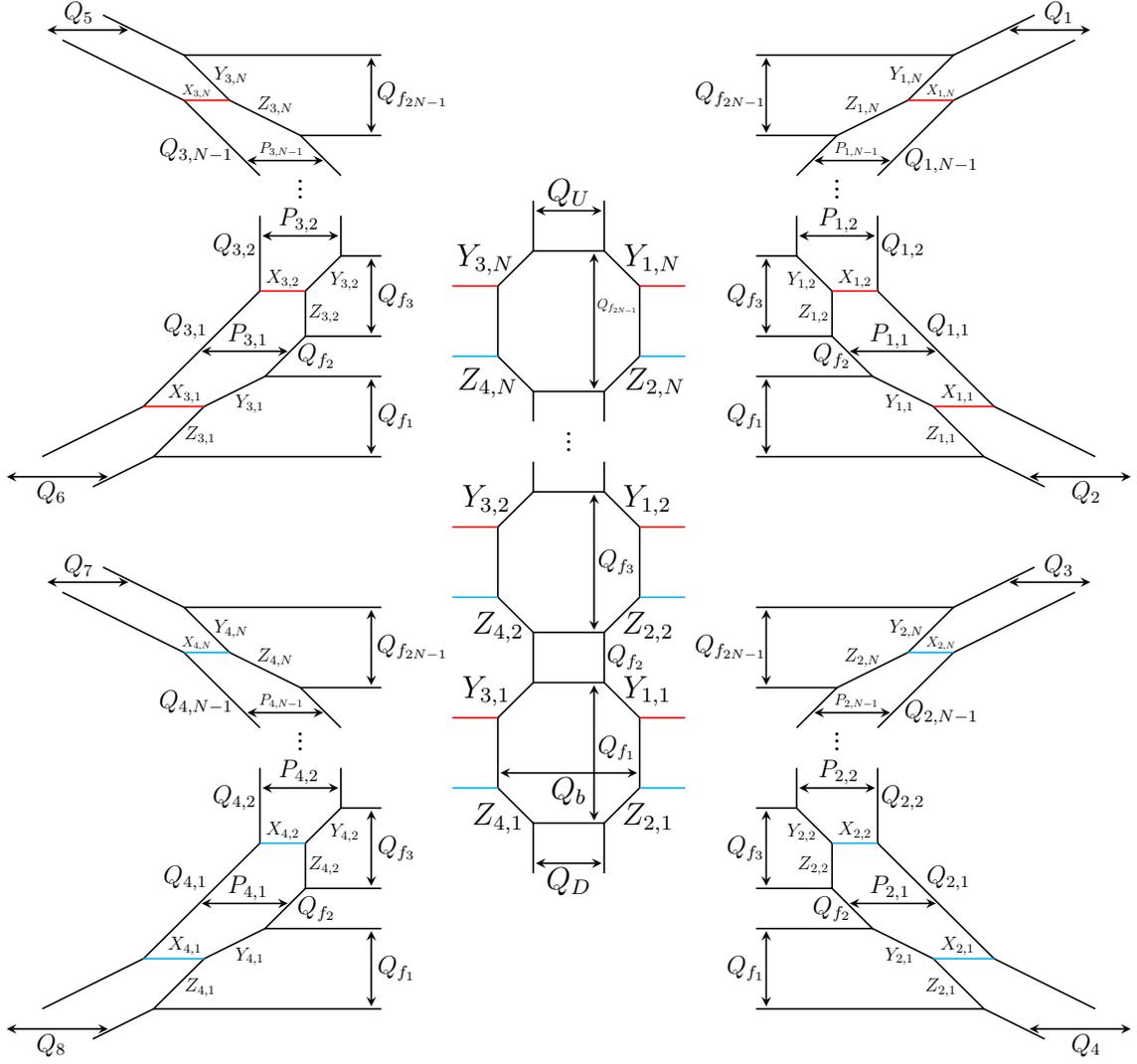
\begin{figure}[htbp]
	\centering
	\begin{tikzpicture}[scale=0.67]
	\draw [semithick]
	(-0.7,0.5)--(-0.7,1.1)
	(0.7,0.5)--(0.7,1.1)
	(-0.7,1.1)--(0.7,1.1)
	(-0.7,1.1)--(-1.4,1.8)
	(0.7,1.1)--(1.4,1.8)
	(1.4,1.8)--(1.4,3.2)
	;
	\draw [semithick]
	(-0.7,4.9)--(-0.7,3.9)
	(0.7,4.9)--(0.7,3.9)
	(-0.7,3.9)--(0.7,3.9)
	(-0.7,3.9)--(-1.4,3.2)
	(0.7,3.9)--(1.4,3.2)
	(-1.4,3.2)--(-1.4,1.8)
	;
	\draw (0,-0.05) node{.};
	\draw (0,0.1) node{.};
	\draw (0,0.25) node{.};
	\draw [semithick]
	(-0.7,0.1-4.8)--(-0.7,1.1-4.8)
	(0.7,0.1-4.8)--(0.7,1.1-4.8)
	(-0.7,1.1-4.8)--(0.7,1.1-4.8)
	(-0.7,1.1-4.8)--(-1.4,1.8-4.8)
	(0.7,1.1-4.8)--(1.4,1.8-4.8)
	(1.4,1.8-4.8)--(1.4,3.2-4.8)
	;
	\draw [semithick]
	(-0.7,4.5-4.8)--(-0.7,3.9-4.8)
	(0.7,4.5-4.8)--(0.7,3.9-4.8)
	(-0.7,3.9-4.8)--(0.7,3.9-4.8)
	(-0.7,3.9-4.8)--(-1.4,3.2-4.8)
	(0.7,3.9-4.8)--(1.4,3.2-4.8)
	(-1.4,3.2-4.8)--(-1.4,1.8-4.8)
	;
	\draw [semithick]
	(-0.7,0.1-8.6)--(-0.7,1.1-8.6)
	(0.7,0.1-8.6)--(0.7,1.1-8.6)
	(-0.7,1.1-8.6)--(0.7,1.1-8.6)
	(-0.7,1.1-8.6)--(-1.4,1.8-8.6)
	(0.7,1.1-8.6)--(1.4,1.8-8.6)
	(1.4,1.8-8.6)--(1.4,3.2-8.6)
	;
	\draw [semithick]
	(-0.7,3.9-8.6)--(0.7,3.9-8.6)
	(-0.7,3.9-8.6)--(-1.4,3.2-8.6)
	(0.7,3.9-8.6)--(1.4,3.2-8.6)
	(-1.4,3.2-8.6)--(-1.4,1.8-8.6)
	;
	\draw [semithick,red]
	(-1.4,3.2)--(-2.1-0.2,3.2)
	(1.4,3.2)--(2.1+0.2,3.2)
	(-1.4,3.2-4.8)--(-2.1-0.2,3.2-4.8)
	(1.4,3.2-4.8)--(2.1+0.2,3.2-4.8)
	(-1.4,3.2-8.6)--(-2.1-0.2,3.2-8.6)
	(1.4,3.2-8.6)--(2.1+0.2,3.2-8.6)
	;
	\draw [semithick,cyan]
	(-1.4,1.8)--(-2.1-0.2,1.8)
	(1.4,1.8)--(2.1+0.2,1.8)
	(-1.4,1.8-4.8)--(-2.1-0.2,1.8-4.8)
	(1.4,1.8-4.8)--(2.1+0.2,1.8-4.8)
	(-1.4,1.8-8.6)--(-2.1-0.2,1.8-8.6)
	(1.4,1.8-8.6)--(2.1+0.2,1.8-8.6)
	;

	\draw [semithick]
	(4+0.5,-1.6+9-2)--(4.8+0.5,-0.8+9-2)
	(4.8+0.5,-0.8+9-2)--(3.2+0.5,-0.8+9-2)
	(4.8+0.5,-0.8+9-2)--(6.2+0.5,-0.1+9-2)
	(6.2+0.5,-0.1+9-2)--(7.1+0.5,0.8+9-2)
	(7.1+0.5,0.8+9-2)--(3.2+0.5,0.8+9-2)
	(7.1+0.5,-0.1+9-2)--(5.6+0.5,-1.6+9-2)
	(7.1+0.5,-0.1+9-2)--(9.5+0.5,1.1+9-2)
	(7.1+0.5,0.8+9-2)--(8.7+0.5,1.6+9-2)
	;
	\draw (4.8+0.5,7.15-2+0.13) node{.};
	\draw (4.8+0.5,7-2+0.13) node{.};
	\draw (4.8+0.5,6.85-2+0.13) node{.};
	\draw [semithick]
	(4+0.5,6.6-2)--(4+0.5,5.8-2)
	(4+0.5,5.8-2)--(3.2+0.5,5.8-2)
	(4+0.5,5.8-2)--(4.7+0.5,5.1-2)
	(4.7+0.5,5.1-2)--(4.7+0.5,4.2-2)
	(4.7+0.5,4.2-2)--(3.2+0.5,4.2-2)
	(5.6+0.5,5.1-2)--(5.6+0.5,6.6-2)
	(5.6+0.5,5.1-2)--(7.9+0.5,2.8-2)
	(4.7+0.5,4.2-2)--(5.5+0.5,3.4-2)
	(5.5+0.5,3.4-2)--(3.2+0.5,3.4-2)
	(5.5+0.5,3.4-2)--(6.7+0.5,2.8-2)
	(6.7+0.5,2.8-2)--(7.7+0.5,1.8-2)
	(7.7+0.5,1.8-2)--(3.2+0.5,1.8-2)
	(7.7+0.5,1.8-2)--(8.9+0.5,1.2-2)
	(7.9+0.5,2.8-2)--(9.9+0.5,1.8-2)
	;
	\draw [<->,semithick] (9.1,-0.6)--(11.1,-0.6);
	\draw [<->,semithick] (6.45+0.3+0.5,4.2-0.3-2)--(4.75+0.3+0.5,4.2-0.3-2);
	\draw [<->,semithick] (3.4+0.5,4.25-2)--(3.4+0.5,5.75-2);
	\draw [<->,semithick] (3.4+0.5,4.25+2)--(3.4+0.5,5.75+2);
	\draw [<->,semithick] (3.4+0.5,4.25-2-2.4)--(3.4+0.5,5.75-2-2.4);
	\draw [<->,semithick](4.55,4.2)--(6.05,4.2);
	\draw [<->,semithick](4.85,5.7)--(6.35,5.7);
	\draw [<->,semithick](8.7,8.3)--(10.3,8.3);
	\node[scale=0.6] at (5.15+0.5,5.34-2) {$X_{1,2}$};
	\node[scale=0.6] at (4.33+0.5,4.65-2) {$Z_{1,2}$};
	\node[scale=0.6] at (4.1+0.5,5.3-2) {$Y_{1,2}$};
	\node[scale=0.8] at (5.84+0.5,4.2-2) {$P_{1,1}$};
	\node[scale=0.6] at (7.18+0.5,3.05-2) {$X_{1,1}$};
	\node[scale=0.6] at (6.85+0.5,2.22-2) {$Z_{1,1}$};
	\node[scale=0.6] at (6.03+0.5,2.85-2) {$Y_{1,1}$};
	\node[scale=0.45] at (7.35,9.08-2) {$X_{1,N}$};
	\node[scale=0.6] at (5.3+0.5,8.8-2) {$Z_{1,N}$};
	\node[scale=0.6] at (6.17+0.5,9.38-2) {$Y_{1,N}$};
	\node[scale=0.8] at (3.2,7) {$Q_{f_{2N-1}}$};
	\node[scale=0.8] at (3.5,3) {$Q_{f_3}$};
	\node[scale=0.8] at (3.5,0.6) {$Q_{f_1}$};
	\node[scale=0.8] at (5.2,1.8) {$Q_{f_2}$};
	\node[scale=0.8] at (10.22,-0.92) {$Q_2$};
	\node[scale=0.8] at (5.33,4.51) {$P_{1,2}$};
	\node[scale=0.5] at (5.68,5.92) {$P_{1,N-1}$};
	\node[scale=0.8] at (9.68,8.63) {$Q_1$};
	\node[scale=0.8] at (7.5,2.4) {$ Q_{1,1}$};
	\node[scale=0.8] at (6.6,4) {$ Q_{1,2}$};
	\node[scale=0.8] at (7.36,5.7) {$ Q_{1,N-1}$};
	\draw [semithick,red]
	(6.2+0.5,-0.1+9-2)--(7.1+0.5,-0.1+9-2)
	(4.7+0.5,5.1-2)--(5.6+0.5,5.1-2)
	(7.9+0.5,2.8-2)--(6.7+0.5,2.8-2)
	;

	\draw [<->,semithick] (9.1,-11.6)--(11.1,-11.6);
	\draw [<->,semithick] (6.45+0.3+0.5,4.2-0.3-2-11)--(4.75+0.3+0.5,4.2-0.3-2-11);
	\draw [<->,semithick] (3.4+0.5,4.25-2-11)--(3.4+0.5,5.75-2-11);
	\draw [<->,semithick] (3.4+0.5,4.25+2-11)--(3.4+0.5,5.75+2-11);
	\draw [<->,semithick] (3.4+0.5,4.25-2-2.4-11)--(3.4+0.5,5.75-2-2.4-11);
	\draw [<->,semithick](4.55,4.2-11)--(6.05,4.2-11);
	\draw [<->,semithick](4.85,5.7-11)--(6.35,5.7-11);
	\draw [<->,semithick](8.7,8.3-11)--(10.3,8.3-11);
	\node[scale=0.6] at (5.15+0.5,5.34-2-11) {$X_{2,2}$};
	\node[scale=0.6] at (4.33+0.5,4.65-2-11) {$Z_{2,2}$};
	\node[scale=0.6] at (4.1+0.5,5.3-2-11) {$Y_{2,2}$};
	\node[scale=0.8] at (5.84+0.5,4.2-2-11) {$P_{2,1}$};
	\node[scale=0.6] at (7.18+0.5,3.05-2-11) {$X_{2,1}$};
	\node[scale=0.6] at (6.85+0.5,2.22-2-11) {$Z_{2,1}$};
	\node[scale=0.6] at (6.03+0.5,2.85-2-11) {$Y_{2,1}$};
	\node[scale=0.45] at (7.35,9.08-2-11) {$X_{2,N}$};
	\node[scale=0.6] at (5.3+0.5,8.8-2-11) {$Z_{2,N}$};
	\node[scale=0.6] at (6.17+0.5,9.38-2-11) {$Y_{2,N}$};
	\node[scale=0.8] at (3.2,7-11) {$Q_{f_{2N-1}}$};
	\node[scale=0.8] at (3.5,3-11) {$Q_{f_3}$};
	\node[scale=0.8] at (3.5,0.6-11) {$Q_{f_1}$};
	\node[scale=0.8] at (5.2,1.8-11) {$Q_{f_2}$};
	\node[scale=0.8] at (10.22,-0.92-11) {$Q_4$};
	\node[scale=0.8] at (5.33,4.51-11) {$P_{2,2}$};
	\node[scale=0.5] at (5.68,5.92-11) {$P_{2,N-1}$};
	\node[scale=0.8] at (9.68,8.63-11) {$Q_3$};
	\node[scale=0.8] at (7.5,2.4-11) {$ Q_{2,1}$};
	\node[scale=0.8] at (6.6,4-11) {$ Q_{2,2}$};
	\node[scale=0.8] at (7.36,5.7-11) {$ Q_{2,N-1}$};

	\draw [<->,semithick] (-9.1,-0.6)--(-11.1,-0.6);
	\draw [<->,semithick] (-6.45-0.3-0.5,4.2-0.3-2)--(-4.75-0.3-0.5,4.2-0.3-2);
	\draw [<->,semithick] (-3.4-0.5,4.25-2)--(-3.4-0.5,5.75-2);
	\draw [<->,semithick] (-3.4-0.5,4.25+2)--(-3.4-0.5,5.75+2);
	\draw [<->,semithick] (-3.4-0.5,4.25-2-2.4)--(-3.4-0.5,5.75-2-2.4);
	\draw [<->,semithick](-4.55,4.2)--(-6.05,4.2);
	\draw [<->,semithick](-4.85,5.7)--(-6.35,5.7);
	\draw [<->,semithick](-8.7,8.3)--(-10.3,8.3);
	\node[scale=0.6] at (-5.15-0.5,5.34-2) {$X_{3,2}$};
	\node[scale=0.6] at (-4.33-0.5,4.65-2) {$Z_{3,2}$};
	\node[scale=0.6] at (-3.93-0.5,5.3-2) {$Y_{3,2}$};
	\node[scale=0.8] at (-5.84-0.5,4.2-2) {$P_{3,1}$};
	\node[scale=0.6] at (-7.08-0.5,3.05-2) {$X_{3,1}$};
	\node[scale=0.6] at (-6.78-0.5,2.22-2) {$Z_{3,1}$};
	\node[scale=0.6] at (-5.8-0.5,2.85-2) {$Y_{3,1}$};
	\node[scale=0.45] at (-7.35,9.08-2) {$X_{3,N}$};
	\node[scale=0.6] at (-5.3-0.5,8.8-2) {$Z_{3,N}$};
	\node[scale=0.6] at (-6.17-0.5,9.38-2) {$Y_{3,N}$};
	\node[scale=0.8] at (-3.08,7) {$Q_{f_{2N-1}}$};
	\node[scale=0.8] at (-3.4,3) {$Q_{f_3}$};
	\node[scale=0.8] at (-3.4,0.6) {$Q_{f_1}$};
	\node[scale=0.8] at (-4.98,1.8) {$Q_{f_2}$};
	\node[scale=0.8] at (-10.22,-0.92) {$Q_6$};
	\node[scale=0.8] at (-5.33,4.51) {$P_{3,2}$};
	\node[scale=0.5] at (-5.68,5.92) {$P_{3,N-1}$};
	\node[scale=0.8] at (-9.68,8.63) {$Q_5$};
	\node[scale=0.8] at (-7.57,2.4) {$ Q_{3,1}$};
	\node[scale=0.8] at (-6.6,4) {$ Q_{3,2}$};
	\node[scale=0.8] at (-7.36,5.9) {$ Q_{3,N-1}$};

	\draw [<->,semithick] (-9.1,-0.6-11)--(-11.1,-0.6-11);
	\draw [<->,semithick] (-6.45-0.3-0.5,4.2-0.3-2-11)--(-4.75-0.3-0.5,4.2-0.3-2-11);
	\draw [<->,semithick] (-3.4-0.5,4.25-2-11)--(-3.4-0.5,5.75-2-11);
	\draw [<->,semithick] (-3.4-0.5,4.25+2-11)--(-3.4-0.5,5.75+2-11);
	\draw [<->,semithick] (-3.4-0.5,4.25-2-2.4-11)--(-3.4-0.5,5.75-2-2.4-11);
	\draw [<->,semithick](-4.55,4.2-11)--(-6.05,4.2-11);
	\draw [<->,semithick](-4.85,5.7-11)--(-6.35,5.7-11);
	\draw [<->,semithick](-8.7,8.3-11)--(-10.3,8.3-11);
	\node[scale=0.6] at (-5.15-0.5,5.34-2-11) {$X_{4,2}$};
	\node[scale=0.6] at (-4.33-0.5,4.65-2-11) {$Z_{4,2}$};
	\node[scale=0.6] at (-3.93-0.5,5.3-2-11) {$Y_{4,2}$};
	\node[scale=0.8] at (-5.84-0.5,4.2-2-11) {$P_{4,1}$};
	\node[scale=0.6] at (-7.08-0.5,3.05-2-11) {$X_{4,1}$};
	\node[scale=0.6] at (-6.78-0.5,2.22-2-11) {$Z_{4,1}$};
	\node[scale=0.6] at (-5.8-0.5,2.85-2-11) {$Y_{4,1}$};
	\node[scale=0.45] at (-7.35,9.08-2-11) {$X_{4,N}$};
	\node[scale=0.6] at (-5.3-0.5,8.8-2-11) {$Z_{4,N}$};
	\node[scale=0.6] at (-6.17-0.5,9.38-2-11) {$Y_{4,N}$};
	\node[scale=0.8] at (-3.08,7-11) {$Q_{f_{2N-1}}$};
	\node[scale=0.8] at (-3.4,3-11) {$Q_{f_3}$};
	\node[scale=0.8] at (-3.4,0.6-11) {$Q_{f_1}$};
	\node[scale=0.8] at (-4.98,1.8-11) {$Q_{f_2}$};
	\node[scale=0.8] at (-10.22,-0.92-11) {$Q_8$};
	\node[scale=0.8] at (-5.33,4.51-11) {$P_{4,2}$};
	\node[scale=0.5] at (-5.68,5.92-11) {$P_{4,N-1}$};
	\node[scale=0.8] at (-9.68,8.63-11) {$Q_7$};
	\node[scale=0.8] at (-7.57,2.4-11) {$ Q_{4,1}$};
	\node[scale=0.8] at (-6.6,4-11) {$ Q_{4,2}$};
	\node[scale=0.8] at (-7.36,5.9-11) {$ Q_{4,N-1}$};

	\node at (0,5.05) {$Q_U$};
	\node[scale=0.45] at (0.96,2.8) {$Q_{f_{2N-1}}$};
	\node at (1.65,3.73) {$Y_{1,N}$};
	\node at (-1.4-0.25,3.73) {$Y_{3,N}$};
	\node at (1.7,1.27) {$Z_{2,N}$};
	\node at (-1.6,1.27) {$Z_{4,N}$};
	\node at (0,1.74-3.8-4.8) {$Q_b$};
	\node[scale=0.8] at (0.4,2.5-4.8) [right] {$Q_{f_3}$};
	\node[scale=0.8] at (0.4,2.6-3.8-4.8) [right] {$Q_{f_1}$};
	\node at (1.55,3.73-4.8) {$Y_{1,2}$};
	\node at (-1.63,3.73-4.8) {$Y_{3,2}$};
	\node at (1.56,1.2-4.8) {$Z_{2,2}$};
	\node at (-1.4,1.2-4.8) {$Z_{4,2}$};
	\node at (0,-3.87-4.8) {$Q_D$};
	\node[scale=0.8] at (1.13,0.6-4.8) {$Q_{f_2}$};
	\node at (1.55,3.73-3.8-4.8) {$Y_{1,1}$};
	\node at (-1.63,3.73-3.8-4.8) {$Y_{3,1}$};
	\node at (1.56,1.2-3.8-4.8) {$Z_{2,1}$};
	\node at (-1.4,1.2-3.8-4.8) {$Z_{4,1}$};	
	\draw [<->,semithick] (0.5,1.15)--(0.5,3.85);
	\draw [<->,semithick] (-0.65,5-0.3)--(0.65,5-0.3);
	\draw [<->,semithick] (-1.35,2.1-3.8-4.8)--(1.35,2.1-3.8-4.8);
	\draw [<->,semithick] (0.5,1.15-4.8)--(0.5,3.85-4.8);
	\draw [<->,semithick] (0.5,1.15-3.8-4.8)--(0.5,3.85-3.8-4.8);
	\draw [<->,semithick] (-0.65,-3.5-4.8)--(0.65,-3.5-4.8);

	\draw [semithick]
	(4+0.5,-1.6+9-2-11)--(4.8+0.5,-0.8+9-2-11)
	(4.8+0.5,-0.8+9-2-11)--(3.2+0.5,-0.8+9-2-11)
	(4.8+0.5,-0.8+9-2-11)--(6.2+0.5,-0.1+9-2-11)
	(6.2+0.5,-0.1+9-2-11)--(7.1+0.5,0.8+9-2-11)
	(7.1+0.5,0.8+9-2-11)--(3.2+0.5,0.8+9-2-11)
	(7.1+0.5,-0.1+9-2-11)--(5.6+0.5,-1.6+9-2-11)
	(7.1+0.5,-0.1+9-2-11)--(9.5+0.5,1.1+9-2-11)
	(7.1+0.5,0.8+9-2-11)--(8.7+0.5,1.6+9-2-11)
	;
	\draw (4.8+0.5,7.15-2-11+0.13) node{.};
	\draw (4.8+0.5,7-2-11+0.13) node{.};
	\draw (4.8+0.5,6.85-2-11+0.13) node{.};
	\draw [semithick]
	(4+0.5,6.6-2-11)--(4+0.5,5.8-2-11)
	(4+0.5,5.8-2-11)--(3.2+0.5,5.8-2-11)
	(4+0.5,5.8-2-11)--(4.7+0.5,5.1-2-11)
	(4.7+0.5,5.1-2-11)--(4.7+0.5,4.2-2-11)
	(4.7+0.5,4.2-2-11)--(3.2+0.5,4.2-2-11)
	(5.6+0.5,5.1-2-11)--(5.6+0.5,6.6-2-11)
	(5.6+0.5,5.1-2-11)--(7.9+0.5,2.8-2-11)
	(4.7+0.5,4.2-2-11)--(5.5+0.5,3.4-2-11)
	(5.5+0.5,3.4-2-11)--(3.2+0.5,3.4-2-11)
	(5.5+0.5,3.4-2-11)--(6.7+0.5,2.8-2-11)
	(6.7+0.5,2.8-2-11)--(7.7+0.5,1.8-2-11)
	(7.7+0.5,1.8-2-11)--(3.2+0.5,1.8-2-11)
	(7.7+0.5,1.8-2-11)--(8.9+0.5,1.2-2-11)
	(7.9+0.5,2.8-2-11)--(9.9+0.5,1.8-2-11);
	\draw [semithick,cyan]
	(6.2+0.5,-0.1+9-2-11)--(7.1+0.5,-0.1+9-2-11)
	(4.7+0.5,5.1-2-11)--(5.6+0.5,5.1-2-11)
	(7.9+0.5,2.8-2-11)--(6.7+0.5,2.8-2-11)
	;

	\draw [semithick]
	(-4-0.5,-1.6+9-2)--(-4.8-0.5,-0.8+9-2)
	(-4.8-0.5,-0.8+9-2)--(-3.2-0.5,-0.8+9-2)
	(-4.8-0.5,-0.8+9-2)--(-6.2-0.5,-0.1+9-2)
	(-6.2-0.5,-0.1+9-2)--(-7.1-0.5,0.8+9-2)
	(-7.1-0.5,0.8+9-2)--(-3.2-0.5,0.8+9-2)
	(-7.1-0.5,-0.1+9-2)--(-5.6-0.5,-1.6+9-2)
	(-7.1-0.5,-0.1+9-2)--(-9.5-0.5,1.1+9-2)
	(-7.1-0.5,0.8+9-2)--(-8.7-0.5,1.6+9-2)
	;
	\draw (-4.8-0.5,7.15-2+0.13) node{.};
	\draw (-4.8-0.5,7-2+0.13) node{.};
	\draw (-4.8-0.5,6.85-2+0.13) node{.};
	\draw [semithick]
	(-4-0.5,6.6-2)--(-4-0.5,5.8-2)
	(-4-0.5,5.8-2)--(-3.2-0.5,5.8-2)
	(-4-0.5,5.8-2)--(-4.7-0.5,5.1-2)
	(-4.7-0.5,5.1-2)--(-4.7-0.5,4.2-2)
	(-4.7-0.5,4.2-2)--(-3.2-0.5,4.2-2)
	(-5.6-0.5,5.1-2)--(-5.6-0.5,6.6-2)
	(-5.6-0.5,5.1-2)--(-7.9-0.5,2.8-2)
	(-4.7-0.5,4.2-2)--(-5.5-0.5,3.4-2)
	(-5.5-0.5,3.4-2)--(-3.2-0.5,3.4-2)
	(-5.5-0.5,3.4-2)--(-6.7-0.5,2.8-2)
	(-6.7-0.5,2.8-2)--(-7.7-0.5,1.8-2)
	(-7.7-0.5,1.8-2)--(-3.2-0.5,1.8-2)
	(-7.7-0.5,1.8-2)--(-8.9-0.5,1.2-2)
	(-7.9-0.5,2.8-2)--(-9.9-0.5,1.8-2);
	\draw [semithick,red]
	(-6.2-0.5,-0.1+9-2)--(-7.1-0.5,-0.1+9-2)
	(-4.7-0.5,5.1-2)--(-5.6-0.5,5.1-2)
	(-7.9-0.5,2.8-2)--(-6.7-0.5,2.8-2)
	;

	\draw [semithick]
	(-4-0.5,-1.6+9-2-11)--(-4.8-0.5,-0.8+9-2-11)
	(-4.8-0.5,-0.8+9-2-11)--(-3.2-0.5,-0.8+9-2-11)
	(-4.8-0.5,-0.8+9-2-11)--(-6.2-0.5,-0.1+9-2-11)
	(-6.2-0.5,-0.1+9-2-11)--(-7.1-0.5,0.8+9-2-11)
	(-7.1-0.5,0.8+9-2-11)--(-3.2-0.5,0.8+9-2-11)
	(-7.1-0.5,-0.1+9-2-11)--(-5.6-0.5,-1.6+9-2-11)
	(-7.1-0.5,-0.1+9-2-11)--(-9.5-0.5,1.1+9-2-11)
	(-7.1-0.5,0.8+9-2-11)--(-8.7-0.5,1.6+9-2-11)
	;
	\draw (-4.8-0.5,7.15-2-11+0.13) node{.};
	\draw (-4.8-0.5,7-2-11+0.13) node{.};
	\draw (-4.8-0.5,6.85-2-11+0.13) node{.};
	\draw [semithick]
	(-4-0.5,6.6-2-11)--(-4-0.5,5.8-2-11)
	(-4-0.5,5.8-2-11)--(-3.2-0.5,5.8-2-11)
	(-4-0.5,5.8-2-11)--(-4.7-0.5,5.1-2-11)
	(-4.7-0.5,5.1-2-11)--(-4.7-0.5,4.2-2-11)
	(-4.7-0.5,4.2-2-11)--(-3.2-0.5,4.2-2-11)
	(-5.6-0.5,5.1-2-11)--(-5.6-0.5,6.6-2-11)
	(-5.6-0.5,5.1-2-11)--(-7.9-0.5,2.8-2-11)
	(-4.7-0.5,4.2-2-11)--(-5.5-0.5,3.4-2-11)
	(-5.5-0.5,3.4-2-11)--(-3.2-0.5,3.4-2-11)
	(-5.5-0.5,3.4-2-11)--(-6.7-0.5,2.8-2-11)
	(-6.7-0.5,2.8-2-11)--(-7.7-0.5,1.8-2-11)
	(-7.7-0.5,1.8-2-11)--(-3.2-0.5,1.8-2-11)
	(-7.7-0.5,1.8-2-11)--(-8.9-0.5,1.2-2-11)
	(-7.9-0.5,2.8-2-11)--(-9.9-0.5,1.8-2-11);
	\draw [semithick,cyan]
	(-6.2-0.5,-0.1+9-2-11)--(-7.1-0.5,-0.1+9-2-11)
	(-4.7-0.5,5.1-2-11)--(-5.6-0.5,5.1-2-11)
	(-7.9-0.5,2.8-2-11)--(-6.7-0.5,2.8-2-11)
	;

	\end{tikzpicture}
\caption{Rank $N$ $(D_4,D_4)$ conformal matter on a circle in terms of brane webs with quadrivalent gluing. }
\label{fig:D4D4RN}
\end{figure}

In this section, we further explore the $(D_4,D_4)$ conformal matter on a circle with rank $N\geq 3$. The quadrivalently glued brane web is shown in figure \ref{fig:D4D4RN} which corresponds to the following affine $D_4$ quiver:
\begin{align*}
	\begin{tikzpicture}
	\node at (0,0) {$SU(2N)$};
	\node at (2.15+0.3,0.9){$SU(N)$};
	\node at (2.15+0.3,-0.9){$SU(N)$};
	\node at (-2.15-0.3,0.9){$SU(N)$};
	\node at (-2.15-0.3,-0.9){$SU(N)$};
	\node at (3+0.3,-1){.};
	\draw[semithick]
	(0.6+0.2,0.2)--(1.5+0.2,0.8)
	(0.6+0.2,-0.2)--(1.5+0.2,-0.8)
	(-0.6-0.2,0.2)--(-1.5-0.2,0.8)
	(-0.6-0.2,-0.2)--(-1.5-0.2,-0.8)
	;
	\end{tikzpicture}
\end{align*}

The independent K\"ahler parameters for the global symmetry are still $Q_1,\cdots,Q_8,Q_b$, the Coulomb branch parameters are $Q_{1,1},\cdots,Q_{1,N-1},Q_{2,1},\cdots,Q_{2,N-1},Q_{3,1},\cdots,$ $\allowbreak Q_{3,N-1},Q_{4,1},\cdots,Q_{4,N-1},Q_{f_1},\cdots,Q_{f_{2N-1}}$. We define 
\begin{equation}
	A_i\equiv\sqrt{Q_{f_i}}\quad \forall i\in\{1,\cdots,2N-1\}. 
\end{equation}

Just like the rank 2 case, we still define the following parameter to replace $Q_b$ as the independent K\"ahler parameter:
\begin{equation}
	Q_0\equiv\frac{Q_b}{Q_{f_1}}\sqrt{\frac{P_{2,1}P_{4,1}}{Q_3Q_7}}. 
\end{equation}
Mimicking the rank 1 case, we can find that for $I\in\{1,2,3,4\}$, $i\in\{1,\cdots,N\}$, 
\begin{equation}
	X_{I,i}=\sqrt{\frac{P_{I,i}P_{I,i-1}}{Q_{f_{2i-1}}}},\quad Y_{I,i}=\sqrt{\frac{P_{I,i}Q_{f_{2i-1}}}{P_{I,i-1}}},\quad Z_{I,i}=\sqrt{\frac{P_{I,i-1}Q_{f_{2i-1}}}{P_{I,i}}},
\end{equation}
in which
\begin{alignat}{4}
    P_{1,0} &\equiv Q_2, \quad & P_{2,0} &\equiv Q_4, \quad & P_{3,0} &\equiv Q_6, \quad & P_{4,0} &\equiv Q_8, \\
    P_{1,N} &\equiv Q_1, \quad & P_{2,N} &\equiv Q_3, \quad & P_{3,N} &\equiv Q_5, \quad & P_{4,N} &\equiv Q_7.
\end{alignat}
Then we can find that
\begin{align}
	Q_D&=\frac{Q_b}{Z_{4,1}Z_{2,1}}=\frac{Q_b}{Q_{f_1}}\sqrt{\frac{P_{2,1}P_{4,1}}{Q_4Q_8}}=Q_0\sqrt{\frac{Q_3Q_7}{Q_4Q_8}},\\
	Q_U&=\frac{Z_{2,1}Z_{4,1}Z_{2,2}Z_{4,2}\cdots Z_{2,N}Z_{4,N}}{Y_{1,1}Y_{3,1}Y_{1,2}Y_{3,2}\cdots Y_{1,N}Y_{3,N}}Q_D\nn\\
	&=\sqrt{\frac{P_{1,0}P_{2,0}P_{3,0}P_{4,0}}{P_{1,N}P_{2,N}P_{3,N}P_{4,N}}}Q_D=Q_0\sqrt{\frac{Q_2Q_6}{Q_1Q_5}}. 
\end{align}
We still define the mass parameters in the following way:
\begin{align}
	&M_1=\sqrt{Q_1Q_2},\quad M_2=\sqrt{\frac{Q_2}{Q_1}},\quad M_3=\sqrt{\frac{Q_4}{Q_3}},\quad M_4=\sqrt{\frac{1}{Q_3 Q_4}},\nn\\
	&M_5=\sqrt{Q_5Q_6},\quad M_6=\sqrt{\frac{Q_6}{Q_5}},\quad M_7=\sqrt{\frac{Q_8}{Q_7}},\quad M_8=\sqrt{\frac{1}{Q_7 Q_8}}. 
	\label{eq:MtoQ_rN}
\end{align}
The period $q$ is still the same as the rank 1, 2 case:
\begin{equation}
	q=Q_2Q_3Q_6Q_7Q_0^2. 
\end{equation}
The ten flop transitions that correspond to $Q_1,\cdots,Q_8,Q_U,Q_D$ are

\begin{align}
	&\mathbf{V}_1=\big{\{}M_1\rightarrow M_2,\ M_2\rightarrow M_1\ \ ; \ \ Q_{1,N-1}\to \frac{M_1 Q_{1,N-1}}{M_2}\big{\}},\nn\\
	&\mathbf{V}_2=\big{\{}M_1\rightarrow M_2^{-1},\ M_2\rightarrow M_1^{-1}\ \ ; \ \ Q_{1,1}\to M_1 M_2 Q_{1,1}\big{\}},\nn\\
	&\mathbf{V}_3=\big{\{}M_3\rightarrow M_4^{-1},\ M_4\rightarrow M_3^{-1}\ \ ; \ \ Q_{2,N-1}\to \frac{Q_{2,N-1}}{M_3 M_4}\big{\}},\nn\\
	&\mathbf{V}_4=\big{\{}M_3\rightarrow M_4,\ M_4\rightarrow M_3\ \ ; \ \ Q_{2,1}\to \frac{M_3 Q_{2,1}}{M_4}\big{\}},\nn\\
	&\mathbf{V}_5=\big{\{}M_5\rightarrow M_6,\ M_6\rightarrow M_5\ \ ; \ \ Q_{3,N-1}\to \frac{M_5 Q_{3,N-1}}{M_6}\big{\}},\nn\\
	&\mathbf{V}_6=\big{\{}M_5\rightarrow M_6^{-1},\ M_6\rightarrow M_5^{-1}\ \ ; \ \ Q_{3,1}\to M_5 M_6 Q_{3,1}\big{\}},\nn\\
	&\mathbf{V}_7=\big{\{}M_7\rightarrow M_8^{-1},\ M_8\rightarrow M_7^{-1}\ \ ;\ \ Q_{4,N-1}\to \frac{Q_{4,N-1}}{M_7 M_8}\big{\}},\nn\\
	&\mathbf{V}_8=\big{\{}M_7\rightarrow M_8,\ M_8\rightarrow M_7\ \ ;\ \ Q_{4,1}\to \frac{M_7 Q_{4,1}}{M_8}\big{\}},\nn\\
	&\mathbf{V}_9=\big{\{}\resizebox{0.85\hsize}{!}{$M_1\rightarrow\left(\frac{M_1^3M_2M_3M_4M_6M_7M_8q}{M_5}\right)^{\frac14},\ M_2\rightarrow\left(\frac{M_1M_2^3M_5}{M_3M_4M_6M_7M_8q}\right)^{\frac14},\ M_3\rightarrow\left(\frac{M_1M_3^3M_5}{M_2M_4M_6M_7M_8q}\right)^{\frac14},$}\nn\\
	&\qquad\quad\,\resizebox{0.85\hsize}{!}{$M_4\rightarrow\left(\frac{M_1M_4^3M_5}{M_2M_3M_6M_7M_8q}\right)^{\frac14},\ M_5\rightarrow\left(\frac{M_2M_3M_4M_5^3M_6M_7M_8q}{M_1}\right)^{\frac14},\ M_6\rightarrow\left(\frac{M_1M_5M_6^3}{M_2M_3M_4M_7M_8q}\right)^{\frac14},$}\nn\\
	&\qquad\quad\,\resizebox{0.84\hsize}{!}{$M_7\rightarrow\left(\frac{M_1M_5M_7^3}{M_2M_3M_4M_6M_8q}\right)^{\frac14},\ M_8\rightarrow\left(\frac{M_1M_5M_8^3}{M_2M_3M_4M_6M_7q}\right)^{\frac14}\ \ ; \ \ A_{2N-1}\rightarrow A_{2N-1}\left(\frac{M_2M_3M_4M_6M_7M_8q}{M_1M_5}\right)^{\frac14}$}\big{\}},\nn\\
	&\mathbf{V}_{10}=\big{\{}\resizebox{0.76\hsize}{!}{$M_1\rightarrow\left(\frac{M_1^3M_4M_8q}{M_2M_3M_5M_6M_7}\right)^{\frac14},\ M_2\rightarrow\left(\frac{M_2^3M_4M_8q}{M_1M_3M_5M_6M_7}\right)^{\frac14},\ M_3\rightarrow\left(\frac{M_3^3M_4M_8q}{M_1M_2M_5M_6M_7}\right)^{\frac14},$}\nn\\
	&\qquad\quad\ \,\resizebox{0.82\hsize}{!}{$M_4\rightarrow\left(\frac{M_1M_2M_3M_4^3M_5M_6M_7}{M_8q}\right)^{\frac14},\ M_5\rightarrow\left(\frac{M_4M_5^3M_8q}{M_1M_2M_3M_6M_7}\right)^{\frac14},\ M_6\rightarrow\left(\frac{M_4M_6^3M_8q}{M_1M_2M_3M_5M_7}\right)^{\frac14},$}\nn\\
	&\qquad\quad\ \,\resizebox{0.84\hsize}{!}{$M_7\rightarrow\left(\frac{M_4M_7^3M_8q}{M_1M_2M_3M_5M_6}\right)^{\frac14},\ M_8\rightarrow\left(\frac{M_1M_2M_3M_5M_6M_7M_8^3}{M_4q}\right)^{\frac14}\ \ ; \ \ A_1\rightarrow A_1\left(\frac{M_4M_8q}{M_1M_2M_3M_5M_6M_7}\right)^{\frac14}$}\big{\}}.
	\label{eq:D4D4quadflopsMRN}
\end{align}

Like the rank 2 theory, we have the following six flop transitions that exchange a single mass parameter between different $SU(N)$ subdiagrams but do not alter the Coulomb branch parameters:
\begin{alignat}{2}
	\mathbf{V}_{\text{I}}&=\big{\{}M_1\rightarrow M_4^{-1},\ M_4\rightarrow M_1^{-1}\big{\}},\quad &\mathbf{V}_{\text{II}}&=\big{\{}M_4\to M_5^{-1},\ M_5\to M_4^{-1}\big{\}},\nn\\
	\mathbf{V}_{\text{III}}&=\big{\{}M_5\to M_8^{-1},\ M_8\to M_5^{-1}\big{\}},\quad &
	\mathbf{V}_{\text{IV}}&=\big{\{}M_1\to M_8^{-1},\ M_8\to M_1^{-1}\big{\}},\nn\\
	\mathbf{V}_{\text{V}}&=\big{\{}M_1\to M_5,\ M_5\to M_1\big{\}},\quad &
	\mathbf{V}_{\text{VI}}&=\big{\{}\ M_4\to M_8,\ M_8\to M_4\big{\}}.  
\end{alignat}
Similar to rank 2 case, we have 64 different ways to form the affine $E_8$ Weyl symmetry by picking six out of the ten flop transitions from equation \eqref{eq:D4D4quadflopsMRN}, the 64 different choices are the same as the rank 2 case which are listed in equation \eqref{eq:64choicesR2}. As an example, we list the standard affine $E_8$ Weyl reflections and affine $E_8$ invariant Coulomb branch parameters from the choice $\mathbf{V}_1,\mathbf{V}_2,\mathbf{V}_3,\mathbf{V}_6,\mathbf{V}_8,\mathbf{V}_9$: 
\begin{align}
	\mathbf{W}_0=&\big\{M_1\to q\,M_8,\ M_8\to q^{-1}M_1\ \ ; \ \ Q_{1,1}\to \frac{M_1 Q_{1,1}}{M_8 q},\ Q_{1,N-1}\to \frac{M_1 Q_{1,N-1}}{M_8 q}\big\}, \nn\\
	\mathbf{W}_1=&\big\{M_1\to M_2,\ M_2\to M_1\ \ ; \ \ Q_{1,N-1}\to \frac{M_1 Q_{1,N-1}}{M_2}\big\},\nn\\
	\mathbf{W}_2=&\big\{M_2\to M_3,\ M_3\to M_2\ \ ; \ \ Q_{1,N-1}\to \frac{M_3 Q_{1,N-1}}{M_2},\ Q_{2,N-1}\to \frac{M_2 Q_{2,N-1}}{M_3}\big\},\nn\\
	\mathbf{W}_3=&\big\{M_3\to M_4,\ M_4\to M_3\ \ ; \ \ Q_{1,1}\to \frac{M_3 Q_{1,1}}{M_4},\ Q_{1,N-1}\to \frac{M_3 Q_{1,N-1}}{M_4},\nn\\
	&\phantom{\big\{}Q_{2,N-1}\to \frac{M_4 Q_{2,N-1}}{M_3}\big\},\nn\\
	\mathbf{W}_4=&\big\{M_4\to M_5,\ M_5\to M_4\ \ ; \ \ Q_{1,1}\to \frac{M_5 Q_{1,1}}{M_4},\ Q_{1,N-1}\to \frac{M_5 Q_{1,N-1}}{M_4}\big\},\nn\\
	\mathbf{W}_5=&\big\{M_5\to M_6,\ M_6\to M_5\ \ ; \ \ Q_{1,1}\to \frac{M_5 Q_{1,1}}{M_6},\ Q_{1,N-1}\to \frac{M_5 Q_{1,N-1}}{M_6},\nn\\
	&\phantom{\big\{}Q_{3,1}\to \frac{M_6 Q_{3,1}}{M_5}\big\},\nn\\
	\mathbf{W}_6=&\big\{M_6\to \frac{1}{M_7},\ M_7\to \frac{1}{M_6}\ \ ; \ \ Q_{1,1}\to \frac{Q_{1,1}}{M_6 M_7},\ Q_{1,N-1}\to \frac{Q_{1,N-1}}{M_6 M_7},\nn\\
	&\phantom{\big\{}Q_{3,1}\to M_6 M_7 Q_{3,1},\ Q_{4,1}\to M_6 M_7 Q_{4,1}\big\},\nn\\
	\mathbf{W}_7=&\big{\{}\resizebox{0.87\hsize}{!}{$M_i\rightarrow \frac{M_i}{(M_1M_2M_3M_4M_5M_6M_7M_8)^{1/4}}\ \forall i\in\{1,\cdots,8\}\ \ ; \ \ 
	A_{2N-1}\rightarrow \frac{A_{2N-1}}{(M_1M_2M_3M_4M_5M_6M_7M_8)^{1/4}},\ $}\nn\\
	&\phantom{\big{\{}}\resizebox{0.72\hsize}{!}{$Q_{1,1}\to \frac{Q_{1,1}}{(M_1M_2M_3M_4M_5M_6M_7M_8)^{1/2}},\ Q_{1,N-1}\to \frac{Q_{1,N-1}}{M_1M_2M_3M_4M_5M_6M_7M_8},\ $}\nn\\
	&\phantom{\big{\{}}\resizebox{0.84\hsize}{!}{$Q_{2,N-1}\to \frac{Q_{2,N-1}}{(M_1M_2M_3M_4M_5M_6M_7M_8)^{1/2}},\ Q_{3,1}\to \left(M_1M_2M_3M_4M_5M_6M_7M_8\right)^{\frac12}Q_{3,1},$}\nn\\
	&\phantom{\big{\{}}\resizebox{0.44\hsize}{!}{$Q_{4,1}\to \left(M_1M_2M_3M_4M_5M_6M_7M_8\right)^{\frac12}Q_{4,1}$}\big{\}},\nn\\
	\mathbf{W}_8=&\big\{M_6\to M_7,\ M_7\to M_6\ \ ; \ \ Q_{3,1}\to \frac{M_6 Q_{3,1}}{M_7},\ Q_{4,1}\to \frac{M_7 Q_{4,1}}{M_6}\big\}.
\end{align}
\begin{alignat}{2}
	&\tilde{Q}_{1,1}=\Theta(q,\boldsymbol{M})^2 \frac{\sqrt{M_1 M_2 M_3 M_5} Q_{1,1}}{\sqrt{M_4 M_6 M_7 M_8^5}},\quad &&\tilde{Q}_{1,N-1}=\Theta(q,\boldsymbol{M})^3\frac{\sqrt{M_1 M_3 M_5} Q_{1,N-1}}{\sqrt{M_2 M_4 M_6 M_7 M_8^7}},\nn\\
	&\tilde{Q}_{2,1}=Q_{2,1}, && \tilde{Q}_{2,N-1}=\Theta(q,\boldsymbol{M})\frac{Q_{2,N-1}}{M_3 M_8},\nn\\
	&\tilde{Q}_{3,1}=\Theta(q,\boldsymbol{M})^{-1}M_6 M_8 Q_{3,1}, && \tilde{Q}_{3,N-1}=Q_{3,N-1},\nn\\
	&\tilde{Q}_{4,1}=\Theta(q,\boldsymbol{M})^{-1}M_7 M_8 Q_{4,1}, && \tilde{Q}_{4,N-1}=Q_{4,N-1},\nn\\
	&\tilde{A}_{2N-1}=\Theta(q,\boldsymbol{M})\frac{A_{2N-1}}{M_8}, && 
\end{alignat}
the Coulomb branch parameters $Q_{I,i},\forall I\in\{1,2,3,4\},\forall i\in\{2,\cdots,N-2\}$ and $A_i,\forall i\in\{1,\cdots,2N-2\}$ are unaffected by the affine $E_8$ Weyl reflections, so they are automatically affine $E_8$ invariant.

The author has made a Mathematica code to compute all the 64 different sets of standard affine $E_8$ Weyl reflections and the corresponding affine $E_8$ invariant Coulomb branch parameters, the code can be found in \cite{WeylSymmetryD4D4}. We list the results of the first three choices of equation \eqref{eq:64choicesR2} in appendix \ref{app:rankN}.

\section{Conclusion}
In this paper, we find that $(D_4,D_4)$ conformal matter theories on a circle with general rank all have affine $E_8$ global symmetry by studying the Weyl symmetry in their quadrivalently glued brane webs. The usual flop transition that corresponds to a Weyl symmetry is exchanging between two parallel branes, but there are more flop transitions that do not belong to this category. We find such nontrivial hidden flop transitions in the rank $N\geq 2$ theories which are crucial in forming the full affine $E_8$ symmetry as well as deriving the affine $E_8$ invariant Coulomb branch parameters. When $N\geq 2$, the theory has 64 different ways to form the affine $E_8$ symmetry due to different transformations on the Coulomb branch parameters, we have computed all the 64 different ways and found out the corresponding Weyl reflections and invariant Coulomb branch parameters.

\acknowledgments
We thank Shi Cheng, Babak Haghighat, Sung-Soo Kim, Yuji Sugimoto and Futoshi Yagi for discussion, we thank Futoshi Yagi for many helpful comments, suggestions and discussions. The work of the author is supported by YMSC, Tsinghua University. 

\appendix

\section{Invariant Coulomb branch parameters for rank 2 theory}\label{app:rank2}
In this section, we list the affine $E_8$ Weyl reflections and the corresponding affine $E_8$ invariant Coulomb branch parameters for the first three choices of the 64 different choices in \eqref{eq:64choicesR2} for the rank 2 theory. As the global symmetry part of these affine $E_8$ Weyl reflections are the same which is just equation \eqref{eq:affineE8Wy} with $y_i$ replaced by $M_i$, we omit the global symmetry part and only list the local symmetry part which is the transformations on Coulomb branch parameters. Among the 64 different choices, when we choose $\mathbf{V}_9$, the affine $E_8$ invariant Coulomb branch parameters for the middle $SU(4)$ node are
\begin{equation}
	\tilde{A}_1=A_1,\quad \tilde{A}_2=A_2,\quad \tilde{A}_3=\frac{\Theta A_3}{M_8}, 
\end{equation}
where $\Theta$ is $\Theta(q,\boldsymbol{M})$ in equation \eqref{eq:JacobiTheta} for short. When we choose $\mathbf{V}_{10}$, the affine $E_8$ invariant Coulomb branch parameters for the middle $SU(4)$ node are
\begin{equation}
	\tilde{A}_1=\frac{\Theta A_1}{M_8},\quad \tilde{A}_2=A_2,\quad \tilde{A}_3=A_3. 
\end{equation}
So in the following list of invariant Coulomb branch parameters, we will also omit $\tilde A_1,\tilde A_2,\tilde A_3$ for simplicity. 

The following are the standard affine $E_8$ Weyl reflections and affine $E_8$ invariant Coulomb branch parameters for the first three choices in equation \eqref{eq:64choicesR2}, where we define the parameter $M\equiv M_1M_2M_3M_4M_5M_6M_7M_8$. 

\paragraph{1. $\mathbf{V}_{1},\mathbf{V}_{2},\mathbf{V}_{3},\mathbf{V}_{5},\mathbf{V}_{7},\mathbf{V}_{9}$}
\begin{align}
	&\mathbf{W}_0=\big\{Q_{1,1}\to \frac{M_1^2 Q_{1,1}}{M_8^2 q^2}\big\},\nn\\
	&\mathbf{W}_1=\big\{Q_{1,1}\to \frac{M_1 Q_{1,1}}{M_2}\big\},\nn\\
	&\mathbf{W}_2=\big\{Q_{1,1}\to \frac{M_3 Q_{1,1}}{M_2},Q_{2,1}\to \frac{M_2 Q_{2,1}}{M_3}\big\},\nn\\
	&\mathbf{W}_3=\big\{Q_{1,1}\to \frac{M_3^2 Q_{1,1}}{M_4^2},Q_{2,1}\to \frac{M_4 Q_{2,1}}{M_3}\big\},\nn\\
	&\mathbf{W}_4=\big\{Q_{1,1}\to \frac{M_5^2 Q_{1,1}}{M_4^2}\big\},\nn\\
	&\mathbf{W}_5=\big\{Q_{3,1}\to \frac{M_5 Q_{3,1}}{M_6}\big\},\nn\\
	&\mathbf{W}_6=\big\{Q_{1,1}\to M_6^2 M_7^2 Q_{1,1},Q_{3,1}\to \frac{Q_{3,1}}{M_6 M_7},Q_{4,1}\to \frac{Q_{4,1}}{M_6 M_7}\big\},\nn\\
	&\mathbf{W}_7=\big\{A_3\to \frac{A_3}{\sqrt[4]{M}},Q_{1,1}\to \sqrt{M} Q_{1,1},Q_{2,1}\to \frac{Q_{2,1}}{\sqrt{M}},Q_{3,1}\to \frac{Q_{3,1}}{\sqrt{M}},Q_{4,1}\to \frac{Q_{4,1}}{\sqrt{M}}\big\},\nn\\
	&\mathbf{W}_8=\big\{Q_{3,1}\to \frac{M_7 Q_{3,1}}{M_6},Q_{4,1}\to \frac{M_6 Q_{4,1}}{M_7}\big\}. 
\end{align}
\begin{equation}
	\tilde{Q}_{1,1}=\frac{\Theta  M_1 M_3 M_5 M_6 M_7 Q_{1,1}}{M_4 M_8^2},\  \tilde{Q}_{2,1}=\frac{\Theta  Q_{2,1}}{M_3 M_8},\
	\tilde{Q}_{3,1}=\frac{\Theta  Q_{3,1}}{M_6 M_8},\  \tilde{Q}_{4,1}=\frac{\Theta  Q_{4,1}}{M_7 M_8}. 
\end{equation}

\paragraph{2. $\mathbf{V}_{1},\mathbf{V}_{2},\mathbf{V}_{3},\mathbf{V}_{5},\mathbf{V}_{8},\mathbf{V}_{9}$}
\begin{align}
	&\mathbf{W}_0=\big\{Q_{1,1}\to \frac{M_1^2 Q_{1,1}}{M_8^2 q^2}\big\},\nn\\
	&\mathbf{W}_1=\big\{Q_{1,1}\to \frac{M_1 Q_{1,1}}{M_2}\big\},\nn\\
	&\mathbf{W}_2=\big\{Q_{1,1}\to \frac{M_3 Q_{1,1}}{M_2},Q_{2,1}\to \frac{M_2 Q_{2,1}}{M_3}\big\},\nn\\
	&\mathbf{W}_3=\big\{Q_{1,1}\to \frac{M_3^2 Q_{1,1}}{M_4^2},Q_{2,1}\to \frac{M_4 Q_{2,1}}{M_3}\big\},\nn\\
	&\mathbf{W}_4=\big\{Q_{1,1}\to \frac{M_5^2 Q_{1,1}}{M_4^2}\big\},\nn\\
	&\mathbf{W}_5=\big\{Q_{3,1}\to \frac{M_5 Q_{3,1}}{M_6}\big\},\nn\\
	&\mathbf{W}_6=\big\{Q_{3,1}\to \frac{Q_{3,1}}{M_6 M_7},Q_{4,1}\to M_6 M_7 Q_{4,1}\big\},\nn\\
	&\mathbf{W}_7=\big\{A_3\to \frac{A_3}{\sqrt[4]{M}},Q_{1,1}\to \frac{Q_{1,1}}{\sqrt{M}},Q_{2,1}\to \frac{Q_{2,1}}{\sqrt{M}},Q_{3,1}\to \frac{Q_{3,1}}{\sqrt{M}},Q_{4,1}\to \sqrt{M} Q_{4,1}\big\},\nn\\
	&\mathbf{W}_8=\big\{Q_{1,1}\to \frac{M_6^2 Q_{1,1}}{M_7^2},Q_{3,1}\to \frac{M_7 Q_{3,1}}{M_6},Q_{4,1}\to \frac{M_7 Q_{4,1}}{M_6}\big\}. 
\end{align}
\begin{equation}
	\tilde{Q}_{1,1}=\frac{\Theta ^3 M_1 M_3 M_5 M_6 Q_{1,1}}{M_4 M_7 M_8^4},\ \tilde{Q}_{2,1}=\frac{\Theta  Q_{2,1}}{M_3 M_8},\
	\tilde{Q}_{3,1}=\frac{\Theta  Q_{3,1}}{M_6 M_8},\  \tilde{Q}_{4,1}=\frac{M_7 M_8 Q_{4,1}}{\Theta }. 
\end{equation}

\paragraph{3. $\mathbf{V}_{1},\mathbf{V}_{2},\mathbf{V}_{3},\mathbf{V}_{6},\mathbf{V}_{7},\mathbf{V}_{9}$}

\begin{align}
	&\mathbf{W}_0=\big\{Q_{1,1}\to \frac{M_1^2 Q_{1,1}}{M_8^2 q^2}\big\},\nn\\
	&\mathbf{W}_1=\big\{Q_{1,1}\to \frac{M_1 Q_{1,1}}{M_2}\big\},\nn\\
	&\mathbf{W}_2=\big\{Q_{1,1}\to \frac{M_3 Q_{1,1}}{M_2},Q_{2,1}\to \frac{M_2 Q_{2,1}}{M_3}\big\},\nn\\
	&\mathbf{W}_3=\big\{Q_{1,1}\to \frac{M_3^2 Q_{1,1}}{M_4^2},Q_{2,1}\to \frac{M_4 Q_{2,1}}{M_3}\big\},\nn\\
	&\mathbf{W}_4=\big\{Q_{1,1}\to \frac{M_5^2 Q_{1,1}}{M_4^2}\big\},\nn\\
	&\mathbf{W}_5=\big\{Q_{1,1}\to \frac{M_5^2 Q_{1,1}}{M_6^2},Q_{3,1}\to \frac{M_6 Q_{3,1}}{M_5}\big\},\nn\\
	&\mathbf{W}_6=\big\{Q_{3,1}\to M_6 M_7 Q_{3,1},Q_{4,1}\to \frac{Q_{4,1}}{M_6 M_7}\big\},\nn\\
	&\mathbf{W}_7=\big\{A_3\to \frac{A_3}{\sqrt[4]{M}},Q_{1,1}\to \frac{Q_{1,1}}{\sqrt{M}},Q_{2,1}\to \frac{Q_{2,1}}{\sqrt{M}},Q_{3,1}\to \sqrt{M} Q_{3,1},Q_{4,1}\to \frac{Q_{4,1}}{\sqrt{M}}\big\},\nn\\
	&\mathbf{W}_8=\big\{Q_{1,1}\to \frac{M_7^2 Q_{1,1}}{M_6^2},Q_{3,1}\to \frac{M_6 Q_{3,1}}{M_7},Q_{4,1}\to \frac{M_6 Q_{4,1}}{M_7}\big\}. 
\end{align}

\begin{equation}
	\tilde{Q}_{1,1}=\frac{\Theta ^3 M_1 M_3 M_5 M_7 Q_{1,1}}{M_4 M_6 M_8^4},\ \tilde{Q}_{2,1}=\frac{\Theta  Q_{2,1}}{M_3 M_8},\
	\tilde{Q}_{3,1}=\frac{M_6 M_8 Q_{3,1}}{\Theta },\  \tilde{Q}_{4,1}=\frac{\Theta  Q_{4,1}}{M_7 M_8}. 
\end{equation}

\section{Invariant Coulomb branch parameters for rank $N\geq 3$ theories}\label{app:rankN}
There are 64 different choices of $\mathbf{V}_{i_1},\mathbf{V}_{i_2},\mathbf{V}_{i_3},\mathbf{V}_{i_4},\mathbf{V}_{i_5},\mathbf{V}_{i_6}$ as listed in equation \eqref{eq:64choicesR2} to form the standard affine $E_8$ Weyl reflections for the rank $N\geq 3$ theories. In this section, we list the results of the first three choices of equation \eqref{eq:64choicesR2}. 
Like in appendix \ref{app:rank2}, we also omit the global symmetry part of these affine $E_8$ Weyl reflections and only list the local symmetry part. For rank $N\geq 3$ theories, only the Coulomb branch parameters that are near the top and bottom of the brane web are affected by the Weyl reflections, the other Coulomb branch parameters are not affected and thus are automatically affine $E_8$ invariant. These invariant Coulomb branch parameters in the four $SU(N)$ gauge nodes are
\begin{align}
	\tilde{Q}_{I,i}=Q_{I,i}\ \ \forall I\in\{1,2,3,4\},\forall i\in\{2,\cdots,N-2\}. 
\end{align}
Among the 64 different choices, when we choose $\mathbf{V}_9$, the affine $E_8$ invariant Coulomb branch parameters for the middle $SU(2N)$ node are
\begin{equation}
	\tilde{A}_i=A_i\ \ \forall i\in\{1,\cdots,2N-2\},\quad \tilde{A}_{2N-1}=\Theta(q,\boldsymbol{M})\frac{A_{2N-1}}{M_8}.
\end{equation}
When we choose $\mathbf{V}_{10}$, the affine $E_8$ invariant Coulomb branch parameters for the middle $SU(2N)$ node are
\begin{equation}
	\tilde{A}_{1}=\Theta(q,\boldsymbol{M})\frac{A_{1}}{M_8},\quad \tilde{A}_i=A_i\ \ \forall i\in\{2,\cdots,2N-1\}. 
\end{equation}
In the following, we list the affine $E_8$ Weyl reflections and affine $E_8$ invariant Coulomb branch parameters for the first three choices of the 64 different choices in equation \eqref{eq:64choicesR2}. 
\paragraph{1. $\mathbf{V}_{1},\mathbf{V}_{2},\mathbf{V}_{3},\mathbf{V}_{5},\mathbf{V}_{7},\mathbf{V}_{9}$}

\begin{align}
	&\mathbf{W}_0=\big\{Q_{1,1}\to \frac{M_1 Q_{1,1}}{M_8 q},Q_{1,N-1}\to \frac{M_1 Q_{1,N-1}}{M_8 q}\big\},\nn\\
	&\mathbf{W}_1=\big\{Q_{1,N-1}\to \frac{M_1 Q_{1,N-1}}{M_2}\big\},\nn\\
	&\mathbf{W}_2=\big\{Q_{1,N-1}\to \frac{M_3 Q_{1,N-1}}{M_2},Q_{2,N-1}\to \frac{M_2 Q_{2,N-1}}{M_3}\big\},\nn\\
	&\mathbf{W}_3=\big\{Q_{1,1}\to \frac{M_3 Q_{1,1}}{M_4},Q_{1,N-1}\to \frac{M_3 Q_{1,N-1}}{M_4},Q_{2,N-1}\to \frac{M_4 Q_{2,N-1}}{M_3}\big\},\nn\\
	&\mathbf{W}_4=\big\{Q_{1,1}\to \frac{M_5 Q_{1,1}}{M_4},Q_{1,N-1}\to \frac{M_5 Q_{1,N-1}}{M_4}\big\},\nn\\
	&\mathbf{W}_5=\big\{Q_{3,N-1}\to \frac{M_5 Q_{3,N-1}}{M_6}\big\},\nn\\
	&\mathbf{W}_6=\big\{Q_{1,1}\to M_6 M_7 Q_{1,1},Q_{1,N-1}\to M_6 M_7 Q_{1,N-1},Q_{3,N-1}\to \frac{Q_{3,N-1}}{M_6 M_7},\nn\\
	&\phantom{\mathbf{W}_6=\big\{}Q_{4,N-1}\to \frac{Q_{4,N-1}}{M_6 M_7}\big\},\nn\\
	&\mathbf{W}_7=\big\{A_{2 N-1}\to \frac{A_{2 N-1}}{\sqrt[4]{M}},Q_{1,1}\to \sqrt{M} Q_{1,1},Q_{2,N-1}\to \frac{Q_{2,N-1}}{\sqrt{M}},Q_{3,N-1}\to \frac{Q_{3,N-1}}{\sqrt{M}},\nn\\
	&\phantom{\mathbf{W}_7=\big\{}Q_{4,N-1}\to \frac{Q_{4,N-1}}{\sqrt{M}}\big\},\nn\\
	&\mathbf{W}_8=\big\{Q_{3,N-1}\to \frac{M_7 Q_{3,N-1}}{M_6},Q_{4,N-1}\to \frac{M_6 Q_{4,N-1}}{M_7}\big\}. 
\end{align}

\begin{alignat}{2}
	&\tilde{Q}_{1,1}=\sqrt{\frac{M_1 M_2 M_3 M_5 M_6 M_7}{M_4 M_8}} Q_{1,1},\quad &&\tilde{Q}_{1,N-1}=\frac{\Theta  \sqrt{M_1 M_3 M_5 M_6 M_7} Q_{1,N-1}}{\sqrt{M_2 M_4 M_8^3}},\nn\\
	&\tilde{Q}_{2,1}=Q_{2,1},&&\tilde{Q}_{2,N-1}=\frac{\Theta  Q_{2,N-1}}{M_3 M_8},\nn\\
	&\tilde{Q}_{3,1}=Q_{3,1},&&\tilde{Q}_{3,N-1}=\frac{\Theta  Q_{3,N-1}}{M_6 M_8},\nn\\
	&\tilde{Q}_{4,1}=Q_{4,1},&&\tilde{Q}_{4,N-1}=\frac{\Theta  Q_{4,N-1}}{M_7 M_8}. 
\end{alignat}

\paragraph{2. $\mathbf{V}_{1},\mathbf{V}_{2},\mathbf{V}_{3},\mathbf{V}_{5},\mathbf{V}_{8},\mathbf{V}_{9}$}
\begin{align}
	&\mathbf{W}_0=\big\{Q_{1,1}\to \frac{M_1 Q_{1,1}}{M_8 q},Q_{1,N-1}\to \frac{M_1 Q_{1,N-1}}{M_8 q}\big\},\nn\\
	&\mathbf{W}_1=\big\{Q_{1,N-1}\to \frac{M_1 Q_{1,N-1}}{M_2}\big\},\nn\\
	&\mathbf{W}_2=\big\{Q_{1,N-1}\to \frac{M_3 Q_{1,N-1}}{M_2},Q_{2,N-1}\to \frac{M_2 Q_{2,N-1}}{M_3}\big\},\nn\\
	&\mathbf{W}_3=\big\{Q_{1,1}\to \frac{M_3 Q_{1,1}}{M_4},Q_{1,N-1}\to \frac{M_3 Q_{1,N-1}}{M_4},Q_{2,N-1}\to \frac{M_4 Q_{2,N-1}}{M_3}\big\},\nn\\
	&\mathbf{W}_4=\big\{Q_{1,1}\to \frac{M_5 Q_{1,1}}{M_4},Q_{1,N-1}\to \frac{M_5 Q_{1,N-1}}{M_4}\big\},\nn\\
	&\mathbf{W}_5=\big\{Q_{3,N-1}\to \frac{M_5 Q_{3,N-1}}{M_6}\big\},\nn\\
	&\mathbf{W}_6=\big\{Q_{3,N-1}\to \frac{Q_{3,N-1}}{M_6 M_7},Q_{4,1}\to M_6 M_7 Q_{4,1}\big\},\nn\\
	&\mathbf{W}_7=\big\{A_{2 N-1}\to \frac{A_{2 N-1}}{\sqrt[4]{M}},Q_{1,N-1}\to \frac{Q_{1,N-1}}{\sqrt{M}},Q_{2,N-1}\to \frac{Q_{2,N-1}}{\sqrt{M}},Q_{3,N-1}\to \frac{Q_{3,N-1}}{\sqrt{M}},\nn\\
	&\phantom{\mathbf{W}_7=\big\{}Q_{4,1}\to \sqrt{M} Q_{4,1}\big\},\nn\\
	&\mathbf{W}_8=\big\{Q_{1,1}\to \frac{M_6 Q_{1,1}}{M_7},Q_{1,N-1}\to \frac{M_6 Q_{1,N-1}}{M_7},Q_{3,N-1}\to \frac{M_7 Q_{3,N-1}}{M_6},\nn\\
	&\phantom{\mathbf{W}_8=\big\{}Q_{4,1}\to \frac{M_7 Q_{4,1}}{M_6}\big\}. 
\end{align}

\begin{alignat}{2}
	&\tilde{Q}_{1,1}=\frac{\Theta  \sqrt{M_1 M_2 M_3 M_5 M_6} Q_{1,1}}{\sqrt{M_4 M_7 M_8^3}},\quad &&\tilde{Q}_{1,N-1}=\frac{\Theta ^2 \sqrt{M_1 M_3 M_5 M_6} Q_{1,N-1}}{\sqrt{M_2 M_4 M_7 M_8^5}},\nn\\
	&\tilde{Q}_{2,1}=Q_{2,1},&&\tilde{Q}_{2,N-1}=\frac{\Theta  Q_{2,N-1}}{M_3 M_8},\nn\\
	&\tilde{Q}_{3,1}=Q_{3,1},&&\tilde{Q}_{3,N-1}=\frac{\Theta  Q_{3,N-1}}{M_6 M_8},\nn\\
	&\tilde{Q}_{4,1}=\frac{M_7 M_8 Q_{4,1}}{\Theta },&&\tilde{Q}_{4,N-1}=Q_{4,N-1}. 
\end{alignat}

\paragraph{3. $\mathbf{V}_{1},\mathbf{V}_{2},\mathbf{V}_{3},\mathbf{V}_{6},\mathbf{V}_{7},\mathbf{V}_{9}$}
\begin{align}
	&\mathbf{W}_0=\big\{Q_{1,1}\to \frac{M_1 Q_{1,1}}{M_8 q},Q_{1,N-1}\to \frac{M_1 Q_{1,N-1}}{M_8 q}\big\},\nn\\
	&\mathbf{W}_1=\big\{Q_{1,N-1}\to \frac{M_1 Q_{1,N-1}}{M_2}\big\},\nn\\
	&\mathbf{W}_2=\big\{Q_{1,N-1}\to \frac{M_3 Q_{1,N-1}}{M_2},Q_{2,N-1}\to \frac{M_2 Q_{2,N-1}}{M_3}\big\},\nn\\
	&\mathbf{W}_3=\big\{Q_{1,1}\to \frac{M_3 Q_{1,1}}{M_4},Q_{1,N-1}\to \frac{M_3 Q_{1,N-1}}{M_4},Q_{2,N-1}\to \frac{M_4 Q_{2,N-1}}{M_3}\big\},\nn\\
	&\mathbf{W}_4=\big\{Q_{1,1}\to \frac{M_5 Q_{1,1}}{M_4},Q_{1,N-1}\to \frac{M_5 Q_{1,N-1}}{M_4}\big\},\nn\\
	&\mathbf{W}_5=\big\{Q_{1,1}\to \frac{M_5 Q_{1,1}}{M_6},Q_{1,N-1}\to \frac{M_5 Q_{1,N-1}}{M_6},Q_{3,1}\to \frac{M_6 Q_{3,1}}{M_5}\big\},\nn\\
	&\mathbf{W}_6=\big\{Q_{3,1}\to M_6 M_7 Q_{3,1},Q_{4,N-1}\to \frac{Q_{4,N-1}}{M_6 M_7}\big\},\nn\\
	&\mathbf{W}_7=\big\{A_{2 N-1}\to \frac{A_{2 N-1}}{\sqrt[4]{M}},Q_{1,N-1}\to \frac{Q_{1,N-1}}{\sqrt{M}},Q_{2,N-1}\to \frac{Q_{2,N-1}}{\sqrt{M}},Q_{3,1}\to \sqrt{M} Q_{3,1},\nn\\
	&\phantom{\mathbf{W}_7=\big\{}Q_{4,N-1}\to \frac{Q_{4,N-1}}{\sqrt{M}}\big\},\nn\\
	&\mathbf{W}_8=\big\{Q_{1,1}\to \frac{M_7 Q_{1,1}}{M_6},Q_{1,N-1}\to \frac{M_7 Q_{1,N-1}}{M_6},Q_{3,1}\to \frac{M_6 Q_{3,1}}{M_7},\nn\\
	&\phantom{\mathbf{W}_8=\big\{}Q_{4,N-1}\to \frac{M_6 Q_{4,N-1}}{M_7}\big\}. 
\end{align}

\begin{alignat}{2}
	&\tilde{Q}_{1,1}=\frac{\Theta  \sqrt{M_1 M_2 M_3 M_5 M_7} Q_{1,1}}{\sqrt{M_4 M_6 M_8^3}},\quad &&\tilde{Q}_{1,N-1}=\frac{\Theta ^2 \sqrt{M_1 M_3 M_5 M_7} Q_{1,N-1}}{\sqrt{M_2 M_4 M_6 M_8^5}},\nn\\
	&\tilde{Q}_{2,1}=Q_{2,1},&&\tilde{Q}_{2,N-1}=\frac{\Theta  Q_{2,N-1}}{M_3 M_8},\nn\\
	&\tilde{Q}_{3,1}=\frac{M_6 M_8 Q_{3,1}}{\Theta },&&\tilde{Q}_{3,N-1}=Q_{3,N-1},\nn\\
	&\tilde{Q}_{4,1}=Q_{4,1},&&\tilde{Q}_{4,N-1}=\frac{\Theta  Q_{4,N-1}}{M_7 M_8}. 
\end{alignat}

\bibliographystyle{JHEP}
\bibliography{ref}
\end{document}